%% file: diss.tex
\documentclass[reqno]{amsart}
\usepackage[T1]{fontenc}
\usepackage[cp850]{inputenc}
\usepackage{eucal}
\usepackage{amsfonts}
\usepackage{amssymb}
\input{mak}
\begin{document}
\numberwithin{equation}{section}
\begin{titlepage}
  \begin{center}
    \vspace*{5cm}
    {\Huge Gauge theories in\vspace{.7cm}\\
    local causal perturbation theory}
    \vfill
    {\Large
    Dissertation\\
    zur Erlangung des Doktorgrades\\
    des Fachbereiches Physik\\
    der Universit\"at Hamburg
    \vfill
    vorgelegt von\\
    Franz-Marc Boas\\
    aus Hildesheim
    \vfill
    Hamburg\\
    1999}
  \end{center}
\end{titlepage}
\setcounter{page}{2}
\begin{center} {\bf Abstract}\end{center}
In this thesis quantum gauge theories are considered in the framework of 
local, causal perturbation theory. Gauge invariance is described in terms 
of the BRS formalism. Local interacting field operators are constructed 
perturbatively and field equations are established. A nilpotent BRS 
transformation is defined on the local algebra of fields. It allows the 
definition of the algebra of local observables as an operator cohomology.
This algebra of local observables can be represented in a Hilbert space.\\
The interacting field operators are defined in terms of time ordered 
products of free field operators. For the results above to hold the time
ordered products must satisfy certain normalization conditions. To 
formulate these conditions also for field operators that contain a 
spacetime derivative a suitable mathematical description of time ordered 
products is developed. \\
Among the normalization conditions are Ward identities for the ghost 
current and the BRS current. The latter are generalizations of a 
normalization condition that is postulated by D\"utsch, Hurth, Krahe and 
Scharf for Yang-Mills theory. It is not yet proven that this condition 
has a solution in every order. All other normalization conditions can be 
accomplished simultaneously.\\
A principle for the correspondence between interacting quantum fields and 
interacting classical fields is established. Quantum electrodynamics and 
Yang-Mills theory are examined and the results are compared with the 
literature.

\vfill
\begin{center} {\bf Zusammenfassung}\end{center}
In dieser Arbeit werden Quanten-Eichtheorien im Rahmen der lokalen, 
kausalen St\"orungstheorie behandelt. Eichinvarianz wird mit Hilfe 
des BRS-For\-ma\-lis\-mus beschrieben. Lokale, wechselwirkende 
Feldoperatoren werden st\"orungstheoretisch konstruiert und 
Feldgleichungen zwischen ihnen werden hergeleitet. Eine nilpotente 
BRS-Transformation wird auf der lokalen Feld-Algebra definiert. Sie 
gestattet die Definition der lokalen Observablen-Algebra als eine
Operator-Kohomologie. Diese lokale Observablen-Algebra besitzt eine
Hilbertraum-Darstellung. \\
Die wechselwirkenden Feldoperatoren werden mit Hilfe zeitgeordneter 
Produkte freier Feldoperatoren definiert. Damit die obigen Resultate 
gelten, m\"ussen die zeitgeordneten Produkte bestimmte 
Normierungsbedingungen erf\"ullen. Um diese Bedingungen 
auch f\"ur Felder mit Raum-Zeit-Ableitungen formulieren zu k\"onnen,
wird eine geeignete mathematische Beschreibung zeitgeordneter Produkte 
entwickelt. \\
Unter den Normierungsbedingungen sind Ward-Identit\"aten f\"ur den
Geist-Strom und den BRS-Strom. Letztere sind Verallgemeinerungen 
einer Normierungsbedingung, die D\"utsch, Hurth, Krahe und Scharf 
f\"ur die Yang-Mills-Theorie fordern. Es ist noch nicht bewiesen, da{\ss}
diese Bedingung in jeder Ordnung eine L\"osung besitzt. Alle anderen
Normierungsbedingungen k\"onnen gleichzeitig erf\"ullt werden. \\
Ein Prinzip f\"ur die Korrespondenz zwi\-schen wechselwirkenden 
Quantenfeldern und wechselwirkenden klassischen Feldern wird 
aufgestellt. Quanten-Elektrodynamik und Yang-Mills-Theorie werden 
untersucht, und die Ergebnisse werden mit der Literatur verglichen. 

\newpage
\tableofcontents
\setlength{\marginparwidth}{3cm}
\include{intro}
\include{alg}
\include{free}
\include{normal}

\include{cpt}
\include{nilpo}
\include{examples}
\include{summary}
\begin{appendix}
  \include{proofN4}
  \include{proofN5}
\end{appendix}

\bibliographystyle{amsalphap}
\bibliography{literatur}

\end{document}

%% file: mak.tex
\newcommand{\Diag}{\operatorname{Diag}}

\chardef\ii="10

\def\operatorname#1{{\mathchoice{\rm #1}{\rm #1}{\rm #1}{\rm #1}}}
\def\bbbone{{\mathchoice {\rm 1\mskip-4mu l} {\rm 1\mskip-4mu l}
{\rm 1\mskip-4.5mu l} {\rm 1\mskip-5mu l}}}
\newcommand{\one}{\bbbone}

\def\bbbr{{\rm I\!R}} 
\def\bbbn{{\rm I\!N}}

\def\pmb#1{\setbox0=\hbox{#1}
    \kern-.025em\copy0\kern-\wd0
    \kern.05em\copy0\kern-\wd0
    \kern-.025em\raise.0433em\box0}
\def\vec#1{\mathchoice{\mbox{\boldmath$\displaystyle\bf#1$}}
{\mbox{\boldmath$\textstyle\bf#1$}}
{\mbox{\boldmath$\scriptstyle\bf#1$}}
{\mbox{\boldmath$\scriptscriptstyle\bf#1$}}}
\def\bbbc{{\mathchoice {\setbox0=\hbox{$\displaystyle\rm C$}\hbox{\hbox
to0pt{\kern0.4\wd0\vrule height0.9\ht0\hss}\box0}}
{\setbox0=\hbox{$\textstyle\rm C$}\hbox{\hbox
to0pt{\kern0.4\wd0\vrule height0.9\ht0\hss}\box0}}
{\setbox0=\hbox{$\rm\scriptstyle C$}\hbox{\hbox
to0pt{\kern0.4\wd0\vrule height0.9\ht0\hss}\box0}}
{\setbox0=\hbox{$\scriptscriptstyle\rm C$}\hbox{\hbox
to0pt{\kern0.4\wd0\vrule height0.9\ht0\hss}\box0}}}}
\def\1#1{{\mathbb #1}}
\def\bbbz{{\mathchoice {\hbox{$\sf\textstyle Z\kern-0.4em Z$}}
{\hbox{$\sf\textstyle Z\kern-0.4em Z$}}
{\hbox{$\sf\scriptstyle Z\kern-0.3em Z$}}
{\hbox{$\sf\scriptscriptstyle Z\kern-0.2em Z$}}}}
\newcommand{\R}{\bbbr}
\newcommand{\C}{\bbbc}
\newcommand{\Z}{\bbbz}
\newcommand{\N}{\bbbn}

\newcommand{\cinfty}{C^{\infty}}

\newcommand{\dd}{\partial}

\newcommand{\half}{\frac{\scriptstyle 1}{\scriptstyle 2}}

\DeclareSymbolFontAlphabet{\mathbm}{AMSb} 

\newcommand{\cD}{{\CMcal D}}
\newcommand{\cF}{{\CMcal F}}
\newcommand{\cG}{{\CMcal G}}\newcommand{\cH}{{\CMcal H}}

\newcommand{\cL}{{\CMcal L}}
\newcommand{\cM}{{\CMcal M}}
\newcommand{\cO}{{\CMcal O}}\newcommand{\cP}{{\CMcal P}}

\newcommand{\cS}{{\CMcal S}}
\newcommand{\cV}{{\CMcal V}}
\newcommand{\cW}{{\CMcal W}}

\newcommand{\eA}{{\mathcal A}}
\newcommand{\eD}{{\mathcal D}}

\newcommand{\eH}{{\mathcal H}}

\newcommand{\eP}{{\mathcal P}}
\newcommand{\eR}{{\mathcal R}}

\newcommand{\fL}{{\mathfrak L}}

\newcommand{\fP}{{\mathfrak P}}

\newcommand{\un}{{\underline{n}}}

\newcommand{\vecp}{{\vec{p}}}

\newcommand{\lra}{\Longrightarrow}
\newcommand{\llra}{\Longleftrightarrow}
\newcommand{\TT}[1]{T\left(#1 \right)}
\newcommand{\TTb}[1]{\overline{T}\left(#1 \right)}
\newcommand{\TTn}[1]{T^0\left(#1 \right)}
\newcommand{\RR}[2]{R\left(#1;#2 \right)}

\renewcommand{\AA}[2]{A\left(#1;#2 \right)}

\newcommand{\DD}[2]{D\left(#1;#2 \right)}

\newcommand{\lm}{\lambda}

\newcommand{\al}{\alpha}
\newcommand{\om}{\omega}
\newcommand{\de}{\delta}
\newcommand{\Lm}{\Lambda}
\newcommand{\De}{\Delta}
\newcommand{\comm}[2]{\left[#1,#2\right]_-}
\newcommand{\acomm}[2]{\left\{#1,#2\right\}_+}
\newcommand{\scomm}[2]{\left[#1,#2\right]_{\mp}}
\newcommand{\norm}[1]{\left\|#1\right\|}
\newcommand{\betr}[1]{\left| #1 \right|}
\newcommand{\set}[1]{\left\{#1\right\}}
\newcommand{\state}[1]{\left|#1\right\rangle}
\newcommand{\costate}[1]{\left\langle#1\right|}

\newcommand{\SP}[2]{\left( #1,#2\right)}
\newcommand{\SPK}[2]{\left\langle #1,#2\right\rangle}

\newcommand{\beq}{\begin{equation}}
\newcommand{\eeq}{\end{equation}}
\newcommand{\beqa}{\begin{eqnarray}}
\newcommand{\eeqa}{\end{eqnarray}}
\DeclareMathOperator{\im}{im}

\DeclareMathOperator{\supp}{supp}

\DeclareMathOperator{\scaldeg}{sd}
\DeclareMathOperator{\End}{End}
\DeclareMathOperator{\Dist}{Dist}
\newcommand{\Ad}[2]{\left(\operatorname{Ad}\,\, #1({#2})\right)}
\newcommand{\funcder}[2]{\frac{\delta #1}{\delta #2}}
\newcommand{\partder}[2]{\frac{\partial #1}{\partial #2}}

\newcommand{\tu}{\tilde{u}}
\newcommand{\tphi}{\tilde{\varphi}}
\newcommand{\tpsi}{\tilde{\psi}}
\newcommand{\tchi}{\tilde{\chi}}
\newcommand{\psiq}{\overline{\psi}}
\newcommand{\vp}{\varphi}

\newcommand{\bT}{\overline{T}}
\newcommand{\ctF}{\widetilde{\cF}}
\newcommand{\tk}{\widetilde{k}}
\newcommand{\tQ}{\widetilde{Q}}
\newcommand{\tB}{\widetilde{B}}
\newcommand{\ts}{\widetilde{s}}
\newcommand{\tj}{\widetilde{\jmath}}
\newcommand{\tC}{\widetilde{\C}}
\newcommand{\tone}{\widetilde{\one}}
\newcommand{\ta}{\widetilde{a}}
\newcommand{\tb}{\widetilde{b}}
\newcommand{\tc}{\widetilde{c}}
\newcommand{\tV}{\widetilde{\cV}}
\newcommand{\tH}{\widetilde{\eH}}

\newcommand{\tA}{\widetilde{A}}

\newcommand{\teA}{\widetilde{\eA}}
\newcommand{\defined}{\stackrel{\text{def}}{=}}
\newcommand{\spacel}{\,\rangle\hspace{-2pt}\langle\,}
\newcommand{\inter}[3]{\left(#1\right)_{\rm int}^{#2}\left(#3\right)}
\newcommand{\class}[2]{\left(#1\right)_{\rm cl}\left(#2\right)}
\newcommand{\classa}[1]{\left(#1\right)_{\rm cl}}

\newcommand{\forlc}{\overline{V}_+}
\newcommand{\backlc}{\overline{V}_-}
\newcommand{\lc}{\overline{V}}
\newcommand{\nth}{\mbox{$n^{\mbox{\scriptsize th}}$ }}

\newcommand{\hp}{\hat{p}}
\newcommand{\hq}{\hat{q}}
\newcommand{\vps}{\vphantom{\sum}}
\newcommand{\vblock}{\vphantom{\begin{split}a\\\ a \end{split}}}
\newcommand{\dddag}{\partial\!\!\!/ }

\newcommand{\kom}[1]{}
\newcommand{\beql}[1]{ \begin{equation} \label{#1}}
\newcommand{\beqal}[1]{\marginpar{{\it \footnotesize #1}} \begin{eqnarray} 
\label{#1}}


%% file: intro.tex
\section{Introduction}

Four fundamental interactions in nature are known today: Gravitation,
electrodynamics, weak and strong nuclear forces. The latter three are
in present day elementary particle physics successfully described by
quantum gauge field theories. Successfully means in this context that
there are no experimental data that do not agree with the predictions 
of these theories and that the agreement is very good e.g.\ in quantum
electrodynamics (QED). 
The general theory of relativity describes gravitation classically. It
is also a gauge theory in a wider sense of the word. A sound quantum 
theory for gravitation is still missing.\\
The distinguishing feature of these gauge theories is their gauge group:
$SU(2)\times U(1)$ for the combined theory of electric and weak 
interactions and $SU(3)$ for the strong interaction. Both gauge groups
are non-Abelian Lie groups. Therefore a comprehensive understanding of
non-Abelian quantum gauge theories is needed to understand nature at 
the quantum level. \\
Originally the conceptual and mathematical framework of quantum field 
theory was developed for Abelian theories and in particular for QED. 
This was already an established theory in 
perfect agreement with the experimental data when physicists directed 
their attention towards non-Abelian gauge theories. They realized that
quantum field theory required a modification of its mathematical 
description before it could be applied to non-Abelian theories.\\
The first study of a non-Abelian model --- motivated by the isospin 
$SU(2)$ group --- which attained wide reception was done by Yang and Mills 
\cite{YanMil54} in 1954\footnote{The first who studied non-Abelian models 
was O. Klein in 1938 \cite{Kle38}}. The
interest of elementary particle physicists in non-Abelian quantum field
theories grew strongly when in the next two decades several other such 
models were proposed to explain various phenomena. These include e.g. 
the Salam--Weinberg model \cite{Sal68,Wei67} and the $SU(3)$ colour model 
for the strong interaction \cite{GM64,Zwe64}, but also attempts to 
quantize gravitation, e.g. \cite{Fey63} or \cite{DeW67a,DeW67b,DeW67c}.\\
There was a series of obstacles to a satisfactory quantum theory for 
non-Abelian gauge theories due to the self coupling of the gauge bosons. 
A naive application of the methods developed for QED leads to serious 
difficulties, like an S-matrix that fails to be unitary 
\cite{Fey63,DeW67b}. \\
A major step to overcome these obstacles was made by Faddeev and 
Popov \cite{FadPo67,Fad69}. They defined a unitary S-matrix in the 
functional 
integral approach, but for that they had to introduce unphysical fields 
that violate the spin-statistics theorem --- the famous Faddeev Popov 
ghosts. In the mid seventies Becchi, Rouet and Stora 
\cite{BecRouSto74,BecRouSto76} and 
independently of them Tyutin \cite{Tyu75} found that the Faddeev Popov 
Lagrangian is invariant under a rigid symmetry transformation that 
mixes the ghosts with the other fields --- the BRS transformation. 
Kugo and Ojima \cite{KugOji79} gave an operator formulation\footnote{Curci
and Ferrari \cite{CurFer75} gave already an operator formulation, but 
they postulated wrong hermiticity properties for the ghosts} for this BRS 
theory, and Scharf and collaborators \cite{SchHuDueKra94a} found with 
the operator gauge invariance a criterion of BRS symmetry for operator 
theories that needs no recurrence to an underlying classical 
theory\footnote{Originally operator gauge invariance was postulated for
theories of the Yang-Mills type. Recent results of Scharf and Wellmann
\cite{SchWel99} that it also a suitable criterion for spin two models}.\\
Quantum field theory is plagued with two sources of possible infinities: 
the ultraviolet and the infrared divergences. Ultraviolet divergences 
are due to the distributional character of the field operators. In 
perturbation theory they can be removed by numerous renormalization
procedures --- so they are under control in this framework. 
Unfortunately these renormalization procedures are not unique --- there 
remains the freedom of finite renormalization. \\
Infrared divergences occur since the asymptotic behaviour of incoming 
and outgoing interacting fields is not under control. This problem is 
particularly severe for non-Abelian gauge theories. It may be overcome 
by a replacement of the coupling constant by a spacetime dependent 
switching function so that the theory becomes free at finite times in 
the past and in the future. But as the real physical coupling is constant,
one must in general perform the adiabatic limit, i.e. let the switching 
function tend to a constant. This limit does not exist in general. \\
In QED the infrared divergences are logarithmic, and Blanchard and 
Seneor \cite{BlaSen75} proved that the adiabatic limit exists for Green's 
and Wightman functions. Unfortunately this is no longer true for 
non-Abelian theories. Their infrared behaviour is in general 
worse. \\
For strongly interacting fields this comes from the experimental 
observation of confinement. This means the fact that strongly interacting 
particles always combine to hadrons. Even after a high energy 
scattering process that breaks up the hadron structure the particles
recombine immediately into new hadrons (hadronization). So the fields 
are not asymptotically free but constitute bound states. Moreover
confinement cannot be described perturbatively. \\
In the electroweak theory confinement does not occur, but the 
model contains unstable, observable particles --- the vector bosons
$W^\pm$ and $Z$. These cannot occur as asymptotic states. \\ 
A solution for the infrared problem is to consider local 
theories, i.e. theories where all fields are localized in 
a finite region of spacetime. If the coupling is constant within this 
region and if the algebra of fields remains unaltered when the coupling 
is modified outside that region, the adiabatic limit needs not to be 
performed. Brunetti and Fredenhagen \cite{BruFre96} proved that such
a modification induces merely a unitary transformation on the 
algebra of fields. So the physical content is not changed by that 
modification and there is consequently no need for the adiabatic limit.
Therefore no infrared problems occur in the construction of the 
local algebras.\\
One common problem of gauge theories --- already encountered in QED ---
is that the algebra of fields must be quantized in an indefinite inner 
product space. Therefore positivity must be assured, i.e. the algebra 
of observables must be non trivially represented in a 
Hilbert space. D\"utsch and Fredenhagen \cite{DueFre98} succeeded in 
proving positivity for perturbation theories quantized in the BRS 
framework, provided the underlying free theory is also positive. In 
their view the interacting theory is regarded as a deformation of that 
underlying free model. They also constructed a local perturbation 
theory for QED. \\
The first to examine Yang-Mills theories in the causal framework
were Scharf and collaborators \cite{SchHuDueKra94a} - 
\cite{SchHuDue95b}, see also \cite{Sch95}. They investigated 
the operator gauge invariance in the Yang-Mills case and found 
that it can be accomplished, provided a weak assumption concerning
the infrared behaviour of the Green's functions is fulfilled.\\
The aim of this thesis is to construct local perturbative gauge 
theories as operator theories in the BRS framework. The design is as 
general as possible, the motivation is Yang-Mills theory which
serves as an example throughout the thesis. \\
The result is that the construction can always be performed, provided
the generalized operator gauge invariance holds. It could not be proven 
that the latter can be accomplished in general models. A similar set
of equations are the {\em descent equations} in the framework of 
algebraic renormalization --- see, e.g. \cite{PigSor95}. It may be 
possible to prove generalized operator gauge invariance by
translating these results into our language, but this seems to be a
tedious task and is not done here.\\
We use the renormalization scheme of causal perturbation theory as it
was developed by Epstein and Glaser \cite{EpGlas73} following ideas 
proposed by Bogoliubov, Shirkov \cite{BogShi59} and St\"uckelberg. 
It avoids divergent expressions throughout the entire procedure. Scharf 
and collaborators as well as D\"utsch and Fredenhagen formulated their 
results in the same framework. This makes it easy to use their results 
for our construction and to compare them with our results. \\
Moreover our approach is local in order to avoid infrared divergences and
to be able to define observables and physical states. \\
Like D\"utsch and Fredenhagen we use normalization conditions for the time
ordered products as an essential tool to establish desired relations in
the field algebra. Their normalization conditions are generalized 
to include fields that contain a spacetime derivative. Ward identities 
for the ghost and BRS 
current are introduced as new normalization conditions with regard to the 
definition of observables and physical states. We introduce an 
algebra of auxiliary variables for the fields containing a spacetime 
derivative and define a 
linear representation of the polynomials in this algebra as operators 
acting on the Fock space. We present a reformulation of time ordering. 
It is formally a multi linear generalization of the linear representation 
mentioned above to multiple arguments. This and the definition of propagator 
functions for the fields with a spacetime derivative allows us to 
generalize the normalization conditions in the desired manner. It is 
proven that all these conditions --- except the BRS Ward identities --- 
can be accomplished simultaneously. The existence of a solution for the 
BRS Ward identities and its compatibility with the other conditions 
must be proven in individual models. The proof for QED is presented.\\
There are relations for the local field algebra that are determined by 
the normalization conditions, e.g.\ renormalized field equations and the 
BRS algebra. The latter allows for a definition of observables and a 
construction of a positive physical state space. \\
The thesis is organized as follows: In chapter (\ref{alg}) we set up the 
algebraic framework of BRS theory, following Kugo and Ojima 
\cite{KugOji79}. The definition of observables and the construction
of the Hilbert space are performed using certain algebraic relations 
between the interacting operators. The rest of the thesis
will be devoted to the construction of models in which these 
relations hold.\\
In chapter (\ref{free}) the free model underlying our perturbation 
theory is put up. The algebra of auxiliary variables is constructed and
its linear representation as Fock space operators is defined. Then the
propagator functions are examined. Finally the proof of Razumov and Rybkin
\cite{RazRyb90} for the positivity of theories with certain BRS charges is 
presented. \\
The new definition of time ordering is given in chapter (\ref{normal}).
It contains also the formulation of six normalization conditions 
and the proof that the first five have simultaneous solutions. 
The sixth, the BRS Ward identities, is shown to be equivalent with a 
generalized version of operator gauge invariance.\\
Local causal perturbation theory is introduced in chapter (\ref{cpt})
along the lines of Epstein, Glaser \cite{EpGlas73}, D\"utsch and 
Fredenhagen \cite{DueFre98}. Conditions for a polynomial to be 
a candidate for a Lagrangian are given. \\
The local field algebra is constructed in chapter (\ref{nilpo}). The 
conserved currents and charges, the ghost number of an interacting field 
and the interacting BRS transformation are defined, field equations 
and the BRS algebra are derived. The chapter concludes with a reflection 
on the correspondence between the quantum theory defined above and its 
classical counterpart. \\
The inspection of gauge theories is deepened in chapter (\ref{examples}) 
for two exemplary models: QED and Yang-Mills theory. The BRS Ward 
identities are proven for QED, and we compare the relations between
the interacting fields with those between the corresponding classical 
fields. \\
At the end a conclusion and an outlook for possible further developments
are included.


%% file: alg.tex

\section{BRS theory --- algebraic considerations}
\label{alg}
In this chapter canonical BRS theory according to Kugo and Ojima
\cite{KugOji79} is carried through on a purely algebraic level. The 
availability of suitable BRS and ghost charges is formulated as 
assumptions. Then perturbative theories --- i.e. theories where the 
operators and the state vectors are formal power series --- are 
examined in this framework. D\"utsch and Fredenhagen 
\cite{DueFre98,DueFre98a} prove that the positivity structure of a 
theory can be maintained during deformation. Their proof is presented 
here. 

\subsection{Why BRS theory?}
All quantum gauge theories share one common difficulty: There is no 
positive definite Hilbert space in which the field algebra can be 
represented and which possesses a nontrivial unitary representation 
of the Poincar\'e group. Nakanishi and Ojima \cite{NakOji90} proved 
that there exists no nontrivial Hilbert space representation for 
manifestly covariant theories with massless gauge bosons. This could 
be circumvented by non covariant gauges, but this means abandoning 
manifest covariance. \\
The field algebra is not observable, so a direct physical interpretation
of the theory which requires a Hilbert space representation is not 
possible. But the algebra of observables must have a Hilbert space 
representation, and the Hilbert space must carry a unitary representation
of the Poincar\'e group. \\
For QED Gupta \cite{Gup50} and Bleuler \cite{Bleu50} found an elegant 
way out of this dilemma. They retain manifest covariance at the 
prize of representing the field algebra in an indefinite inner product 
space. Then there exists a non trivial, pseudo unitary\footnote{Pseudo 
unitary, pseudo hermitian etc. means unitary, hermitian etc. w.r.t. the 
indefinite inner product.} representation of the Poincar\'e group. 
This space is too big: It contains vectors 
with negative norm that have no physical interpretation --- they would
lead to negative transition probabilities. Consequently the physical 
state vectors form a distinguished proper subspace of the inner product 
space. This subspace is selected by a linear subsidiary condition, and
it is found to be positive semidefinite. It becomes a Hilbert space 
with unitary action of the Poincar\'e group when all state vectors 
differing by a zero norm vector are identified with each other and the 
space is subsequently completed. \\
Unfortunately this strategy breaks down in non Abelian gauge theories 
because there is no appropriate subsidiary condition available. This
is due to the nonlinear self interaction of the gauge fields.\\
BRS theory is a solution for that problem. The canonical BRS 
formalism of Kugo and Ojima \cite{KugOji79} follows the same ideas as
Gupta and Bleuler but it can also be applied to non-Abelian theories.  
Initially the algebra of fields is again represented in an indefinite 
inner product space. The presence of the ghosts in the BRS approach
makes it possible to define a suitable subsidiary condition for the 
physical subspace which is a Hilbert space. The formalism provides also 
a definition of an algebra of observables that is represented in this 
Hilbert space. There 
exists a pseudo unitary action of the Poincar\'e group on the indefinite 
space. This action is lifted to a unitary one on the Hilbert space. 

\subsection{Canonical BRS theory}
The construction starts in the following situation: There is an initial 
Hilbert space $\set{\cV,\SP{\,\cdot\,}{\,\cdot\,}}$ with a positive
scalar product $\SP{\,\cdot\,}{\,\cdot\,}$ that 
encompasses all fields including the unphysical ones (scalar vector 
bosons, ghosts etc.). This scalar product has no direct physical
meaning. It does not describe the transition amplitudes, in particular it 
is not
Poincar\'e covariant. The adjoint in this Hilbert space is denoted as 
${}^+$, i.e. $\SP{\phi}{A\psi} = \SP{A^+\phi}{\psi}$ for 
every\footnote{$\End(\cV)$ is the space of endomorphisms on $\cV$} 
$A \in \End{\cV}$. It is possible to find a Krein operator 
$J \in \End(\cV)$ in the Hilbert space with the following three 
properties:
\begin{itemize}
  \item $J$ is hermitian, i.e. $J^+= J$ 
  \item It is idempotent, i.e. $J^2= \one$
  \item It defines a new inner product on $\cV$ via
  \beq
    \SPK{\phi}{\psi} \defined  \SP{\phi}{J\psi}
  \end{equation}
  such that the new inner product is Poincar\'e covariant.
\end{itemize}
The new inner product is assumed to describe the correct transition 
probabilities. Therefore it is referred to as the physical inner 
product. The vector space $\cV$ forms a Krein space with the physical 
product $\SPK{\,\cdot\,}{\,\cdot\,}$. Since $\SP{\,\cdot\,}{\,\cdot\,}$ 
was not covariant while $\SPK{\,\cdot\,}{\,\cdot\,}$ was, $J=\one$ can
be excluded. Then the physical inner product is always indefinite, 
because there must exist a vector $\state{\phi}$ such
that $(\one - J)\state{\phi} \neq 0$, and then $(\one - J)\state{\phi}$ 
has negative norm. The adjoint w.r.t. the physical inner product is 
defined as an involution denoted by ${}^*$, namely $A^*\defined JA^+J$, 
such that $\SPK{\phi}{A\psi} = \SPK{A^*\phi}{\psi}$ for every 
$A \in \End{\cV}$. \\
For the canonical BRS theory the following assumption is essential:\\
{\bf A1}: There exists an operator $Q_B \in \End(\cV)$ --- the BRS charge 
--- with the following properties:
\begin{itemize}
  \item $Q_B$ is a conserved charge.
  \item It is pseudo hermitian, i.e. $(Q_B)^* = Q_B$.
  \item It is nilpotent\footnote{Nilpotent means throughout this
  thesis nilpotent of order two.}, i.e. $(Q_B)^2 = 0$.
  \item It annihilates the vacuum, i.e. $Q_B \state{\om} = 0$ 
\end{itemize}
where $\state{\om}$ is the vacuum vector. This assumption is highly non 
trivial, and the appearance of ghosts in $\cV$ is necessary for it. It 
has to be verified in the concrete model.\\
It is easily verified that the image of $Q_B$ contains only zero norm
vectors w.r.t. the physical scalar product:
\begin{equation}
  \SPK{Q_B \phi}{Q_B \phi} = \SPK{\phi}{(Q_B)^2 \phi} = 0 .
\end{equation}
With the second assumption a grading is introduced on $\cV$ by means of
the ghost charge $Q_c$.\\
{\bf A2}: There exists an operator $Q_c\in \End(\cV)$ --- the ghost charge 
--- with the following properties: 
\begin{itemize}
  \item $Q_c$ is a conserved charge.
  \item It is anti pseudo hermitian, i.e. $(Q_c)^* = - Q_c$.
  \item It has integer eigenvalues, i.e. 
  $Q_c \state{\psi} = q \state{\psi}\,\,\lra\,\,q\in \Z$.
  \item It satisfies the commutator relation
  $\comm{Q_c}{Q_B} = Q_B$.
  \item It annihilates the vacuum, i.e. $Q_c \state{\om} = 0$ .
\end{itemize}
The eigenvalue of a state vector w.r.t. the ghost charge is called its
ghost number. For the physical inner product of two vectors to be non 
zero they must have opposite ghost numbers: Let $Q_c \,\psi = q \,\psi$ 
and $Q_c \,\phi = p \,\phi$, then
\begin{equation}
  0 = \SPK{\psi}{Q_c\phi} - \SPK{\psi}{Q_c\phi}
  = \SPK{\psi}{Q_c\phi} + \SPK{Q_c\psi}{\phi}
  = (q+p)\SPK{\psi}{\phi},
\end{equation}
so $(q+p)=0$ or $\SPK{\psi}{\phi}=0$. This implies in particular that 
only states with vanishing ghost number can have non zero norm
w.r.t. the physical inner product.\\
The commutator relation $\comm{Q_c}{Q_B} = Q_B$ forms together with the 
nilpotency of the BRS charge, $(Q_B)^2 = 0$, the {\em BRS algebra}.\\ 
Like in the Gupta-Bleuler scheme the negative norm states are excluded 
by a subsidiary condition. The kernel of $Q_B$ is regarded as
a candidate for the physical Hilbert space. It contains 
necessarily zero norm states from the image of $Q_B$ --- due to 
$(Q_B)^2 = 0$ we have $\im Q_B \subset \ker Q_B$ --- and possibly also 
vectors with non vanishing ghost number. Therefore the following 
definition for the Hilbert space $\cH_{\rm ph}$ of physical 
state vectors is given:
\begin{equation}
  \cH_{\rm ph} \defined 
  \overline{(\ker Q_B,\cV)/(\im Q_B,\cV)
  }^{\norm{\cdot}}.
\end{equation}
Completion is understood in the norm topology. Now it must be verified
that the physical state vectors form a positive definite inner product
space. This is guaranteed if the following positivity assumption is
valid.\\
{\bf A3}: 
\begin{itemize}
  \item The kernel of $Q_B$ contains only positive semidefinite vectors, 
  i.e. $Q_B \state{\phi} = 0 \,\,\lra \,\,\SPK{\phi}{\phi} \geq 0$
  \item Its image encompasses all zero norm vectors in its kernel, 
  i.e.
  $\state{\phi} \in \ker (Q_B,\cV)$ and $\SPK{\phi}{\phi} = 0 
  \quad \lra \quad \state{\phi} \in (\im Q_B,\cV)$.
\end{itemize}
The second point guarantees in particular that all elements in 
$(\ker Q_B,\cV)$ with nonvanishing ghost number are in $(\im Q_B,\cV)$.
The scalar product is well defined on these equivalence classes, so it 
does not depend on the representative of a class:
\begin{equation}
  \SPK{\phi + Q_B\chi}{\psi} = \SPK{\phi}{\psi} + \SPK{\chi}{Q_B\psi}
  = \SPK{\phi}{\psi} . 
\end{equation}
It is also positive definite by construction --- if assumption {\bf A3}
holds ---, so the quotient space is a pre Hilbert space and becomes 
a Hilbert space after completion. The structure above is called a state 
cohomology.\\
The ghost charge induces a derivation on $\End(\cV)$,
\begin{equation}
  s_c(A) \defined \comm{Q_c}{A} \qquad \forall A \in \End(\cV).
\end{equation}
Its eigenvalue for an operator $A\in \End(\cV)$ is called the ghost 
number of $A$ and is always an integer. \\
The BRS charge induces an graded derivation on $\End(\cV)$, namely the 
BRS transformation\footnote{Here $\scomm{\,\cdot\,}{\,\cdot\,}$ denotes
the graded commutator. Suppose, $A, B \in \End{\cV}$ have ghost numbers
$a,b \in \Z$. Then $\scomm{A}{B} \defined AB - (-1)^{ab} BA$.}
\begin{equation}
  s(A) \defined \scomm{Q_B}{A} \qquad \forall A \in \End(\cV).
\end{equation}
It is nilpotent because $Q_B$ is also nilpotent and the Jacobi-identity 
holds for the graded commutators. \\
With these definitions the algebra of observables $\eA_{\rm ph}$ can be 
defined as
\begin{equation}
  \eA_{\rm ph} \defined 
  \left(\vps (\ker s,\End(\cV))\cap(\ker s_c,\End(\cV))\right)/
  \left(\vps (\im s,\End(\cV))\cap(\ker s_c,\End(\cV))\right).
\end{equation}
This structure is called an operator cohomology. Its elements are well
defined operators on $\cH_{\rm ph}$, i.e. 
$\eA_{\rm ph} (\ker Q_B,\cV) \subset (\ker Q_B,\cV)$ and 
$\eA_{\rm ph}\, [0] = [0]$, where $[0]$ is the equivalence class of 
zero.\\
There is a ${}^*$-involution induced on the algebra of observables by 
the ${}^*$-involution on the representatives. But unlike the original 
involution this one acts on operators on a Hilbert space, so the notions 
hermitian, unitary and so on must be used without the prefix pseudo. \\
There is also a unitary action of the Poincar\'e group defined on 
$\cH_{\rm ph}$, namely the lift of the initial pseudo-unitary action
on the representatives to the equivalence classes. This induces a 
unitary representation on $\cH_{\rm ph}$. \\
There is a physical interpretation available for the cohomologies.
Initially the model is not characterized in terms of the algebra of 
field operators described here but in terms of the sub algebra without
the ghosts --- these were only introduced to make possible the definition 
of the BRS charge. In the picture above physics is invariant under
local gauge transformations, i.e. gauge transformations generated by
spacetime dependent functions. Then the BRS transformation, 
restricted to the sub algebra, may be regarded as the infinitesimal 
local gauge transformation. The role of the spacetime dependent 
functions is played by the ghosts. For them the BRS transformation
is defined such that it is nilpotent on the entire algebra.\\
So the restriction to the kernel of $s$ singles out fields that are
invariant under infinitesimal gauge transformations. Fields in the same 
equivalence class are regarded as physically indistinguishable. In this
interpretation the physical Hilbert space contains equivalence 
classes of states that are invariant under infinitesimal gauge 
transformations.
\subsection{Interacting theories and deformation stability}\label{defstab}
In perturbation theory field operators are represented by 
formal power series of linear operators. This makes it necessary to 
recapitulate the BRS formalism for formal power series of state vectors
and operators, since e.g. the notion of positivity is not defined a
priori for formal power series. This situation has been examined by 
D\"utsch and Fredenhagen \cite{DueFre98a,DueFre98} and we present their 
results here. \\
In their picture the interacting theory is a deformation of an 
underlying free theory. In some models positivity --- i.e. assumption 
{\bf A3} --- can be proven by direct computation for the underlying 
free theory. D\"utsch and Fredenhagen found a construction for the 
deformed --- i.e. interacting --- state space such that positivity holds 
also there in a sense defined below. \\
In the interacting theory both the state space and the operators acting
on it are modules over the ring $\tC$ of formal power series of complex
numbers:
\begin{equation} 
  \tC \defined \set{\ta = \sum_{n=0}^{\infty} g^n a_n:\quad a_n \in \C}
\end{equation}
where $g$ is the deformation parameter. The element 
$\tone \defined (1,0,0,\dots)$ is the identity in this ring. An element 
$\ta \in \tC$ is only invertible\footnote{Bordemann and Waldmann 
\cite{BorWal96} consider Laurent series instead. These are invertible
if $\ta \neq 0$, so they form a field.} if $a_0\neq 0$. The interacting 
indefinite inner product space is defined as the $\tC$-module
$\tV\defined \set{\tpsi = \sum_n g^n\,\psi_n\!: \,\,\, \psi_n 
\in \cV}$ 
which has the inner product $\SPK{\cdot}{\cdot}$ induced from $\cV$.
For $\tpsi = \sum_n g^n\,\psi_n$ and $\tchi = \sum_n g^n\,\chi_n$ this
means
\begin{equation}
  \begin{split}
    &\SPK{\cdot}{\cdot}: \quad \tV \times \tV \to \tC \\
    &\SPK{\tpsi}{\tchi} = \sum_n g^n \left(\sum_{k=1}^n 
    \SPK{\psi_k}{\chi_{n-k}} \right).
  \end{split}
\end{equation}
This is sesquilinear in $\tC$, i.e. $\SPK{\ta \tchi}{\tb \tpsi} =
\ta^*\tb \SPK{\tchi}{\tpsi}$. The ${}^*$ means complex conjugation, 
where the deformation parameter $g$ is real, so
\begin{equation}
  \ta^* = \sum_{n=0}^{\infty} g^n \overline{a}_n
\end{equation}
where $\overline{\phantom{a}}$ denotes complex conjugation in $\C$. \\
The operators in $\End(\tV)$ acting on $\tV$ can be written as
\begin{equation}
  \End(\tV) = \set{\tA = \sum_n g^n A_n: \quad A_n \in \End(\cV)}
\end{equation}
and form a $\tC$-module, too. The multiplication law in this algebra
is
\begin{equation}
  \tA \cdot \tB = \sum_n g^n 
  \left( \sum_{k=1}^n A_k\cdot B_{n-k}\right) 
  \qquad \tA,\tB \in \End(\tV).
\end{equation}
The interacting BRS-charge and the interacting ghost charge are 
such operators,
\begin{equation}
  \begin{split}
    &\tQ_B = \sum_n g^n\,Q_{B,n}, \qquad Q_{B,n} \in \End(\cV) \\
    \text{and}\qquad 
    &\tQ_c = \sum_n g^n\,Q_{c,n}, \qquad Q_{c,n} \in \End(\cV),
  \end{split}
\end{equation}
where $\tQ_{B,0}$ and $\tQ_{c,0}$ agree with the free charges. $\tQ_B$ 
must be chosen such that it is nilpotent, $\tQ_B^2=0$, and pseudo 
hermitian, $(\tQ_B)^* = \tQ_B$, and $\tQ_c$ must be anti pseudo 
hermitian, $(\tQ_c)^* = -\tQ_c$. The involution $^*$ is the one induced 
from $\End(\cV)$. The charges must satisfy the BRS algebra 
$\comm{\tQ_c}{\tQ_B}=\tQ_B$.\\ 
The interacting state space can be defined as in the general case,
\begin{equation}
  \tH_{\rm ph} \defined 
  (\ker \tQ_B,\tV)/(\im \tQ_B,\tV) ,
\end{equation}
with the only difference that the space is not completed since
there is no convenient topology in the space of formal power 
series. \\
The question is whether this space has a positive scaler product, 
and above all what positivity means for formal power series. \\
Following D\"utsch and Fredenhagen \cite{DueFre98} we adopt here 
Steinmann's \cite{Ste89} point of view\footnote{Here Bordemann and 
Waldmann \cite{BorWal96} follow again a different prescription: They 
define a real formal power series as positive if its first non vanishing 
coefficient is a positive number. With this definition the field of real
Laurant series becomes ordered. The notion of positivity presented here 
is a stricter one: Every positive series in Steinmann's sense is also 
positive in their sense, but not converse.} that a formal power series 
$\tb = \sum_{n} b_n g^n \in \tC$ is {\em positive} if it is the 
absolute square of another power series $\tc \in \tC$, i.e.
$\tb = \tc^*\tc$. D\"utsch and Fredenhagen define also that a class of 
state vectors $[\tphi] \in \tH_{\rm ph}$ can be normalized if there 
exists an $\ta \in \tC$ and $[\tpsi] \in \tH_{\rm ph}$ such that 
$[\tphi] = \ta \,[\tpsi]$ and $\SPK{[\tpsi]}{[\tpsi]}= \tone$. \\
With these notions of positivity and normalizability they prove in 
\cite{DueFre98} the following results:\\
Let the positivity assumption {\bf A3} be fulfilled for the undeformed
theory. Then
\begin{eqnarray}
  (i)&& \SPK{\tpsi}{\tpsi} \geq 0 \qquad \forall \,\tpsi 
  \in (\ker \tQ_B,\tV)\\
  (ii)&& \tpsi \in \left(\ker \tQ_B,\tV\right) \,\wedge \,
  \SPK{\tpsi}{\tpsi} = 0 \quad \lra \quad 
  \tpsi \in \left(\im \tQ_B,\tV \right) ,\\
  (iii)&& \forall \,\psi \in (\ker Q_B,\cV)\qquad 
  \exists \tpsi \in (\ker \tQ_B,\tV): \quad (\tpsi)_0 = \psi
  \vphantom{\SPK{\tpsi}{\tpsi}}\\
  (iv)&& \text{Every }[\tpsi] \neq 0 \in \tH_{\rm ph} \text{ is 
  normalizable in the sense above}. \vphantom{\SPK{\tpsi}{\tpsi}}
\end{eqnarray}
For the proofs of these results we refer to their article. \\
So assumption {\bf A3} is fulfilled for the interacting theory if it
is fulfilled for the free theory underlying it. Therefore the 
interacting physical state space defined above is a pre Hilbert space. 
Result $(iii)$ implies that an interacting vacuum state 
$\state{\tilde{\om}}$ can be defined that is annihilated by $\tQ_B$ 
such that $\state{\tilde{\om}}_0 = \state{\om}$, provided that 
the free charge annihilates the free vacuum. \\
The interacting BRS-transformation is the formal power series
\begin{equation}
  \ts = \sum_n g^n \,s_n ,\qquad
  \ts (\tA) \defined \scomm{\tQ_B}{\tA} \quad \forall \tA \in \End{\tV}.
\end{equation}
Each $s_n$ is an anti-derivation on $\End{\tV}$ and $s_0$ agrees with 
the free BRS-trans\-for\-ma\-tion. $\ts_c$ is analogously defined as 
\begin{equation}
  \ts_c = \sum_n g^n\,s_{c,n} ,\qquad
  \ts_c (\tA) \defined \scomm{\tQ_c}{\tA} \quad \forall \tA \in 
  \End{\tV}
\end{equation}
where each $s_{c,n}$ is a derivation on $\End{\tV}$ and $s_{c,0}$ agrees 
with $s_c$. The interacting observable algebra is
defined as
\begin{equation}
  \teA_{\rm ph} \defined 
  \left(\vps (\ker \ts,\End{\tV})\cap(\ker \ts_c,\End{\tV})\right)/
  \left(\vps (\im \ts,\End{\tV})\cap(\ker \ts_c,\End{\tV})\right) .
\end{equation}
So in the framework of BRS theory an algebra of interacting observables
can be defined and represented in a (pre) Hilbert space if the following 
conditions can be accomplished:
\begin{enumerate}
  \item In the underlying free theory a ghost charge $Q_c$ and a BRS 
  charge $Q_B$ can be defined that fulfill the assumptions $\bf A1$ -
  $\bf A3$.
  \item A conserved interacting BRS charge $\tQ_B$ can be constructed 
  such that $(\tQ_B)_0 = Q_B$ with the 
  properties $\tQ_B^2=0$ and $(\tQ_B)^*= \tQ_B$. 
  \item A conserved interacting ghost charge $\tQ_c$ with integer 
  eigenvalues can be constructed such that $(\tQ_c)_0 = Q_c$ with the 
  property $(\tQ_c)^*= - \tQ_c$ .
  \item The BRS algebra $\comm{\tQ_c}{\tQ_B}=\tQ_B$ holds.
\end{enumerate}

%% file: free.tex
\section{The free theory}
\label{free}

We start our considerations concerning BRS theory with free theories. 
The treatment of free theories in the BRS framework is not a goal in 
its own but provides us with definitions that will become important for 
the interacting theory in the next chapters. Furthermore positivity is 
proven for the underlying free model in order to take advantage of 
deformation stability for the interacting theory. \\
We already pointed out the essential significance of normalization 
conditions for the time ordered products in our construction. For some 
of these normalization conditions it is necessary to give a precise 
meaning to expressions like $\partder{A}{\vp_j}(x)$, the derivative of 
a Wick monomial $A$ w.r.t.\ a free field operator $\vp_j$. D\"utsch and
Fredenhagen \cite{DueFre98} solve this problem for QED by an implicit 
definition,
\begin{equation}
  \scomm{A(x)}{\vp_j(y)} = i \sum_{k} \De_{jk} (x-y) 
  \, \partder{A}{\vp_k} (x)  
\end{equation}
where $\De_{jk}(x)$ is a commutator function. This equation is indeed
a definition for the partial derivative on the right hand side if 
the theory contains no derivated fields\footnote{Derivated fields means 
here and below fields containing a spacetime derivative.} like QED. But 
for theories that do contain such derivated fields --- like Yang-Mills 
theory --- there is no such definition available. \\
The natural attempt to include derivated fields would be the replacement 
of the partial derivative by a functional derivative, where the latter 
would be defined by means of
\begin{equation}
  \scomm{A(x)}{\vp_j(y)} = i \sum_{k} \int d^4z \,\De_{jk} (z-y)\,
  \funcder{A(x)}{\vp_k(z)} .
\end{equation}
Unfortunately the equation above is no definition. This can be seen as 
follows: Let $D$ be the differential operator that implements the field 
equations for the free fields such that
\begin{equation}
  \sum_{j} D_{ij}^x \De_{jk} (x) = 0.
\end{equation}
Such an operator exists in general, it will be explicitely constructed
later in this chapter. We can define an operator $D^*$ according 
to 
\begin{equation}
  \int d^4z f(z) \left(D^{*,z}_{ij} g(z)\right) = 
  \int d^4z \left(D^{z}_{ij} f(z)\right) g(z). 
\end{equation}
Then an expression of the form $\sum_m D^{*,z}_{mj} \Phi_m (x,z)$
with arbitrary operators $\Phi_m (x,z)$ can be added to the functional 
derivative without altering the equation. \\
Our strategy to solve this problem is the following: We introduce 
an algebra that is generated by symbols for the basic and derivated 
fields. These symbols serve as auxiliary variables. For this algebra 
the derivative w.r.t.\ a generator is defined. The polynomials in this 
algebra are then linearly represented as operator valued distributions 
acting on the Fock space. Time ordering is defined as a multi linear 
representation of several such polynomials as distributional Fock space 
operators in the next chapter. The derivative that we needed to formulate
the normalization condiditons occurs only in the 
arguments of time ordered products. With the definition of time ordered 
products introduced above these arguments are polynomials in the 
algebra. For the algebra the derivative is well defined, and therefore 
the normalization conditions can be formulated.\\
The chapter is organized as follows: In the first section we 
define the algebra $\eP$ of auxiliary variables. In the second section 
the Fock space $\cF$ and operators therein are constructed. This 
construction will be completely standard and is included here to 
establish our notation. \\
In the third section the linear representation of the algebra $\eP$ as 
(distributional) Fock space operators is defined. This definition 
includes commutator functions for basic and derivated fields.\\
In the fourth section propagator functions for basic and derivated fields
are constructed that have a differential operator as their inverse. \\
The chapter concludes with a section concerning the free model
underlying Yang-Mills theory where the essential operators --- ghost 
charge, BRS charge etc. --- are defined. In particular we present 
Razumovs and Rybkins \cite{RazRyb90} proof for the positivity of that 
theory. 

\subsection{The algebra of auxiliary variables}\label{defP}
The algebra $\eP$ is the graded commutative $\C$-algebra 
generated by auxiliary variables for the basic and derivated fields. 
At first we specify its generators. Therefore we determine which 
basic fields and which derivatives of the basic fields we wish to 
deal with in the model to be defined. For example, with respect to 
Yang-Mills theory we include Lie algebra valued vector bosons $A_\mu$ 
and its first derivatives $(\dd_\nu A_\mu)$, since in the interaction 
Yang-Mills Lagrangian both non derivated and derivated vector bosons 
appear. For the same reason ghosts and anti-ghosts $u, \tu$ and their 
derivatives $(\dd_\nu u), (\dd_\nu \tu)$ are added. If we whish to 
include fermionic matter, coloured spinor fields $\psi, \psiq$ must be 
incorporated, but no derivated spinors because these do not appear in 
the interaction Lagrangian. The set of fields is then completed by the 
double derivated vector bosons $(\dd_\nu \dd_\rho A_\mu)$ which do not 
appear in the Lagrangian but in the BRS current (see below). The non 
derivated fields are referred to as {\em basic fields}. \\
Then we define one symbol for each of these fields --- with a distinct 
symbol for each derivative of the basic fields that is included in the
list above. These symbols are the generators of $\eP$. The generators 
corresponding to the basic fields are called the {\em basic generators},
those corresponding to derivated fields are called the {\em higher
generators}. We adopt the following notation: the generators are written 
as $\vp_i$ where the index $i$ numerates the basic and higher generators.
Sometimes it is desirable to distinguish basic and higher generators. 
Then the generators are denoted as $\vp_{i}^{\alpha}$, where the index
$i$ numerates here the {\em basic} generators, and $\alpha$ is a multi 
index, 
\begin{equation}
  \al = (\betr{\al}, \mu_1,\dots , \mu_{\betr{\al}}) .
\end{equation}
The {\em degree} $\betr{\al}$ of a generator $\vp_{i}^{\alpha}$ is
is the number of spacetime derivatives on the corresponding field 
operator. Basic generators are therefore denoted as $\vp_i^{(0)}$. 
The indices $\mu_1,\dots , \mu_{\betr{\al}}$ are Lorentz indices 
corresponding to the Lorentz indices of the spacetime derivatives on
the field operator. The symbols $\vp_i$ may carry additional Lorentz
(e.g. if $\vp_i = A_\mu$) or spinor (e.g. if $\vp_i = \psi$) indices. 
We will define a representation of the Lorentz group on $\eP$ at the end
of this section. To give an example for the multi indices, we relate some
generators $\vp_i$ to the corresponding field operators:
\begin{equation}
  \begin{split}
    \vp_i^{(0)} \quad\longleftrightarrow \quad \vp_i(x), \qquad\qquad 
    &\vp_i^{(1,\mu)} \quad\longleftrightarrow \quad \dd^\mu_x \vp_i(x) \\
    &\vp_i^{(2,\mu\nu)} \quad\longleftrightarrow \quad
    \dd^\mu_x \dd^\nu_x \vp_i(x) \dots \quad .
  \end{split}
\end{equation}
The symbols are symmetric under permutation of the Lorentz-indices stemming 
from the multi-indices, e.g. $\vp_i^{(2,\mu\nu)} = \vp_i^{(2,\nu\mu)}$. \\
The set $\cG$ of all generators of $\eP$ is defined as 
\begin{equation}
  \cG \defined \set{\vp_i^{\alpha}: \quad \vp_i^{\alpha} \text{ has a 
  counterpart in the desired set of fields}}.
\end{equation}
Sometimes the set of basic generators will become important:
\begin{equation}
  \cG_b \defined \set{\vp_i^{\alpha} \in \cG: \quad \alpha = (0)}
  \subset \cG .
\end{equation}
Now $\eP$ is defined as the unital\footnote{This means that there is an 
identity operator $\one$ included in $\eP$} algebra generated by $\cG$. 
In addition, $\eP$ is graded symmetric. There are two gradings involved 
here: the ghost number $g$ and the (physical) fermion number $f$,
\begin{equation}
  f,g: \quad \set{\text{monomials in } \eP} \to \Z .
\end{equation}
They are additive quantum numbers,
\begin{equation}
  g(AB) = g(A) + g(B) \quad \mbox{ and } \quad f(AB) = f(A) + f(B) 
  \qquad \forall \, A,B \in \eP,
\end{equation}
and are defined as
\begin{equation}
  \begin{split}
    g(u^{\alpha}) = 1, \qquad g(\tu^{\alpha}) = -1,\qquad
    g(\vp_i^{\alpha}) = 0 \quad\mbox{otherwise}& \\
    f(\psi^{\alpha}) = 1, \qquad f(\psiq^{\alpha}) = -1, \qquad
    f(\vp_i^{\alpha}) = 0 \quad\mbox{otherwise}&. 
  \end{split}
\end{equation}
Polynomials in $\eP$ that have a definite ghost or fermion number are
called homogeneous w.r.t.\ the ghost or fermion number.\\ 
Graded symmetric means that for any two elements of the algebra 
$A, B \in \eP$ the following commutation relation holds:
\begin{equation}
  AB = (-1)^{g(A)g(B)+f(A)f(B)} BA \qquad \forall A,B \in  \eP .
\end{equation}
This means that $\eP$ is the unital algebra freely generated by $\cG$
with the equivalence relation $AB \sim (-1)^{g(A)g(B)+f(A)f(B)} BA$ 
divided out. The commutation relation above implies that ghosts 
fulfill {\em commutation} relations with physical fermions. \\
It is important that the elements of $\eP$ are only symbols. In 
particular they are no operators in a Hilbert space and no functions 
on a manifold. The higher generators have no relation with the basic 
ones and the symbols do not satisfy field equations --- e.g.\  
$g_{\mu\nu} u^{(2,\mu\nu)} \neq 0$, where $g_{\mu\nu}$ is the metric 
tensor, although the ghost $u$ is a massless Klein-Gordon field in our 
example. Only the representation of the symbols as operator valued 
distributions in Fock space will restore these relations.\\
On $\eP$ a derivative w.r.t.\ its generators is defined as a graded 
derivation according to  
\begin{equation}
  \begin{split}
    \partder{}{\vp_{i}} \left(A \cdot B\vps\right)  =  
    \left(\partder{A}{\vp_{i}}\right) \cdot B + 
    (-1)^{f(A)f(\vp_{i}) + g(A)g(\vp_{i})} A \cdot 
    \left(\partder{B}{\vp_{i}}\right) \qquad&\\
    \partder{\vp_{i}}{\vp_{j}} = \de_{ij} \one \qquad 
    \forall A,B \in \eP,\,\, \vp_{i},\vp_{j} \in \cG &.
  \end{split}
\end{equation}
The representation $\eR$ of the Lorentz group (or its covering group
$SL(2,\C)$ for the spinors) on $\eP$ is defined as follows: It acts as 
an algebra homomorphism, i.e.\ a linear mapping for which
\begin{equation}
  \eR_{\Lm}\left( \prod_i \vp_i\right) = \prod_i  
  \left( \eR_\Lm (\vp_i) \vps\right), \qquad \Lm \in \fL^\uparrow_+,\quad
  \vp_i \in \cG.
\end{equation}
where $\fL^\uparrow_+$ is the homogeneous proper Lorentz group. The 
action of $\eR$ on the generators is the same as for the corresponding 
field operators. For the basic generators this means the following: 
Suppose the generator $(\vp_i)^{(0)}$ corresponds to the basic field
$\vp_i(x)$, $(\vp_i)^{(0)} \longleftrightarrow \vp_i(x)$, and the basic
field transforms according to\footnote{The field operators and the action 
$U$ of the Lorentz group on them are constructed in the next chapter} 
\begin{equation}
  U(\Lm)\vp_i(x) U^{-1}(\Lm) = \sum_j \left( \eR_\Lm \right)_{ij} 
  \vp_j(\Lm^{-1}x) ,\qquad \Lm \in \fL^\uparrow_+
\end{equation}
for some numerical matrix $\left( \eR_\Lm \right)$. Then the basic 
generator transforms according to
\begin{equation}
  \eR_\Lm \left((\vp_i)^{(0)}\right) \defined \sum_j 
  \left( \eR_\Lm \right)_{ij} (\vp_j)^{(0)} 
\end{equation}
with the same numerical matrix $\left( \eR_\Lm \right)$. In our standard 
example of Yang-Mills theory we have e.g. 
\begin{equation}
  \eR_\Lm (A^\mu) = (\Lm)^\mu_\nu A^\nu , \qquad \eR_\Lm (u) = u,
  \qquad \eR_\Lm (\tu) = \tu,\qquad \Lm \in \fL^\uparrow_+
\end{equation}
Here $(\Lm)^\mu_\nu$ is the representative of $\Lm$ in the defining
representation of $\fL^\uparrow_+$. The higher generators transform 
according to
\begin{equation}
  \eR_\Lm \left((\vp_i)^{(n,\mu_1\dots \mu_n)}\right) \defined \sum_j
  \sum_{\nu_1\dots \nu_n}  (\Lm)^{\mu_1}_{\nu_1}\cdots (\Lm)^{\mu_n}_{\nu_n} 
  \left( \eR_\Lm \right)_{ij} (\vp_j)^{(n,\nu_1\dots \nu_n)} . 
\end{equation}
with $(\Lm)$ like above, and this completes the definition of $\eR$. \\
There is also an anti-linear involution ${}^*$ defined on $\eP$. It acts 
on products according to
\begin{equation}
  (a AB)^* = \overline{a} B^* A^* \qquad \forall A,B \in \eP,
  \quad a \in \C  ,
\end{equation}
where $\overline{\phantom{a}}$ denotes complex conjugation in $\C$.\\
The involution is to implement the Krein adjoint for the fields in $\eP$.
So take a basic generator $\vp_i$ and a basic field $\vp_i(x)$ like above
and let $\left(\vp_i(x)\right)^{*} = \sum_j a_{ij} \vp_j(x)$, where the
$^*$-operation on the left hand side is the Krein adjoint on the fields.
Then we define for this basic generator and its corresponding higher 
generators
\begin{equation}
  \left( (\vp_i)^{(n,\mu_1\dots \mu_n)} \right)^* \defined
  \sum_j a_{ij} (\vp_j)^{(n,\mu_1\dots \mu_n)} .
\end{equation}
In anticipation of the results presented in the next section we state 
what this means for the basic generators in the standard example:
\begin{equation}
  (A^\mu)^* = A^\mu, \quad (u)^* = u,\quad (\tu)^* = - \tu \quad
  (\psi)^* = \psiq \gamma^0, \quad (\psiq)^* = \gamma^0 \psi.
\end{equation}

\subsection{Fock space and Fock space operators}\label{fock}
In this section we will construct the field operators already mentioned 
as operator valued distributions in the Fock space. We begin with some 
notations: A four-vector $p$ on the 
forward light cone $\forlc$ will be denoted as\footnote{The construction 
is outlined here for massless fields, for simplicity.}
\beq
  \hp \defined \left(E_p, \vecp  \right), \qquad E_p \defined 
  \sqrt{\vecp^2} .
\end{equation}
The invariant volume measure on the light cone and its Dirac 
distribution are defined as usual:
\beq
  d\hp \defined \frac{d^3\vecp}{2(2\pi)^3 E_p} \qquad 
  \de (\hp) \defined 2 (2\pi)^3 E_p \de (\vecp) .
\end{equation}
At first we must construct the Fock space $\cF_{\vp_i}$ for each basic 
field that corresponds to a basic generator $\vp_i \in \cG_b$. That 
means for our standard example $\vp_i = (A_\mu^a),(u^a,\tu_a)$ or 
$(\psi^r,\psiq_r)$, where $a$ and $r$ are possible internal indices. To 
this end we begin with the $n$-particle Hilbert space $\eH^n_{\vp_i}$. 
It is the Hilbert space of $L^2(d\hp_1\cdots d\hp_n, M^n)$ functions of 
$n$ momenta and $n$ sets of indices (group-, colour- and Lorentz indices, 
for example) which are collectively written as $a_i$:
\beq
  \vp^n_{(a_1\dots a_n)} (\hp_1,\dots ,\hp_n) \in \eH^n_{\vp_i}
\end{equation}
These functions are completely symmetric or antisymmetric 
under transposition of momenta and indices, 
$(\hp_i,a_i) \leftrightarrow (\hp_j,a_j)$, depending on 
the bosonic or fermionic character of $\vp_i$. \\
The scalar product on $\eH^n_{\vp_i}$ is then defined as
\begin{equation}
  \SP{\psi^n}{\phi^n} \defined \sum_{a_1\dots a_n} 
  \int d\hp_1\cdots d\hp_n\,\, 
  \overline{\psi}^n_{(a_1\dots a_n)}(\hp_1,\dots ,\hp_n)
  \phi^n_{(a_1\dots a_n)}(\hp_1,\dots ,\hp_n)
\end{equation}
This scalar product is positive and allows to define a norm 
$\norm{\phi^n}\defined \SP{\phi^n}{\phi^n}^{\half}$.  \\
With $\eH^0_{\vp_i}\defined \C$ and 
$\SP{\phi^0}{\psi^0} \defined \overline{\phi^0} \psi^0$ we can define
the Fock space $\cF_{\vp_i}$ for the field $\vp_i$ as
\beql{defF}
  \cF_{\vp_i} \defined \bigoplus_{n=0}^{\infty} \eH^n_{\vp_i} ,\qquad
  \SP{\phi}{\psi} = \sum_{n=0}^{\infty} \SP{\phi^n}{\psi^n} ,
\end{equation}
where $\cF_{\vp_i}$ contains only sequences $\phi$ with  
$\SP{\phi}{\phi} < \infty$. The vector 
$\state{\omega_{\vp_i}} \defined (1,0,0,\dots)$ is the vacuum for this
Fock space. \\
Next we define $\eD_{\vp_i}$ as the dense subspace of $\eH_{\vp_i}$ that 
includes only elements with a finite particle number and whose wave 
functions are Schwartz' test functions:
\beql{defD}
  \phi \in \eD_{\vp_i}\subset \cF_{\vp_i} \qquad \llra \qquad 
  \exists \,m \in \N : \quad \phi \in \bigoplus_{n=0}^{m} \cS(M^n) 
  \subset  \cF_{\vp_i}
\end{equation}
where $\cS(M^n)$ is the space of Schwartz' test functions on $M^n$. 
This subspace has the advantage that Wick products are well defined 
operators acting on it \cite{GarWigh64}. It is the common domain of
all operators on $\cF_{\vp_i}$ defined below. Recently Brunetti and 
Fredenhagen \cite{BruFre99} have found a definition of Wick products that 
is well posed on a bigger dense subspace than $\eD_{\vp_i}$, 
but we will stick in this thesis to the space $\eD_{\vp_i}$ defined above.\\
Annihilation operators may be defined on $\eD_{\vp_i}$ according to 
\begin{equation}
  \begin{split}
    &v_{a}(\hp): \quad \eD_{\vp_i} \to \eD_{\vp_i}, \\
    &[v_{a}(\hp)\phi]^{(n)}_{(a_1\dots a_n)}
    (\hp_1,\dots ,\hp_n) = 
    \sqrt{n+1} \phi^{(n+1)}_{(a,a_1\dots a_n)}
    (\hp, \hp_1,\dots ,\hp_n) .
  \end{split}
\end{equation}
Their adjoint --- w.r.t.\ the scalar product defined above --- operators 
$v^+_{i}(p)$, the creation operators $v^+_{a}(\hp)$, are defined as 
\begin{equation}
  \begin{split}
  &[v^+_{a}(\hp)\phi]^{(n)}_{(a_1\dots a_n)}
  (\hp_1,\dots ,\hp_n) = \\
  &\qquad = \sqrt{n}
  \left( 
    \de_{a,a_1} \de(\hp_1-\hp)
    \phi^{(n-1)}_{(a_2\dots a_n)}(\hp_2,\dots ,\hp_n)
    \vphantom{\sum_{k=2}^{n}} \right.\\
    &\qquad\qquad\qquad
    \left. 
    \pm \sum_{k=2}^{n} \de_{a,a_k} \de(\hp_k-\hp)
    \phi^{(n-1)}_{(a_1\dots \check{a}_k\dots a_n)}
    (\hp_2\dots \check{p}_k \dots \hp_n) 
  \right).
  \end{split}
\end{equation}
Here $\check{\phantom{m}}$ means omission of the corresponding argument 
and the plus sign occurs if the field is bosonic, the minus sign if it is 
fermionic. The creation operators are no endomorphisms of $\eD_{\vp_i}$ 
but map $\eD_{\vp_i}$ to $\eD'_{\vp_i}$, the dual space of $\eD_{\vp_i}$. 
This is due to the appearance of the delta function in their 
definition. \\
Creation and annihilation operators fulfill the usual (anti-) 
commutation relations,
\begin{equation}
  \scomm{v^+_{a}(\hp)}{v_{b}(\hq)} = \de_{ab} \de(\hp-\hq),
  \qquad \scomm{v^+_{a}(\hp)}{v^+_{b}(\hq)} = 
  \scomm{v_{a}(\hp)}{v_{b}(\hq)} = 0
\end{equation}
for bosons and ghosts (where the commutator above is the graded one) and 
\begin{equation}
  \acomm{v^+_{a}(\hp)}{v_{b}(\hq)} = \de_{a,-b} \de(\hp-\hq),
  \qquad \!\acomm{v^+_{a}(\hp)}{v^+_{b}(\hq)} 
  = \acomm{v_{a}(\hp)}{v_{b}(\hq)} = 0
\end{equation}
for spinors. Here $v_{-a}(\hp)$ is the annihilator for the field
that is conjugate to the field with the annihilator $v_{a}(\hp)$.\\
The normal ordering --- or Wick ordering --- of an arbitrary product of 
creation and annihilation operators is defined as the same product with 
all the annihilation operators on the right and all the creation operators 
on the left. The normal product of a product $W$ is denoted as $:W:$.\\
Operators on the Fock space can be defined from these distributional 
operators according to
\begin{equation}
  v_{a}(f) = \int d\hp \,\overline{f(\hp)}\, v_{a}(\hp),\qquad
  v^+_{a}(f) = \int d\hp \,f(\hp) \,v^+_{a}(\hp) .
\end{equation}
With this smearing also the Wick products become operators in 
$\End(\eD_{\vp_i})$.\\
The field operators defined below are operator valued distributions
acting on the dense subspace $\eD_{\vp_i}$. To give a precise meaning to 
that expression, we define the $n^{\rm th}$ order operator valued 
distributions on an arbitrary subspace $\eD$ of a Fock space, abbreviated 
as $\Dist_n(\eD)$, as $\C$-linear strongly continuous mappings
\begin{equation}
  \Dist_n(\eD) \defined \set{A: \quad \cD(M^n) \to \End(\eD)}.
\end{equation}
where $M$ is the Minkowski space and $\cD{M^n}$ the space of test functions
on $M^n$ with compact support.\\
The field operators defined below are in $\Dist_1(\eD_{\vp_i})$.\\ 
We begin the definition of the field operators that correspond to
the basic generators with the vector bosons. The corresponding Fock
space is denoted as $\cF_A$, its dense subspace as $\eD_A$. The creation 
and annihilation operators are denoted as $a^{a,+}_\mu (\hp)$ and 
$a^{a}_\mu (\hp)$. They fulfill the commutation relations
\begin{equation}
  \comm{a^{+,a}_{\mu}(\hp)}{a_{\nu}^b(\hq)} = \de^{ab} \de_{\mu\nu}
  \de(\hp-\hq),
  \qquad \comm{a^{+,a}_{\mu}(\hp)}{a^{+,b}_{\nu}(\hq)} = 
  \comm{a_{\mu}^a(\hp)}{a_{\nu}^b(\hq)} = 0 .
\end{equation}
The vector boson field is defined as
\begin{equation}
  \begin{split}
    &A^a_{0}(x) \defined 
    \int d\hp 
    \left[
      a_0^a(\hp) e^{-i\hp x} - a_0^{a,+}(\hp) e^{i\hp x}
    \right]\quad \in \Dist_1(\eD_A),\\ 
    &A^a_{i}(x) \defined 
    \int d\hp 
    \left[
      a_i^a(\hp) e^{-i\hp x} + a_i^{a,+}(\hp) e^{i\hp x}
    \right]\quad \in \Dist_1(\eD_A) .
  \end{split}
\end{equation}
It satisfies the commutation relation
\begin{equation}
  \comm{A^a_{\mu}(x)}{A^b_{\nu}(y)} = i \de^{ab} g_{\mu\nu} D(x-y) 
\end{equation}
and the massless Klein-Gordon equation
\begin{equation}
  \square^x A^a_{\mu}(x) .
\end{equation}
Here $D(x)$ is the massless Pauli-Jordan function
\begin{equation}
  D(x) \defined 2i \int d\hp \,\sin(\hp x).
\end{equation}
It has causal support. It may be split into a positive and a negative
frequency part according to 
\begin{equation}
  D^+(x) \defined \int d\hp \,e^{i\hp x}, \qquad D^-(x) \defined 
  - D^+(-x).
\end{equation}
Its corresponding retarded, advanced and Feynman propagators $D^R, D^A$ 
and $D^F$ are defined as
\begin{equation}
  D^R(x) \defined \theta(x^0)D(x), \qquad\!\! D^A(x) \defined -\theta 
  (-x^0)D(x), \qquad\!\! D^F(x) \defined D^R(x) - D^-(x) .
\end{equation}
They are the inverse of the massless Klein-Gordon operator:
\begin{equation}
  \square^x D^{R,A,F}(x) = \de (x) .
\end{equation}
Clearly $D^R$ has retarded and $D^A$ has advanced support.\\
The $0$-component of the vector bosons is anti hermitian, 
$(A^a_{0})^+ = - A^a_0$. 
Furthermore the scalar product is not Lorentz invariant as can be easily 
verified already in the one-particle space. This is a typical situation 
in gauge theories as described in the last chapter. To find a physical 
inner product on $\cF_A$ one must find a suitable Krein operator $J_A$
acting on it and define
\begin{equation}
  \SPK{\phi}{\psi} \defined  \SP{\phi}{J_A\psi} .
\end{equation}
This suitable Krein operator is
\begin{equation}
  J_A \defined (-1)^{N_0}, \qquad 
  N_0 = \sum_b \int d\hp \,a^{b,+}_0(\hp)a^{b}_0(\hp),
\end{equation}
where $N_0$ is the number operator for $A^0(x)$ with eigenvalues in $\N$. 
It is obviously hermitian, $J=J^+$, and idempotent, $J^2 =\one$. With the 
$^*$-involution 
\begin{equation}
  B^* \defined J_A B^+J_A,\qquad \forall B\in \End(\eD_A),
\end{equation}
also called the Krein adjoint, the vector bosons become pseudo-hermitian, 
$(A^a_{\mu})^* =  A^a_\mu$. Furthermore we find the inner product 
$\SPK{\cdot}{\cdot}$ to define a Lorentz invariant norm, but it is
indefinite.\\
The definition of the spinor Fock space $\cF_\psi$ and the field operators 
$\psi(x), \psiq(x)$ acting therein proceeds in the same way and can be 
found in textbooks on quantum field theory. The fermions satisfy the 
commutation relations
\begin{equation}
  \comm{\psi(x)}{\psiq(y)} = -i (i\dddag_x+m) D(x-y) 
\end{equation}
and the field equations
\begin{equation}
  (i\dddag_x-m) \psi = 0, \qquad \psiq (-i\overleftarrow{\dddag}_x-m) = 0. 
\end{equation}
The Krein operator $J_\psi$ on the spinor Fock space is trivial, 
$J_\psi = \one$. \\
On the Fock space for the ghosts, $\cF_u$ with its dense subspace 
$\eD_u$, creation and annihilation operators are denoted by 
$b^{a,+} (\hp)$, $c^{a,+} (\hp)$, $b^{a} (\hp)$ and $c^{a} (\hp)$,
respectively. They fulfill the anti-commutation relations
\begin{equation}
  \acomm{b^{+,a}(\hp)}{b^b(\hq)} = \de^{ab} \de(\hp-\hq),
  \qquad \acomm{c^{+,a}(\hp)}{c^b(\hq)} = \de^{ab} \de(\hp-\hq)
\end{equation}
and all other anti-commutators vanish. The ghost field $u^a(x)$ and the 
anti-ghost field $\tu^a(x)$ are defined as
\begin{equation}
  \begin{split}
    &u^a(x) \defined 
    \int d\hp 
    \left[
      b^a(\hp) e^{-i\hp x} + c^{a,+}(\hp) e^{i\hp x}
    \right]\quad \in \Dist_1(\eD_u),\\ 
    &\tu^a(x) \defined 
    \int d\hp 
    \left[
      - c^a(\hp) e^{-i\hp x} + b^{a,+}(\hp) e^{i\hp x}
    \right]\quad \in \Dist_1(\eD_u) .
  \end{split}
\end{equation}
Then we get for the anti-commutators of the ghosts
\begin{equation}
  \begin{split}
    &\acomm{u^a(x)}{\tu^b(y)} = - i \de^{ab} D(x-y) .\\
    &\acomm{u^a(x)}{u^b(y)} = \acomm{\tu^a(x)}{\tu^b(y)} = 0. 
  \end{split}
\end{equation}
The Krein operator for the ghosts was explicitely determined by
Krahe \cite{Kra95} and reads
\begin{equation}
  J_u = \exp
  \left(
    \frac{i\pi}{2} \int d\hp 
    \left[ 
      b^+(\hp) b(\hp) -  
      b^+(\hp) c(\hp) +
      c^+(\hp) c(\hp) -   
      c^+(\hp) b(\hp)
    \right]
  \right).
\end{equation}
For us it is only important that this implies for the field 
operators
\begin{equation}
  \left(u^a(x)\right)^* = u^a(x) \qquad\text{and} \qquad 
  \left(\tu^a(x)\right)^* = - \tu^a(x) ,
\end{equation}
so the ghosts are pseudo-hermitian and the anti-ghosts are 
anti-pseudo-hermitian. \\
Now we introduce the pseudo-unitary representation of the proper 
Poincar\'e group $\fP^\uparrow_+$ in the individual Fock spaces. It 
reads for scalar fields like the ghosts
\begin{equation}
  \begin{split}
    &[U(p)\phi]^{(0)} = \phi^{(0)},\qquad p = (a,\Lm) \in \fP^\uparrow_+\\
    &
    \begin{split}
      [U(p)\phi]^{(n)}_{(a_1\dots a_n)}(\hq_1,\dots ,\hq_n) 
      \,=\, \exp\left( -i (\hq_1 +\dots +\hq_n)\cdot a \right)\times 
      &\\
      \times \phi^{(n)}_{(a_1\dots a_n)}
      (\Lm\hq_1,\dots ,\Lm\hq_n) &.
    \end{split}
  \end{split}
\end{equation}
For vector fields like the vector bosons it reads
\begin{equation}
  \begin{split}
    &[U(p)\phi]^{(0)} = \phi^{(0)},\qquad p = (a,\Lm) \in \fP^\uparrow_+\\
    &
    \begin{split}
      \left([U(p)\phi]^{(n)}\right)_{(a_1\dots a_n)}^{(\mu_1\dots \mu_n)}
      (\hq_1,\dots ,\hq_n) 
      \,=&\, \exp\left( -i (\hq_1 +\dots +\hq_n)\cdot a \right)\times 
      \\
      &\times (\Lm)^{\mu_1}_{\nu_1}\cdots (\Lm)^{\mu_n}_{\nu_n}
      \left(\phi^{(n)}\right)_{(a_1\dots a_n)}^{(\nu_1\dots \nu_n)}
      (\Lm\hq_1,\dots ,\Lm\hq_n) 
    \end{split}
  \end{split}
\end{equation}
where  the Lorentz indices $\mu_i$ have been separated from the other 
indices $a_i$ and summation over repeated indices is understood. The 
matrices $(\Lm)$ are the representatives of $\Lm$ in
the defining representation of $\fL^\uparrow_+ \subset \fP^\uparrow_+$, 
like above. For the spinors an analogous definition holds. The vacuum 
vector $\state{\om_{\vp_i}}$ is clearly Poincar\'e invariant. As was 
pointed out by Krahe \cite{Kra95}, it is also cyclic w.r.t. the field 
operators defined above.\\
The field operators transform according to
\begin{equation}
  \begin{split}
    &U(p) u^a(x) U^{-1}(p) = u^a(\Lm^{-1}x - a),\qquad
     U(p) \tu^a(x) U^{-1}(p) = \tu^a(\Lm^{-1}x - a)\\
    &U(p) A^a_{\mu}(x)U^{-1}(p) = \left(\Lm\right)^{\nu}_{\mu} A^a_{\nu}
    (\Lm^{-1}x - a) .
  \end{split}
\end{equation}
With the Fock spaces for the individual fields the Fock space of the
entire theory $\cF$, its dense subspace $\eD$ and the Krein operator $J$
acting on $\cF$ are defined as
\begin{equation}
  \cF \defined \bigotimes_i \cF_{\vp_i}  
  \qquad \eD \defined \bigotimes_i \eD_{\vp_i}
  \qquad J \defined \bigotimes_i J_{\vp_i}  .
\end{equation}
The vacuum vector of the Fock space $\cF$ is denoted by $\state{\om}$. 
We introduce the notation $\om_0\left(A \right)$
for $\costate{\om} A \state{\om}$ for every $A \in \End(\eD)$. Here
$\End (\eD)$ is the algebra of endomorphisms of $\eD$. An important 
fact concerning this algebra is that it has trivial centre. Even more,
for an arbitrary element $A \in \End(\eD)$ the following equivalence 
holds:
\beql{trivcent}
  \begin{split}
    \scomm{A}{\TT{\vp_i}(x)} &= 0 \qquad \forall 
    \vp_i \in \cG_b,  \\
    \llra \qquad A &= a \cdot \one, \quad a \in \C .
  \end{split}
\end{equation}
For the proof of this assertion see Scharf \cite{Sch95}, for example. \\
In chapter (\ref{nilpo}) it will turn out that spacetime must be 
compactified in spacelike directions for the BRS charge to be a well 
defined operator. Therefore it is important to construct the Fock space 
and the operators acting on it also for a quantum field theory in the
compactified spacetime. This has been done by D\"utsch and Fredenhagen in 
\cite[appendix A]{DueFre98}. We refer to their results, especially 
concerning the choice of boundary conditions, but we do not go here into 
details.  

\subsection{The linear representation of $\eP$}\label{defrep}
In this section we define the $\C$-linear representation $T$ of the 
polynomials in $\eP$ as operator valued distributions 
\begin{equation}
  T: \quad \eP \to \Dist_1(\eD) .
\end{equation}
Linear representation means that the linear structure of $\eP$ is 
preserved, but not its structure as an algebra. This comes from the
fact that a pointwise product of distributions is in general no 
well defined operation. \\
The precise definition of $T$ will take three steps: at first it is 
defined for the basic generators, then for the higher generators and
finally for composed elements of $\eP$. \\
The first end has already been achieved with the definition of an 
operator valued distribution $\vp_i(x) \in \Dist_1(\eD)$ for each basic 
generator $\vp_i \in \cG_b$. The representative of the basic
generator is defined as: 
\beql{defgen}
  \TT{\vp_i}(x) \defined \vp_i (x), \qquad \vp_i \in \cG_b, \quad 
  \vp_i(x) \in \Dist_1(\eD) . 
\end{equation}
This definition can work only for the basic generators since for the 
higher ones there are no corresponding free field operators. \\
For these generators we define
\begin{equation}
  \TT{(\vp_i)^{(n, \nu_1\dots \nu_n)}}(x) \defined \dd_x^{\nu_1}\cdots 
  \dd_x^{\nu_n} \vp_i(x) , \qquad (\vp_i)^{(\dots)} \in \cG .
\end{equation}
We remind the reader that there are no relations between the basic 
generators and the higher generators in $\eP$, and that there are no 
field equations in $\eP$. But with the definition above there is a
relation established between the representatives of the basic and those
of the higher generators, and the former clearly satisfy field 
equations. So the linear representation is not faithful. \\
For the representation of the composed elements in $\eP$ we define 
at first the commutator function
\beql{commfunc}
  i  \De_{ij}(x-y) =  \scomm{T(\vp_i)(x)}{T(\vp_j)(y)} , 
  \qquad \vp_i, \vp_j \in \cG .
\end{equation}
Here $i$ and $j$ take on values also for the higher generators. With 
this commutator function we give an implicit definition of the 
representation of monomials in $\eP$, namely
\beql{highprod}
  \begin{split}
    \scomm{T(W)(x)}{T(\vp_i)(y)} &= 
    i \sum_j \TT{\partder{W}{\vp_j}}(x)\De_{ij}(x-y) \\
    \om_0\left( T(W)(x)\vps \right) &= 0 \qquad  W\in \eP. 
  \end{split}
\end{equation}
The existence of a solution is guaranteed by the observation that the 
normal products solve both equations. Suppose 
$A= \prod_i \vp_i  \in \eP,\,\, \vp_i \in \cG$, then the normal product 
$:\prod_i \TT{\vp_i}(x):\,\, \in \Dist_1(\eD)$ is indeed a searched for 
solution. \\
The uniqueness of this solution can be seen inductively. Suppose, the
representation for all monomials containing at most $k-1$ generators
is defined. Then the commutator condition determines the solution
for monomials of $k$ generators up to a $\C$-number distribution --- 
this is due to eqn. \eqref{trivcent}. The $\C$-number 
part is determined by the second condition --- it is zero. So the 
equations above give indeed a definition for the representation of 
monomials in $\eP$. This completes the definition of the
representation. \\
As for the Pauli-Jordan function $D(x)$ we can find a positive and a 
negative frequency solution for the commutator function $\De_{ij}(x)$.
The two point function --- or positive frequency part of $\De$ ---
is denoted as $\De^+$ and defined as
\begin{equation}
  i \De_{ij}^+ (x-y)  \defined 
  \om_0\left( \TT{\vp_i}(x)\,\, \TT{\vp_j}(y)\right) ,
\end{equation}
the negative frequency part of $\De$ is defined as 
\begin{equation}
  \De_{ij}^- (x) \defined \De_{ij} (x) - \De_{ij}^+ (x) .
\end{equation}

\subsection{The propagator functions}\label{defpropfun}
In this section we define propagator functions 
$\De_{ij}^R(x), \De_{ij}^A(x)$ analogous to $D^R(x)$ and $D^A(x)$ that 
are restrictions of $\De_{ij}(x)$ to the past and future light cone,
such that $\De_{ij}^R(x) - \De_{ij}^A(x) = \De_{ij}(x)$. Simultaneously 
we search for a differential operator $D^x_{ij}$ that takes over the 
part of the Klein-Gordon operator, i.e. that fulfills the equations 
\begin{equation}
  \sum_j D^x_{ij} \De^{R,A}_{jk} (x) = \de_{ik} \de(x) 
  \quad\lra\quad 
  \sum_j D^x_{ij} \De_{jk} (x) = 0 .
\end{equation}
This means in particular that the propagators must be invertible with
$D^x_{ij}$ as their inverse. \\
To see what form the propagators might have we take a closer 
look at the commutator function. If $\cG$ contains $r$ generators,  this 
is an $r\times r$-matrix. It has the following block diagonal structure: 
\begin{equation}
  \De(x) = 
  \begin{pmatrix}
    \De^{\vp_1}(x) & 0 & 0 & \dots \vblock\\
    0 & \De^{\vp_2}(x) & 0 & \dots \vblock\\
    0 & 0 & \De^{\vp_3}(x) & \dots \vblock\\
    \vdots & \vdots & \vdots & \ddots \vblock
  \end{pmatrix} .
\end{equation}
Here the matrices $\De^{\vp_i}$ are of two different types. The first 
type corresponds to field operators that have no distinct conjugate
field like the uncharged vector bosons. Then the index $\vp_i$ 
corresponds to the field, e.g. $\vp_i=A_\mu$. The other type 
corresponds to field operators $\vp_i$ that do have such a distinct
conjugate field $\tphi_i$ like ghosts with the anti-ghosts. In this case
the field and the conjugated field form one common block in the 
matrix and the index $\vp_i$ corresponds to the field and its
conjugated field, e.g. $\vp_i=u,\tu$. All blocks include also the 
commutators of the derivatives as far as higher generators exist in 
$\cG$ that correspond to these derivatives. In our standard example it 
has the form
\begin{equation}
  \De(x) = 
  \begin{pmatrix}
    \De^{A}(x) & 0 & 0  \vblock\\
    0 & \De^{u,\tu}(x) & 0  \vblock\\
    0 & 0 & \De^{\psi,\psiq}(x) \vblock
  \end{pmatrix} 
\end{equation}
if QED is treated where no internal indices appear. In Yang-Mills theory,
where internal indices do appear, there is an individual block for each
index on the diagonal. From now on we will disregard internal indices for
their inclusion is straightforward. The vector boson part has the form
\beql{DelA}
  \De^{A}(x) \defined
  g_{\mu\nu}
  \begin{pmatrix}
    D(x) & -\dd^{\nu_1}_x D(x) & \dd^{\nu_1}_x \dd^{\nu_2}_x D(x) 
    \vblock\\
    \dd^{\rho_1}_x D(x) & -\dd^{\nu_1}_x\dd^{\rho_1}_x D(x) & 
    \dd^{\nu_1}_x\dd^{\rho_1}_x \dd^{\nu_2}_x D(x) \vblock\\
    \dd^{\rho_2}_x\dd^{\rho_1}_x D(x) & -\dd^{\nu_1}_x\dd^{\rho_2}_x
    \dd^{\rho_1}_x D(x) & \dd^{\nu_1}_x\dd^{\rho_2}_x\dd^{\rho_1}_x 
    \dd^{\nu_2}_x D(x) \vblock
  \end{pmatrix},
\end{equation}
the ghost part
\beql{Delu}
  \De^{u,\tu}(x) \defined
  \begin{pmatrix}
    0 & 0 & -D(x) & \dd^{\nu_1}_x D(x) \vblock\\
    0 & 0 & -\dd^{\rho_1}_x D(x) & \dd^{\rho_1}_x \dd^{\nu_1}_x D(x) 
    \vblock\\
    D(x) & -\dd^{\nu_1}_x D(x) & 0 & 0 \vblock\\
    \dd^{\rho_1}_x D(x) & -\dd^{\rho_1}_x \dd^{\nu_1}_x D(x) & 0 & 0 
    \vblock
  \end{pmatrix}
\end{equation}
and the spinor part
\beql{Delpsi}
  \De^{\psi,\psiq}(x) \defined
  \begin{pmatrix}
    0 & -i(i\dddag + m )D_m(x) \vblock\\
    i(i\dddag - m )D_m(x) & 0 \vblock
  \end{pmatrix}.
\end{equation}
The distribution $D_m$ is the Pauli-Jordan function for mass $m$.
The matrices are given here in the basis 
\beql{basA}
  \left( (A_\mu)^{(0)}, (A_\mu)^{(1,\nu_1)}, 
  (A_\mu)^{(2,\nu_1\nu_2)} \vps \right)^t 
\end{equation}
for the vector bosons, 
\beql{basu}
  \left( (u)^{(0)}, (u)^{(1,\nu_1)},(\tu)^{(0)}, (\tu)^{(1,\nu_1)} \vps 
  \right)^t 
\end{equation}
for the ghosts and anti-ghosts and
\beql{baspsi}
  \left( (\psi)^{(0)},(\psiq)^{(0)} \vps \right)^t, 
\end{equation}
for the spinors, where $^t$ denotes transposition. \\
The natural attempt would be to replace the Pauli-Jordan function $D(x)$
by its retarded, $D^R(x)$, or advanced, $D^A(x)$, propagator in each 
entry to define the matrices $\De_{ij}^R(x)$ and $\De_{ij}^A(x)$. 
These would clearly be well defined distributions with the desired 
support properties, but they would not be invertible. This comes from the 
fact that with this definition each row would be the derivative of the 
row above, and therefore the determinant --- w.r.t.\ convolution --- of 
these matrices would vanish. \\
To improve the definition above we observe that the matrices 
$\De_{ij}^{R,A}(x)$ are defined by their desired support properties
--- $\supp \De_{ij}^{R}(x) \subset \forlc$ and 
$\supp \De_{ij}^{A}(x) \subset \backlc$ --- and their relation to the
commutator function everywhere but in the origin. That means that
we may alter the propagator functions only at the origin, i.e. by
delta distributions or its derivatives at the individual entries. As a 
further restriction of possible propagators we demand that this 
modification does not increase the scaling degree (see below) of the 
individual entries and that it does not change the Lorentz 
transformation property of that entry. \\
Scaling degree means the following: For every numerical distribution 
$d$ one can define a dilated distribution
\begin{equation}
  d_\lm (x) = d(\lm x) \qquad \lm \in \R^+\setminus \set{0}.
\end{equation}
Clearly $d_\lm$ is a numerical distribution, too. Then the {\em scaling 
degree} $\scaldeg (d)$ of $d$ w.r.t. the origin is defined, according
to Steinmann \cite{Ste71}, as
\begin{equation}
  \scaldeg (d) \defined \inf \set{\beta \in \R: \quad 
  \lim_{\lm \searrow 0} \lm^\beta d_\lm = 0},
\end{equation}
where the equation in the bracket holds in the sense of 
distributions. \\
The restriction on the scaling degree fixes some entries uniquely,
e.g. $\De^R_{ij}(x) = D^R(x)$ if $\vp_i = u$ and $\vp_j = \tu$. For 
others there remains a certain ambiguity, e.g. 
\begin{equation}
  \De^R_{ij}(x) = - g_{\mu\nu} \dd^\rho\dd^\sigma D^R(x) - 
  C g_{\mu\nu}g^{\rho\sigma} \de(x)
\end{equation}
if $\vp_i = (A_\mu)^{(1,\rho)}$ and $\vp_j = (A_\nu)^{(1,\sigma)}$.
The numerical constant $C$ is then arbitrary. In the following we define
propagator functions with an inverse that is a differential operator and
we will give later the explicit form of these differential operators. \\
The propagators have the same block diagonal structure as the 
commutator function:
\begin{equation}
  \De^{R,A}(x) = 
  \begin{pmatrix}
    \De^{\vp_1}_{R,A}(x) & 0 & 0 & \dots \vblock\\
    0 & \De^{\vp_2}_{R,A}(x) & 0 & \dots \vblock\\
    0 & 0 & \De^{\vp_3}_{R,A}(x) & \dots \vblock\\
    \vdots & \vdots & \vdots & \ddots \vblock
  \end{pmatrix} .
\end{equation}
In the following we will consider only the construction of the 
retarded propagator. The advanced propagator is defined as
$\De^{A}= \De^{R} - \De$. For the determination of the individual 
blocks we notice that usually the $(0,0)$-component\footnote{We start 
the numbering of columns and rows with zero, such that the index of a
column or row agrees with the degree of the corresponding generator} of 
the commutator function has a scaling degree smaller than the spacetime 
dimension, so that its retarded solution is uniquely determined by the 
following condition: 
\begin{equation}
  \De_{00}^{R,\vp_i} (x) = \De_{00}^{\vp_i} (x) \qquad
  x \not\in \forlc , \quad\supp \De^R_{00} \in \forlc .
\end{equation}
With this the general matrix element of a block $\De^{\vp_i}_{R}$
of the retarded propagator can be written as
\beql{generalelem}
  \De_{jk}^{R,\vp_i} (x) = (-1)^k \dd^{\mu_1} \cdots \dd^{\mu_k}
  \dd^{\nu_1} \cdots \dd^{\nu_j} \De_{00}^{R,\vp_i} (x)
  + (-1)^k \de_{jk} C_{\vp_i,k}\, \de(x) .
\end{equation}
(no summation over $k$ in the last term). The constants $C_{\vp_i,k}$ are 
non zero real numbers, $C_{\vp_i,k} \in \R \setminus \set{0}$. As these 
constants will determine the normalization of higher order time ordered 
products (c.f. next chapter), they will be called normalization 
constants. \\
In our standard example the propagator has the form 
\begin{equation}
  \De^R(x) = 
  \begin{pmatrix}
    \De^{A}_R(x) & 0 & 0  \vblock\\
    0 & \De^{u,\tu}_R(x) & 0  \vblock\\
    0 & 0 & \De^{\psi,\psiq}_R(x) \vblock
  \end{pmatrix} .
\end{equation}
For the vector boson block of the retarded propagator, $\De^A_R$, we 
omit all spacetime arguments because it otherwise would not fit 
into the line. Then it reads
\beql{DelAprop}
  \De^{A}_{R} =
  g_{\mu\nu}
  \begin{pmatrix}
    D^{R} & -\dd^{\nu} D^{R} & \dd^{\nu}\dd^\rho D^{R} \vblock\\
    \dd^{\sigma} D^{R} & -\dd^{\nu}\dd^{\sigma} D^{R} - 
    C_{A,1} g^{\nu\sigma} \de
    & \dd^\nu\dd^\rho\dd^\sigma D^{R} \vblock\\
    \dd^\sigma\dd^\tau D^{R} & -\dd^\nu\dd^\sigma\dd^\tau D^{R} & 
    \dd^\nu\dd^\rho\dd^\sigma\dd^\tau D^{R} - C_{A,2} g^{\nu\sigma} 
    g^{\rho\tau} \de\vblock
  \end{pmatrix},
\end{equation}
For the ghosts we get the contribution,
\beql{Deluprop}
  \De^{u,\tu}_{R}(x) =
  \begin{pmatrix}
    0 & -d^{u}_{R}(x) \\
    d^{u}_{R}(x) & 0 
  \end{pmatrix}
\end{equation}
with the $2\times 2$-matrices
\begin{equation}
  d^{u}_{R}(x) = 
  \begin{pmatrix}
      D^{R}(x) & -\dd^{\nu_1}_x D^{R}(x)  \vblock\\
      \dd^{\rho_1}_x D^{R}(x) & -\dd^{\nu_1}_x\dd^{\rho_1}_x 
      D^{R}(x) - C_{u,1} g^{\nu_1\rho_1} \de(x) \vblock 
  \end{pmatrix} .
\end{equation}
The spinors finally give 
\beql{Delpsiprop}
  \De^{\psi,\psiq}_{R}(x) =
  \begin{pmatrix}
    0 & -(i\dddag + m )D^R_m(x) \vblock\\
    (i\dddag - m )D^R_m(x) & 0 .\vblock
  \end{pmatrix}
\end{equation}
$D^R_m$ is the retarded part of $D_m$. 
All the matrices are given in the same basis as for the commutator 
function. The retarded propagator function has obviously retarded 
support and agrees with the commutator function outside the forward 
light cone. \\
The advanced propagator $\De^A = \De^R - \De$ has advanced support 
and agrees with the commutator function outside the backward light 
cone. In the example above the respective advanced propagators 
can be derived from the retarded propagators by a substitution of 
$D^R$ with $D^A$.\\
We define also a Feynman propagator 
\begin{equation}
  \De^F: \quad \De^F_{ij}(x) \defined \De^R_{ij}(x) - \De^-_{ij}(x) .
\end{equation}
Now we come to the differential operator valued matrix $D^x$ that 
inverts the propagators defined above, i.e. for which the equation
\beql{invert}
  \sum_j D^x_{ij} \De^{R,A,F}_{jk} (x) = \de_{ik} \de(x) 
\end{equation}
holds. It is an $r\times r$ matrix, where $r$ was the number of generators 
in $\cG$. It has the usual block diagonal form:
\beql{defDx}
  D^x = 
  \begin{pmatrix}
    D^{\vp_1,x} & 0 & 0 & \dots \vblock\\
    0 & D^{\vp_2,x} & 0 & \dots \vblock\\
    0 & 0 & D^{\vp_3,x} & \dots \vblock\\
    \vdots & \vdots & \vdots & \ddots \vblock
  \end{pmatrix} 
\end{equation}
or, in our standard example,
\begin{equation}
  D^x = 
  \begin{pmatrix}
    D^{A,x} & 0 & 0 \vblock\\
    0 & D^{u,\tu,x} & 0 \vblock\\
    0 & 0 & D^{\psi,\psiq,x} \vblock
  \end{pmatrix} .
\end{equation}
Like for the propagators the individual blocks correspond to field
operators or pairs of conjugated fields. We define the blocks 
for single fields as $(s+1)\times(s+1)$-matrices if higher generators 
up to degree $s$ are included in $\cG$ for that field. Let 
$K^{\vp_i,x}$ be the differential operator that defines the field 
equation for $\vp_i(x)$, i.e. which fulfills the equation 
\beql{defK}
  K^{\vp_i,x} \,\De^{\vp_i}_R(x) = \de(x) ,
\end{equation}
e.g. $K^{A,x}=\square^x$. Then the corresponding block is written in the 
basis
\begin{equation}
  \left( (\vp_i)^{(0)}, (\vp_i)^{(1,\nu_1)}, \dots, 
  (\vp_i)^{(s,\nu_1\dots \nu_s)} \vps \right)^t
\end{equation}
as the matrix with the components
\beql{generaldA}
  \begin{split}
    D^{\vp_i,x}_{00} &= \left( K^{\vp_i,x}  - \sum_{k=1}^{n} 
    (-1)^k C_{\vp_i,k}^{-1}\square^k \right) \\
    D^{\vp_i,x}_{0k} &= (-1)^k C_{\vp_i,k}^{-1}\left( \dd^{\nu_1} \cdots 
    \dd^{\nu_k} \right) \\
    D^{\vp_i,x}_{j0} &= C_{\vp_i,j}^{-1}\left( \dd^{\sigma_1} \cdots 
    \dd^{\sigma_j} \right) \\
    D^{\vp_i,x}_{kk} &= - C_{\vp_i,k}^{-1}\left( g^{\nu_1\sigma_1} \cdots 
    g^{\nu_k\sigma_k} \right) \\
    D^{\vp_i,x}_{jk} &= 0 \quad \text{otherwise}.
  \end{split}
\end{equation}
where the constants $C_{\vp_i,k}$ are those determined in the propagator
functions.  \\
Again we exemplify the definition above for our standard example. The 
only uncharged fields there are the vector bosons. In the same basis as
for the commutator function, the block $D^{A,x}$ has the form
\beql{DA}
  D^{A} \defined
  \begin{pmatrix}
    (1+C_{A,1}^{-1})\square - C_{A,2}^{-1}\square^2 & 
    -C_{A,1}^{-1}\dd^{\nu_1} & C_{A,2}^{-1}\dd^{\nu_1} \dd^{\nu_2} \vblock \\
    C_{A,1}^{-1}\dd^{\rho_1} & - C_{A,1}^{-1}g^{\rho_1\nu_1} & 0  \vblock\\
    C_{A,2}^{-1}\dd^{\rho_1}\dd^{\rho_2} & 0 & 
    - C_{A,2}^{-1}g^{\rho_1\nu_1} g^{\rho_2\nu_2}  \vblock
  \end{pmatrix} .
\end{equation}
For the charged fields we construct according to the rules above one
block $D^{\vp_i,x}$ for the fields $\vp_i$ and one block $D^{\tphi_i,x}$
for the conjugated fields $\tphi_i$. Like for the propagators, the 
combined block for the fields and conjugated fields reads then
\beql{Du}
  D^{\vp_i,\tphi_i,x} \defined
  \begin{pmatrix}
    0 & D^{\tphi_i,x} \vblock\\
    - D^{\vp_i,x} & 0 \vblock
  \end{pmatrix}
\end{equation}
in the basis
\begin{equation}
  \left( 
    (\vp_i)^{(0)}, \dots, (\vp_i)^{(s,\nu_1\dots \nu_s)},
    (\tphi_i)^{(0)}, \dots, (\tphi_i)^{(s,\nu_1\dots \nu_s)}\vps 
  \right)^t,
\end{equation}
if higher generators up to degree $s$ are included. The expressions for 
our standard example, i.e. for the ghosts and the spinors, read then
\beql{du1}
  D^{u,\tu} \defined
  \frac{1}{C_{u,1}}
  \begin{pmatrix}
    0 & 0 & (1+C_{u,1})\square & -\dd^{\nu_1} \vblock\\
    0 & 0 & \dd^{\rho_1} & - g^{\nu_1\rho_1} \vblock\\
    - (1+C_{u,1})\square & \dd^{\nu_1} & 0 & 0\vblock\\
    - \dd^{\rho_1} & g^{\nu_1\rho_1} & 0 & 0 \vblock
  \end{pmatrix}.
\end{equation}
For the spinors no higher generators are included in our example,
so they contribute the expression
\beql{dpsi1}
  D^{\psi} \defined
  \begin{pmatrix}
    0 & -(i\dddag + m) &  \vblock\\
    (i\dddag - m) & 0  \vblock
  \end{pmatrix}.
\end{equation}
where the operator in the second line acts from the right. \\
It is easily verified by direct calculation that this differential 
operator really inverts the propagators. Furthermore, the 
representatives of the generators satisfy the following {\em free 
field equations}:
\beql{fe}
  \sum_{j}  D_{ij}^x T(\vp_j)(x) = 0  .
\end{equation}
Here the sum runs over {\em all} generators. This equation holds 
independently of the choice of the normalization constants
$C_{\vp_i,k}$. From its definition it is already clear that the 
commutator function is annihilated by $D^x$:
\begin{equation}\label{Dunique}
  \sum_j D_{ij}^x \De_{jk} (x) =  0 .
\end{equation}
If $D^x$ is determined, the propagator functions $\De^R, \De^A$ and 
$\De^F$ are {\em uniquely} determined by the following conditions:
\begin{itemize}
  \item $\De^{R,A,F}(x)$ must fulfil eqn. \eqref{invert}
  \item $\De^{R}(x) = \De(x) \quad 
  \forall x \not\in \backlc$ and $\De^{R}(x) = 
  0 \quad \forall x \in \backlc\setminus \set{0}$
  \item $\De^{A}(x) = \De^{R}(x) - \De(x)$
  \item $\De^{F}(x) = \De^{R}(x) - \De^{+}(x) $ .
\end{itemize}
So $D^x$ is a relativistically covariant, hyperbolic differential 
operator with a unique solution for the Cauchy problem. In particular
the normalization constants $C_{\vp_i,k}$ that appear in the propagators 
are uniquely determined by their choice in the differential operator 
$D^x$. \\
We do not claim that our choice for the operator $D^x$ or the propagators
is the most general one. But we point out that there are serious 
restrictions to the choice of the propagators. As we already saw, the
apparently easiest choice is not invertible, and all other choices we
tried proved to be invertible, but with pseudodifferential operators 
as their inverse instead of differential operators. We do not examine
the question whether field equations with pseudodifferential operators 
are suitable choices within the general framework of quantum field theory.
Instead we stick to differential operators as one is used to, the more
so as the propagators we have defined above are completely sufficient 
for our purposes.

\subsection{The free BRS theory}\label{freeBRS}
We examine in this section the free theory that includes vector bosons, 
spinors and ghosts. This is the theory that served as an example 
throughout the considerations above. The generators for the algebra
$\eP$ are in this model $A^a_\mu,(A^a_\mu)^{(1,\nu)}$ and 
$(A^a_\mu)^{(2,\nu\rho)}$ for the Lie algebra valued vector bosons, 
$u^a,\tu^a,(u^a)^{(1,\mu)}$ and $(\tu^a)^{(1,\mu)}$ for 
the respective ghosts and anti-ghosts and $\psi^r$ and $\psiq^r$ for the 
coloured spinors. The field operators that correspond to the generators 
$A^a_\mu, u^a, \tu_a,\psi^r$ and $\psiq^r$ are already constructed as
operators in the Fock space $\cF$ with a common dense domain $\eD$.
As we already mentioned when we constructed the Fock space, the inner 
product $\SPK{\cdot}{\cdot}$ is indefinite. To perform the BRS 
construction, we must define a BRS charge and a ghost charge and prove
that the state space is positive. \\
At first we define the ghost current
\begin{equation}
  k^\mu \defined i \sum_a \left[(u^a)^{(0)} (\tu^a)^{(1,\mu)} 
  - (u^a)^{(1,\mu)} (\tu^a)^{(0)} \right]\qquad \in\eP .
\end{equation} 
and the BRS current
\begin{equation}
  j^\mu_B \defined \sum_a \left[ (u^a)^{(1,\mu)} (A_\nu^a)^{(1,\nu)}
  - (u^a)^{(0)} (A_\nu^a)^{(2,\nu\mu)}\right] \qquad \in\eP .
\end{equation}
as elements of $\eP$. Then their definitions as operators in the 
Fock space follow immediately as
\begin{equation}
  k^\mu(x) = \TT{k^\mu}(x) \qquad \text{and} \qquad j_B^\mu(x) = 
  \TT{j_B^\mu}(x) .
\end{equation}
Taking into account the field equations, we note that both operators 
are conserved,
\begin{equation}
  \dd_\mu^x k^\mu(x) = \dd_\mu^x j_B^\mu(x) = 0.
\end{equation}
Now it is possible to define the corresponding charges, the ghost
charge $Q_c$ and the BRS charge $Q_B$, as
\begin{equation}
  Q_c \defined \lim_{\lm\searrow 0} \int d^4x \,h_\lm(x)\, k^0(x) \qquad 
  \text{and} \qquad 
  Q_B \defined \lim_{\lm\searrow 0} \int d^4x \,h_\lm(x)\, j_B^0(x) .
\end{equation}
Here $h_\lm \in \cD(M), \,\, \lm \in \R^+\setminus \set{0}$ is a test
function that has the following structure: 
\beql{defh}
  \begin{split}
    h_\lm(x) = \lm h^t(\lm\cdot x_0)b(\lm\vec{x}) , \qquad &h^t \in \cD(\R)  
    \qquad b \in \cD(\R^3), \\
    &\int dx_0 \,h^t(x_0) = 1, 
  \end{split}
\end{equation}
with $b=1$ on an open domain including the origin of $\R^3$. Due to a 
general argument of Requardt \cite{Req76} the limit $\lm \searrow 0$ 
exists and it is independent of the choice of $h_\lm$. So the
charges define well posed operators in the Fock space. The 
charges have no counterpart in the symbolic algebra, because the 
integrals would make no sense there. \\
The ghost transformation and the BRS transformation are (anti-) 
derivations on the algebra $\End(\eD)$:
\begin{equation}
  s_c (A) \defined \comm{Q_c}{A}, \qquad \text{and} \qquad 
  s_0 (A) \defined \scomm{Q_B}{A} \qquad  \forall A \in \End(\eD).
\end{equation}
The derivations give for the basic fields the following results:
\begin{equation}
  \begin{split}
    &s_c (u^a(x)) = u^a(x), \qquad s_c (\tu^a(x)) = - \tu^a(x),\\
    &s_c (\vp_i(x)) = 0 \quad \text{otherwise},\\
    &s_0 (A^a_\mu(x)) = i \dd_\mu^x u^a(x), \qquad 
    s_0 (\tu^a(x)) = - i \dd^\mu_x A_\mu^a(x),\\
    &s_0 (\vp_i(x)) = 0 \quad \text{otherwise}. \\
  \end{split}
\end{equation} 
Finally we must prove that for the physical state space, defined as
the state cohomology of $\cF$ w.r.t.\ the BRS charge $Q_B$ above, the 
positivity assumption holds. This has already been done by Kugo and
Ojima \cite{KugOji79}, but we present here a modern version that is 
due to Razumov and Rybkin \cite{RazRyb90}. We collect here only the 
essential points of their proof. \\
At first they note that the entire space $\eD$ can be decomposed 
as
\begin{equation}
  \cF = \im Q_B \oplus \left( \im Q_B \cap \im Q^+_B\right) 
  \oplus \im Q^+_B
\end{equation}
with
\begin{equation}
  \begin{split}
    &\im Q_B \oplus \left( \im Q_B \cap \im Q^+_B\right) 
    = \ker Q_B\\ 
    \text{and}\qquad &  \left( \im Q_B \cap \im Q^+_B\right) 
    \oplus \im Q^+_B = \ker Q^+_B.
  \end{split}
\end{equation}
Then they propose an alternative definition of the physical (pre-) 
Hilbert space according to
\begin{equation}
  \eH_{\rm phys} = \left( \im Q_B \cap \im Q^+_B\right) 
\end{equation}
or, which is the same,
\begin{equation}
  \eH_{\rm phys} = \ker \left( \acomm{Q_B}{Q^+_B} \right).  
\end{equation}
This definition of the physical (pre-) Hilbert space deviates from 
the original one in the way that it selects from each equivalence
class there exactly one representative. Now a direct calculation 
of the operator $\acomm{Q_B}{Q_B^+}$ reveals
\begin{equation}
  \acomm{Q_B}{Q_B^+} = N_0 + N_L + N_g
\end{equation}
where $N_0$ is the number operator of scalar vector bosons introduced
above, $N_L$ is the corresponding operator for the longitudinal vector 
bosons and $N_g$ the operator that counts the total number ghosts and
anti-ghosts. Comparison with the definition of the Krein operator
\begin{equation}
  J = (-1)^{N_0} \otimes \one \otimes J_g
\end{equation}
reveals that $J = \one$ on 
$\eH_{\rm phys} = \ker \left( \acomm{Q_B}{Q^+_B} \right)$. Therefore
the inner product must be positive on $\eH_{\rm phys}$ since the
original scalar product was. It is not necessary to restrict the 
physical Hilbert space to the kernel of $Q_c$ since this Hilbert
space is already contained in $\ker N_g \subset \ker Q_c$.\\
The result ensuring positivity holds also for our definition of 
$\eH_{\rm phys}$ as
\begin{equation}
  \eH_{\rm phys} = \overline{(\ker Q_B,\eD) / (\im Q_B,\eD)}^{\norm{\cdot}}  ,
\end{equation}
since in this definition each equivalence class modulo $(\im Q_B)$ 
corresponds to exactly one element of 
$\ker \left( \acomm{Q_B}{Q^+_B} \right)$, 
and the inner product does not depend on the choice of the representative 
within the equivalence class. \\
Then the algebra of observables is defined as usual, 
\begin{equation}
  \eA_{\rm ph} \defined 
  \left(\vps (\ker s,\End(\eD))\cap(\ker s_c,\End(\eD))\right)/
  (\im s,\End(\eD)).
\end{equation}
As was pointed out by D\"utsch and Fredenhagen, the algebra is 
faithfully represented in the physical Hilbert space. 


%% file: normal.tex
\section{Time ordered products and their normalization}
\label{normal}

In this chapter the construction of time ordered products, 
antichronological products and their respective properties are presented. 
Since the construction is in general not unique, normalization 
conditions are postulated that restrict the ambiguity. The construction
of time ordered products is the central point for causal perturbation 
theory, which is presented in the next chapter. In particular this is the 
point where renormalization takes place in this framework.  \\
The time ordering of $n$ arbitrary Wick polynomials 
$W_1(x_1),\dots ,W_n(x_n)$, $W_i \in \Dist_1(\eD)$, can be done by the 
following prescription 
\beql{naiveT}
\begin{split}
  T(W_1(x_1) \cdots W_n(x_n)) 
  \defined & (-1)^{f(\pi)+g(\pi)} \sum_{\pi \in \eP_\un} 
  \theta(x_{\pi(1)}^0- x_{\pi(2)}^0) \cdots \qquad \\
  &\cdots \theta(x_{\pi(n-1)}^0- x_{\pi(n)}^0)
  W_{\pi(1)}(x_{\pi(1)})\cdots W_{\pi(n)}(x_{\pi(n)}) 
\end{split}
\end{equation}
if all the points $x_i$ are different. Here $\eP_\un$ is the set of 
permutations of $\un\defined \set{1,\dots, n}$, $f(\pi)$ is the number 
of transpositions in $\pi \in \eP_\un$ that involve arguments with an 
odd fermion number and $g(\pi)$ is the number of those that involve 
arguments with odd ghost number. $\theta$ is the Heaviside step 
function,
\beql{Heaviside}
  \theta (x) = 
  \begin{cases}
    1 & \text{if } x^0 > 0 \\
    0 & \text{otherwise.} 
  \end{cases}
\end{equation}
The crucial point is that this prescription is not defined for coinciding 
points, because the Wick polynomials $W_i$ are distributions that ``do not 
like to be multiplied by discontinuous functions'' \cite{Sto93}. This is
the origin of the ultraviolet divergences of quantum field theories. 
The prescription above gives, as it stands, well defined distributions 
only on a smaller space of test functions than $\cD(M^n)$. This is the space
of test functions in $\cD(M^n)$ that vanish with all their derivatives 
if two or more of their spacetime arguments coincide. To form time
ordered products these distributions on the smaller space of test 
functions must be extended to elements of $\Dist_n(\eD)$.  \\
The time ordering of $n$ arguments is usually regarded as a mapping of 
$n$ operator valued distributions in $\Dist_1(\eD)$ to an operator 
valued distribution in $\Dist_n(\eD)$. We however define the time 
ordering of $n$ arguments as a mapping of $n$ polynomials in $\eP$ to 
an operator valued distribution in $\Dist_n(\eD)$. As already mentioned 
this has the advantage that the normalization conditions can be 
formulated also for derivated fields. Beside that technical point the
extension of the distributions follows the method of Epstein and Glaser
\cite{EpGlas73}. The extension exists always but is in general not unique. 
Therefore for each combination of arguments one element in $\Dist_n(\eD)$ 
must be chosen as the time ordered product of these arguments. This 
choice is called the {\em normalization} of that time ordered product 
according to Scharf \cite{Sch95}.\\
The normalization conditions implement various properties of the time 
ordered products that are desired from the physical point of view. The 
postulation of the normalization conditions restricts the number of 
possible normalizations, but the extension is in general still not 
unique. \\
This chapter is organized as follows: The first section presents the 
properties of time ordered products that are required for their
construction. In the next section this construction is performed.  
Antichronological products are defined in the third section. The 
chapter concludes with a section in which the normalization conditions 
are formulated. 

\subsection{Properties of time ordered products}
The construction of time ordered products proceeds by induction. The 
time ordered products of a number of arguments are built out of the 
time ordered products with fewer arguments. This construction works 
only if the time ordered products with fewer arguments have certain 
properties. These properties are presented here. They are \\
{\bf P1} (Well posedness): The time ordering operator for $n$ 
arguments, $T_n$, is a multi linear mapping of $n$ polynomials
in $\eP$ to the operator valued distributions of order $n$ on the 
dense subspace $\eD\subset \cF$:
\begin{equation}
  T_n: \qquad \underbrace{\eP \times\dots \times \eP}_{n \,{\rm times}} 
  \to \Dist_n(\eD) .
\end{equation}
If the arguments are explicitely given, the index $n$ indicating the 
number of arguments will be omitted. \\
From the physical interpretation of time ordering we would expect 
that the time ordering operator must have at least two arguments, for 
otherwise there is nothing to be put in order. But it turns out to 
be useful to extend the mapping defined above formally also to the
cases $n=0$ and $n=1$. This is achieved by the following definitions:
\begin{equation}
  T_0 \defined \one, \qquad \one \in \End(\eD)
\end{equation}
and 
\begin{equation}
  T_1(W)(x) \defined T(W)(x), \qquad \forall W \in \eP.
\end{equation}
Here $T$ on the right hand side is the linear representation defined in 
the last chapter. The 
operator valued distributions obtained by the time ordering are called 
{\em time ordered products} or {\em $T$- products}. The time ordered 
product of the polynomials $W_1,\dots, W_n$ is written as 
\begin{equation}
  \TT{W_1,\dots ,W_n}(x_1,\dots,x_n). 
\end{equation}
The definition above implies that the tensor product of two time ordered 
products with $m$ and $n$ arguments is a well defined operator valued 
distribution in $\Dist_{m+n}(\eD)$. Arguments that are multiples of the 
identity can be removed according to 
\begin{equation}
  \TT{W_1,\dots, W_n, a\cdot \one}(x_1, \dots , x_n,y) = 
  a\cdot \TT{W_1,\dots ,W_n}(x_1, \dots , x_n) \qquad 
  \forall a \in \C.
\end{equation}
{\bf P2} (Graded symmetry): Time ordered products are {\em totally 
graded symmetric} under permutations of their indices. That means
\beql{grsymm}
  \begin{split}
    &\TT{W_{\pi(1)},\dots ,W_{\pi(n)}}(x_{\pi(1)}, \dots, x_{\pi(n)})\\
    &\qquad = (-1)^{f(\pi)+g(\pi)} \TT{W_1,\dots ,W_n}(x_1, \dots, x_n)
    \qquad \forall\,\pi \in \eP_\un ,
  \end{split}
\end{equation}
where the integers $f(\pi)$, $g(\pi)$ were defined in eqn. 
\eqref{naiveT}.\\
{\bf P3} (Causality): Time ordered products are {\em causal}, that 
means they fulfill eqn. \eqref{naiveT} for non coinciding points. 
Even more, outside the total diagonal $\Diag_n$ (see below) the time 
ordered product of $n$ arguments is completely determined by those that
have fewer arguments. The total diagonal $\Diag_n \subset M^n$ is the 
set where {\em all} points coincide:
\begin{equation}
  \Diag_n = \set{(x_1,\dots,x_n)\in M^n: \quad x_1=\dots =x_n} .
\end{equation}
If not all points $x_i$ coincide there exists a spacelike surface 
$\Sigma \subset M$ that separates the points $X=\set{x_1,\dots,x_n}$ into 
a future subset $Z$ and a past subset $Z^c = X \setminus Z$ such that 
\begin{equation}
    \Sigma \cap X = \emptyset, \qquad Z \subset (\Sigma + \forlc), 
    \qquad Y \subset (\Sigma + \backlc).
\end{equation}
This situation will be denoted as $Z\gtrsim Z^c$. Furthermore we introduce 
the abbreviation
\beql{shortnot}
  \TT{W_Z}(x_Z) \defined \TT{W_1,\dots ,W_k}(x_1,\dots x_k) \qquad
  \text{if } Z = \set{x_1,\dots x_k} .
\end{equation}
Causality means that the time ordered product $\TT{W_X}(x_X)$ is required 
to satisfy {\em causal factorization}:
\beql{noncoT}
  \TT{W_X}(x_X) = \TT{W_Z}(x_Z) \TT{W_{Z^c}}(x_{Z^c}) \qquad \text{if }
  Z\gtrsim Z^c.
\end{equation}
It provides a recursive definition of the time ordered products up to 
the diagonal $\Diag_n$. There the separation into future and past subsets 
is impossible and therefore no causal factorization exists. Validity of 
causal factorization for every number of arguments implies that spacelike 
separated time ordered products (anti-) commute\footnote{The notation 
$Z \spacel Z^c$ means that $Z$ and $Z^c$ are spacelike separated, i.e.\ 
$Z\gtrsim Z^c$ and $Z^c\gtrsim Z$.}:
\beql{causality}
  \scomm{\TT{W_Z}(x_Z)}{\TT{W_{Z^c}}(x_{Z^c})} = 0
  \qquad \text{if } Z \spacel Z^c . 
\end{equation}
{\bf P4} (Translational invariance): Time ordered products are 
{\em translationally invariant}, that means that for every $a \in M$ 
the following equation holds:
\beql{trinv}
  \begin{split}
    \Ad{U}{p} \,&\TT{W_1,\dots, W_n\vps}(x_1, \dots, x_n) =\\
    & = \TT{W_1,\dots, W_n\vps}(x_1 - a, \dots, x_n - a) 
    \qquad \forall \,p=(a,\one) \in \cP_+^{\uparrow} .
  \end{split}
\end{equation}
Here $U$ is the representation of the Poincar\'e group in the Fock
space introduced in the last chapter. 

\subsection{Inductive construction of time ordered products}\label{indkon}
In this section the inductive construction of the time ordered 
products is outlined. It goes back to Epstein and Glaser \cite{EpGlas73}. 
We use a formulation of their procedure proposed by Stora \cite{Sto93} 
and recently elaborated by Brunetti and Fredenhagen in \cite{BruFre99}. 
This section will not contain the proofs of the theorems. For them we 
refer to the latter article. \\
Formally the time ordering is also defined for a single argument by the 
linear representation $T$. The latter is uniquely defined for all 
$W\in \eP$. This will serve as a starting point for the induction. 
Obviously the representation satisfies properties {\bf P1} - 
{\bf P4}. \\ 
We suppose that all $T$-products for up to $n-1$ arguments are already
constructed and satisfy properties {\bf P1} - {\bf P4}. Due to property
{\bf P3} the time ordered products for $n$ arguments are therefore 
completely determined on $M^n \setminus \Diag_n$, i.e.\ for all test
functions in $\cD(M^n \setminus \Diag_n)$. To construct the distributions
off the diagonal we introduce at first a partition of $M^n\setminus \Diag_n$
into the spaces
\begin{equation}
  \begin{split}
    \complement_Z \defined \set{(x_1,\dots ,x_n) \in M^n: \quad 
    x_i\not\in (x_j+\backlc),\quad  \forall\,i\in Z, \,j\in Z^c}& \\
    \text{for every } Z \neq \emptyset, Z \neq X &.
  \end{split}
\end{equation}
It is easy to see (and has been proven in \cite[Lemma 4.1]{BruFre99}) 
that
\begin{equation}
  \bigcup_{\stackrel{Z\neq \emptyset,}{Z \neq X}} \complement_Z = 
  M^n\setminus \Diag_n .
\end{equation}
Furthermore we define 
\beql{defT1}
  T_Z(W_X)(x_X) \defined 
  \begin{cases}
    \TT{W_Z}(x_Z) \TT{W_{Z^c}}(x_{Z^c}) \qquad \text{if } (x_1,\dots ,x_n) 
    \in \complement_Z,\\
    0 \qquad \text{otherwise}.
  \end{cases}
\end{equation}
$T_Z(W_X)(x_X)$ is a well defined operator valued distribution in
$\Dist_n(\eD)$. Finally we choose an arbitrary locally finite 
$\cinfty$-partition of unity for $M^n\setminus \Diag_n$,
\beql{partunity}
  \begin{split}
    \set{f_Z}:\qquad 
    &\sum_{Z} f_{Z} = 1 \text{ on } 
    M^n\setminus \Diag_n, \\
    &\supp f_{Z} \in \complement_Z ,\qquad
    f_Z \in \cinfty(M^n\setminus \Diag_n) .
  \end{split}
\end{equation}
The restriction of $\TT{W_X}(x_X)$ to $M^n\setminus \Diag_n$\footnote{That 
means $\TTn{W_X}(x_X):\quad \cD(M^n\setminus \Diag_n) \to \End(\eD)$} can 
now be defined as 
\begin{equation}
  \label{0T}
  \TTn{W_X}(x_X) \defined \sum_{Z} 
  f_Z(x_X) \cdot T_Z(W_X)(x_X). 
\end{equation}
This definition does not depend on the choice of $\set{f_Z}$ because 
we assumed that eqns. \eqref{noncoT} and \eqref{causality} hold for the
$T$-products with fewer arguments. This makes the $T^0$-products well 
defined operator valued distributions on 
test functions in $\cD(M^n\setminus \Diag_n)$ that satisfy the 
properties {\bf P1} - {\bf P4}. For the proofs see \cite{BruFre99}. \\
For the construction of the time ordered products with $n$ arguments
the $T^0$-products must be extended to the diagonal. They are linear 
combinations of products of numerical distributions $t^{\,0}$ with 
Wick products $:W_1(x_1)\cdots W_n(x_n):$, where the 
$W_i(x_i)$ are Wick monomials in $\Dist_1(\eD)$. It is not trivial
that these products exist, because distributions are multiplied at the 
same spacetime point, but it was shown by Epstein and Glaser that 
translational invariance implies that this product is indeed well 
defined --- this result is referred to as ``Theorem 0'' 
in \cite[p. 229]{EpGlas73}. From the modern point of view the product 
exists because the wave front sets of the distributions do not linearly 
combine to zero in the cotangent spaces, see \cite{BruFre99}.\\
For the extension of the operator valued distributions to the diagonal it 
suffices to extend each numerical distribution $t^{\,0}$ and to prove that 
the resulting product is well defined. The latter is no problem here 
because the ``Theorem 0'' applies also to the extended distributions.\\
Translational invariance ({\bf P4}) implies that the numerical 
distributions $t^{\,0}$ depend only on the relative coordinates 
$(y_1,\dots ,y_{n-1}) \defined (x_1-x_n,\dots ,x_{n-1}-x_n)$ such that 
$\Diag_n$ is the origin in the space of the $y$'s. This allows us to give 
a further restriction to the extension of the numerical distributions 
to the diagonal: Each distribution $t^{\,0}$ is regarded as a 
distribution in the space of relative coordinates. Then the scaling 
degree of the extended distribution $t$ must not exceed that of the 
original distribution $t^{\,0}$ in relative coordinates. Brunetti and 
Fredenhagen \cite{BruFre99} prove 
that such an extension always exists as a well defined distribution for 
test functions in $\cD(M^{n-1})$ --- or in $\cD(M^n)$ if one returns to 
the original coordinates. It is unique only if the original distribution 
has a scaling degree $\scaldeg(t^{\,0})$ that satisfies the following 
inequality:
\beql{ineq}
  \scaldeg(t^{\,0}) < (n-1) \times d
\end{equation}
where $n-1$ is the number of relative coordinates and $d$ the spacetime 
dimension. This can be seen as follows: The distribution $t$ is already 
determined up to the diagonal $\Diag_n$. In other words, two extensions 
may differ only by a delta distribution with support at the origin of the 
relative 
coordinates or by a derivative of it. If the scaling degree of $t^{\,0}$ 
satisfies the inequality above, it is not possible to add a delta 
distribution or a derivative of it without violating the restriction
on the scaling degree. Therefore the solution is unique then. In general
the inequality does not hold and the extension is therefore ambiguous,
corresponding to the freedom of finite renormalization in other 
renormalization procedures. 
 
\subsection{Antichronological products}
In this section we define antichronological products. This definition 
can be given recursively as $\bT_0=\one$ and\footnote{The notation is the 
same as in eqn. \eqref{shortnot}} 
\beql{unit}
\begin{split}
  &\TTb{W_X}(x_X) \,\defined \\
  &\qquad = - \sum_{Y\subset X, Y\neq \emptyset} 
  (-1)^{\betr{Y}} \TT{W_Y}(x_Y) \TTb{W_{Y^c}}(x_{Y^c})  \\
  &\qquad = - \sum_{Y\subset X, Y\neq X} 
  (-1)^{\betr{Y^c}} \TTb{W_Y}(x_Y) \TT{W_{Y^c}}(x_{Y^c}) 
\end{split}
\end{equation}
for $n\geq 1$. Here possible signs that come from changes in the order
of the arguments are neglected for simplicity. They can be easily 
recovered using {\bf P2}, which holds for the $\bT$-products, too (see 
below).\\
Iterating the recursive definition above one finds the following explicit 
expression for the $\bT$-products:
\begin{equation}
  \TTb{W_X}(x_X) 
  = \sum_{P} (-1)^{\betr{P}+\betr{X}} \prod_{p\in P} \TT{W_p}(x_p) .
\end{equation}
Here the sum runs over all partitions $P$ of $X$ into $\betr{P}$ 
nonempty subsets. With this definition the antichronological products 
become for non coinciding points, i.e.\ $x_i\neq x_j \,\,\forall i\neq j$,
\begin{equation}
\begin{split}
  \TTb{W_X}(x_X) 
  = &\sum_{\pi \in \eP_\un} 
  \theta(x^0_{\pi(1)}- x^0_{\pi(2)}) \cdots \\
  &\cdots \theta(x^0_{\pi(n-1)}- x^0_{\pi(n)})
  \TT{W_{\pi(n)}}(x_{\pi(n)})\cdots  \TT{W_{\pi(1)}}(x_{\pi(1)}).
\end{split}
\end{equation}
The antichronological products satisfy properties {\bf P1, \bf P2} 
and {\bf P4}. Property {\bf P3} holds for them in the reverse order.
That means that  
under the same conditions and with the same notation as in {\bf P3} 
antichronological products satisfy
\begin{equation}
  \TTb{W_X\vps}(x_X) = \TT{W_{Z^c}\vps}(x_{Z^c}) \TT{W_Z\vps}(x_Z) 
  , \qquad Z\gtrsim Z^c
  \tag{{\bf P3'}}
\end{equation}
justifying their name since they are defined like the time ordered 
products but with the opposite order.
\subsection{Normalization conditions}\label{normcon}
In this section we formulate the normalization conditions that restrict 
the ambiguity in the extension of the $\bT$-products to the diagonal. 
They implement Poincar\'e covariance \eqref{N1} and unitarity \eqref{N2}.
They define the time ordered products up to a $\C$-number distribution
\eqref{N3} and determine them uniquely if at least one argument is a 
generator from $\cG$ \eqref{N4}. Finally they determine Ward 
identities for the ghost current \eqref{N5} and the BRS current 
\eqref{N6}. It is proven that the conditions \eqref{N1} - \eqref{N5}
have common solutions. For condition \eqref{N6} this must be done for
the individual models.\\
The first normalization condition establishes Poincar\'{e} covariance 
w.r.t.\ the representation $U$ of the Poincar\'{e} group 
$\cP_+^{\uparrow}$ introduced in chapter (\ref{free}). It reads
\beql{N1}
  \begin{split}
    &\Ad{U}{p} \,\TT{W_1,\dots , W_{n}\vps}(x_1,\dots ,x_n) = \\
    &\qquad = \TT{\eR_\Lm(W_1),\dots , \eR_\Lm(W_n)}(\Lm^{-1} x_1-a,\dots ,
    \Lm^{-1} x_n-a)
  \end{split}
  \tag{{\bf N1}}
\end{equation}
for every $p=(a,\Lm) \in \cP_+^{\uparrow}$ and all monomials $W_i \in \eP$. 
Here $\eR_\Lm$ is the representation of the Lorentz group on $\eP$
introduced in section (\ref{defP}). Property $\bf P4$ is in view of 
\eqref{N1} only the special case with $p=(a,\one)$. \\
Popineau and Stora \cite{PopSto82} have proven that this condition has 
always a solution, but their article is unfortunately not published.
So we refer the reader to Scharf \cite[p. 282]{Sch95} for the proof. 
Recently Prange, Bresser and Pinter, \cite{BrePraPin99} and \cite{Pra99}, 
have found even a general construction prescription for covariant 
normalizations.\\
The second normalization condition establishes pseudo-unitarity by 
means of 
\beql{N2}
  \TT{W_1,\dots , W_n}^{*}(x_1,\dots ,x_n)  \,=\,
  \TTb{W_n^{*},\dots , W_1^{*}}(x_n,\dots ,x_1) 
  \quad \forall \,W_i \in \eP, \tag{\bf N2}
\end{equation}
where the $^*$-involution on the left hand side is the Krein adjoint on 
$\End(\eD)$, while the $^*$-involution on the right hand side is the 
adjoint operation in $\eP$ defined in section (\ref{defP}). Note that 
the order of the arguments is reversed. It can of be put into the 
original order by means of {\bf P2}.\\
It was already shown by Epstein and Glaser \cite{EpGlas73} that 
eqn. \eqref{N2} can always be accomplished.  Their argument and 
the compatibility of \eqref{N2} with \eqref{N1} can be easily 
understood: Suppose, \eqref{N2} holds for all integers $m<n$ 
simultaneously with eqn. \eqref{N1}. Then for every normalization 
$T' = \TT{W_1,\dots , W_n}$ that is compatible with eqn. 
\eqref{N1} the distribution $T=\half(T' +  {T'}^*)$ satisfies 
eqn. \eqref{N2} and will also be an extension of $T^0$ because 
\eqref{N2} holds for the $T^0$-products by induction. It will 
automatically be a solution of eqn. \eqref{N1}  
since the representation $U$ was chosen to be pseudo-unitary, 
i.e. $U(p)^* = U(p)^{-1}$.\\
To formulate the third normalization condition we remind the reader
of the commutator function $\De_{jk}(x)$, eqn. \eqref{commfunc} in 
section [\ref{defrep}]. The normalization condition reads:
\beql{N3}
\begin{split}
  &\scomm{\TT{W_1 ,\dots , W_n}(x_1,\dots ,x_n)}{\vp_i(y)\vps} = \\
  &\qquad = i \sum_{k=1}^n \sum_j \Delta_{ij}(x_k-y)\cdot 
  \TT{W_1 ,\dots , \partder{W_k}{\vp_j},\dots , W_n}(x_1,\dots ,x_n),
\end{split}
\tag{\bf N3}
\end{equation}
for every $W_i \in \eP, \quad \vp_i(y) = \TT{\vp_i}(y), \quad \vp_i \in \cG$. 
The second sum runs over all generators in $\cG$, not only the basic
generators. \\
Since an element of $\End(\eD)$ is a multiple of the identity if it 
(anti-) commutes with all the $\vp_i(y)$ --- see eqn. \eqref{trivcent} 
in section (\ref{fock}) ---, this condition determines the time ordered 
products uniquely up to a $\C$-number, provided the time ordered products 
that involve the sub monomials are known. This can be explicitly seen in 
an equivalent equation, the {\em causal Wick expansion}
\beql{CWE}
\begin{split}
  \TT{W_1,\dots ,W_n\vps}(x_1, \dots, x_n) =
  \sum_{\gamma_1,\dots , \gamma_n} \om_0
  \left(
    \TT{W_1^{(\gamma_1)},\dots, W_n^{(\gamma_n)}\vps}(x_1, \dots, x_n)
  \right) &\\
  \times \frac{:\vp^{\gamma_1}(x_1) \cdots \vp^{\gamma_n}(x_n):}{
  \gamma_1!\cdots \gamma_n!} &.
\end{split}
\end{equation}
Here the $\gamma_i \in \N^r$ are multi indices, vectors with one entry 
for each of the $r$ generators in $\cG$, i.e. 
\begin{equation}
  \gamma_i = ((\gamma_i)_1,\dots , (\gamma_i)_r) \quad \in \N^r
\end{equation}
The $W^{(\gamma_i)}$ are derivatives,
\begin{equation}
  W^{(\gamma_i)} \defined \partder{^{\betr{\gamma_i}}W}{^{(\gamma_i)_1} 
  \vp_1\cdots \dd^{(\gamma_i)_r} \vp_r} ,
\end{equation}
where $\betr{\gamma_i} = \sum_{k=1}^r (\gamma_i)_k$. The $\vp^{\gamma_i}$ 
are defined as
\begin{equation}
  \vp^{\gamma_i}(x) \defined \TT{\prod_{k=1}^r \vp_k^{(\gamma_i)_k}}(x) .
\end{equation}
Finally 
\begin{equation}
  (\gamma_i)! \defined \prod_{k=1}^r (\gamma_i)_k ! . 
\end{equation}
It is shown in the appendix, section \ref{proofN3}, that the causal 
Wick expansion is indeed equivalent with \eqref{N3}. Compatibility 
with eqn. \eqref{N1} is easily verified since \eqref{N3} respects the
Poincar\'e transformation properties. With the same construction as
after eqn. \eqref{N2} one can show that for every common solution of 
\eqref{N1} and \eqref{N3} a normalization can be constructed that is also 
a solution of \eqref{N2}. \\
In particular in the formulation \eqref{CWE} of \eqref{N3} it is 
immediately clear that only $\om_0 \left(\TT{W_1,\dots, W_n\vps}(x_1, 
\dots, x_n) \right)$ --- the term with $\gamma_1 = \dots =\gamma_n = 0$
in \eqref{CWE} --- is left open to be normalized, since all other terms 
are determined by the time ordered products for the sub monomials. These 
distributions correspond to the vacuum diagrams 
of the respective time ordered product in the Feynman graph picture.
So condition \eqref{N3} has the consequence that only vacuum diagrams 
need to be (re-) normalized, a fact that is well known from other 
renormalization procedures. \\
The fourth normalization condition is a differential equation that
uniquely determines time ordered products with at least one generator
$\vp_i \in \cG$ among its arguments. This assertion holds under the 
assumption that the time ordered products for fewer arguments are
already known. The condition reads:
\beql{N4}
\begin{split}
  &\sum_j D^y_{ij} \TT{W_1,\dots , W_n,\vp_j\vps}(x_1,\dots ,x_n,y) = \\
  &\qquad = i\sum_{k=1}^n \TT{W_1 ,\dots , \partder{W_k}{\varphi_i} 
  ,\dots , W_n}(x_1,\dots ,x_n) \,\delta (x_k-y)  ,
\end{split}
\tag{\bf N4}
\end{equation}
where $W_i\in \eP, \vp_j\in \cG$. It is proven in the appendix, 
section \ref{proofN4}, that condition \eqref{N4} has common solutions 
with condition \eqref{N3}. Compatibility with condition \eqref{N1}
is again immediate since \eqref{N4} is Poincar\'e covariant. A
solution of \eqref{N1}, \eqref{N3} and \eqref{N4} that satisfies
also \eqref{N2} can be found by the same procedure as above. \\
In the chapter concerning the interacting theory we will see that 
eqn. \eqref{N4} already implies the interacting field equations. \\
Condition \eqref{N4} possesses an alternative formulation, like \eqref{N3}. 
Its integrated version reads
\beql{N4int}
  \begin{split}
    &\TT{W_1,\dots ,W_n,\vp_i\vps}(x_1,\dots ,x_n,y) = \\
    &\quad = i \sum_{k=1}^{n} \sum_{j} \De_{ij}^F(y-x_k) 
    \TT{W_1,\dots , \partder{W_k}{\vp_j},\dots, W_n}(x_1,\dots ,x_n)\\
    &\qquad + \hspace{-5pt}\sum_{\gamma_1\cdots\gamma_n}\om_0
    \left(
      \TT{W_1^{(\gamma_1)},\cdots, W_n^{(\gamma_n)}\vps}(x_1, \dots, x_n)
    \right)
    \frac{:\vp^{\gamma_1}(x_1) \cdots \vp^{\gamma_n}(x_n)
    \vp_i(y):}{\gamma_1!\cdots \gamma_n!} .
  \end{split}
\end{equation}
The sum over $j$ runs again over all generators including the higher
ones.\\
This formulation shows explicitely that with eqn. \eqref{N4} the time 
ordered products with at least one generator among its arguments are 
already determined. In appendix (\ref{proofN4}) the equivalence of the 
two formulations is proven.\\
Eqn. \eqref{N4} uniquely fixes the Feynman propagators for derivated 
fields. These in turn determine all tree level diagrams. Comparing 
\eqref{N4} with results from other renormalization procedures shows an  
important difference between the causal approach and other approaches: 
The definition of the propagators for the derivated fields differ between
the causal approach and other approaches. Therefore also the Green's 
functions at tree level are different. The difference between
the conventional propagators and our prescription is labelled by the
normalization constants $C_{\vp_i,k}$. Only if all these constants are 
set to zero the difference disappears. But we saw already that the 
propagators are then no longer invertible. For example, in the 
conventional renormalization procedures we have
\begin{equation}
  \om_0\left(\TT{\dd^\mu_x A_\nu(x), \dd^\nu_y A_\rho(y)}\right) 
  = - i \dd_\rho^x \dd^\mu_x D^F(x-y),
\end{equation}
while the corresponding propagator in our causal theory reads
\begin{equation}
  \om_0\left(\TT{(A_\nu)^{1,\mu}, (A_\rho)^{1,\nu}}(x,y)\right) 
  = - i \dd_\rho^x \dd^\mu_x D^F(x-y) - i C_{A,1} \de_{\rho}^{\mu} 
  \de(x-y) .
\end{equation}
Now we come to the Ward identities for the ghost current. This is a 
normalization condition for time ordered products that contain a ghost 
current $k^{\mu}$ --- see section (\ref{freeBRS}) --- as an argument. It 
reads 
\beql{N5}
\begin{split}
  &\dd^{y}_{\mu}\TT{W_{1} ,\dots , W_{n}, k^{\mu}}(x_1,\dots ,x_n,y) = \\
  &\qquad = \sum_{k=1}^n g(W_k)\,\delta(y-x_k) \, 
  \TT{W_{1} ,\dots , W_{n}}(x_1,\dots ,x_n) 
\end{split}
\tag{\bf N5} 
\end{equation}
for all monomials $W_i \in \eP$. It holds if none of the arguments 
contains a generator $(u^a)^{(\al)}$ or $(\tu^a)^{(\al)}$ with 
$\betr{\al}\geq 2$. \\
The proof that this normalization condition has common solutions with
the conditions \eqref{N1} - \eqref{N4} is given in appendix (\ref{proofN5}). 
For a technical reason that will be explained there this normalization 
condition can only be proven for arguments $W_i$ that do not contain 
$k^{\mu}$ as a sub monomial --- in particular $k^{\mu}$ itself is 
excluded. In the examples where \eqref{N5} is applied in the following 
chapters this limitation will not be relevant.\\
We state here one particular fact that will come out in the proof: There 
exists exactly one choice for the normalization constant $C_{u,1}$ such
that condition \eqref{N5} has common solutions with \eqref{N1} - 
\eqref{N4}. This choice is $C_{u,1} = -1$.\\
Following D\"utsch and Fredenhagen \cite{DueFre98} who made the calculation 
for the Ward identities for the electric current (see below) we prove in
appendix (\ref{proofN5}) that there exists an integrated version 
of \eqref{N5}, namely
\beql{N5int}
\begin{split}
  &s_c\TT{W_{1} ,\dots , W_{n}}(x_1,\dots ,x_n) = \\
  &\qquad = \left(\sum_{k=1}^n g(W_k)\right) 
  \TT{W_{1} ,\dots , W_{n}}(x_1,\dots ,x_n) .
\end{split} 
\end{equation}
So as a consequence of \eqref{N5} the ghost number of a time ordered 
product is simply the sum of the ghost numbers of its arguments.\\
Eqn. \eqref{N5} and \eqref{N5int} are equivalent in 
the following sense: If \eqref{N5} holds then \eqref{N5int} is 
automatically valid, too. If \eqref{N5int} holds, then a normalization
can be found that is compatible with \eqref{N5}. For details see 
appendix (\ref{proofN5}).\\
D\"utsch and Fredenhagen \cite{DueFre98} proved an analogous Ward identity 
for the electric current $j^\mu_{\rm el} = \psiq \gamma^\mu \psi$. 
Here $\psi$ and $\psiq$ are the electron and the positron field, 
respectively. Their Ward identity reads in our language
\beql{N5'}
\begin{split}
  &\dd^{y}_{\mu}\TT{W_{1} ,\dots , W_{n}, j_{\rm el}^{\mu}}
  (x_1,\dots ,x_n,y) = \\
  &\qquad = i\left(\sum_{k=1}^n f(W_k)\,\delta(y-x_k) \right) 
  \TT{W_{1} ,\dots , W_{n}}(x_1,\dots ,x_n) .
\end{split}
\tag{\bf N5'} 
\end{equation}
It holds if the monomials $W_i$ do not contain generators $\psi^{(\al)}$ 
or $\psiq^{(\al)}$ with $\betr{\al} \geq 1$. The existence of common 
solutions of \eqref{N5'} with the other normalization conditions
can be proven along the same lines as for the ghost current Ward 
identities, provided that either none of the arguments $W_i$ contains 
$j^{\mu}_{\rm el}$ as a sub monomial or that all the $W_i$ are the QED 
Lagrangian $\cL_{QED} = A_\mu j^\mu_{\rm el}$ or sub monomials of it. \\
To formulate the Ward identity for the BRS current we anticipate here a
condition for the Lagrangian that will be illuminated more closely
in section (\ref{interlag}). In QED and Yang-Mills theory there exist
so called $Q(n)$-vertices for the Lagrangians. These are polynomials 
$\cL_1^\mu, \cL_2^{\mu\rho}, \dots \in \eP$ totally antisymmetric in 
their Lorentz-indices for which the following identities hold:
\begin{equation}
  s_c \TT{\cL_i^{\mu_1,\dots ,\mu_i}}(x) = i \dd_\rho^x
  \TT{\cL_{i+1}^{\mu_1,\dots ,\mu_i,\rho}}(x) .
\end{equation}
We admit only polynomials $\cL$ as Lagrangians if there exist such 
$Q(n)$-vertices and in addition so called $R(n)$-vertices 
$\cM_1, \cM_2, \dots$ that are polynomials in $\eP$ which satisfy 
the following condition: There exists a normalization of 
$\TT{\cL_i^{\mu_1,\dots ,\mu_i},j^\mu}$ that is compatible with the
normalization conditions \eqref{N1} - \eqref{N4} and for which the 
equation
\begin{equation}
  \begin{split}
    \dd_\mu^y \TT{\cL_i^{\mu_1,\dots ,\mu_i},j^\mu}(x,y)\,\, = \,\,\,
    &i \dd_{\nu}^x 
    \left(\de(x-y) \cL^{\mu_1,\dots ,\mu_i,\nu}_{i+1}(x)\vps \right)\\ 
    &+ i \left(\dd_{\nu}^x \de(x-y)\vps\right) 
    \cM^{\mu_1,\dots ,\mu_i,\nu}_{i+1} (x)
  \end{split}
\end{equation}
holds. The series of equations terminates at a certain point i.e. there 
exists an $m\in \N$ with $\cL_m = 0, \,\,\cM_m = 0$. This is the condition 
\eqref{C4} in section (\ref{interlag}). The $R(n)$-vertices are totally 
antisymmetric in their Lorentz indices, too.\\
With the notion of $Q(n)$-vertices and the $R(n)$-vertices we can give the 
next normalization condition, the Ward identities for the BRS current:
\beql{N6}
\begin{split}
  &\dd^{y}_{\mu}
  \TT{\cL_{i_1} ,\dots , \cL_{i_n}, j^{\mu}\vps}(x_1,\dots ,x_n,y) = \\
  &\qquad = i \sum_{k=1}^n \dd_\nu^k
  \left(
    \delta(y-x_k) \,
    \TT{\cL_{i_1} ,\dots , \cL^\nu_{i_k+1} ,\dots , \cL_{i_n}\vps}
    (x_1,\dots ,x_n) 
  \right) \\
  &\qquad \quad + i \sum_{k=1}^n \left(\dd_\nu^k \de(y-x_k)\vps\right)
  \TT{\cL_{i_1} ,\dots , \cM^\nu_{i_k+1} ,\dots , \cL_{i_n}\vps}
  (x_1,\dots ,x_n) 
\end{split}
\tag{\bf N6} 
\end{equation}
where $i \in \N$ and we define $\cL_0 = \cL$. \\
The same calculation leading to eqn. \eqref{N5int} can also be applied 
to condition \eqref{N6} and gives the {\em generalized operator gauge 
invariance} 
\beql{GOGI}
\begin{split}
  &s_0
  \TT{\cL_{i_1}, \dots ,\cL_{i_n}}(x_1,\dots ,x_n) = \\
  &\qquad = i \sum_{k=1}^n \dd_\nu^k
  \TT{\cL_{i_1}, \dots ,\cL^\nu_{i_k+1}, \dots ,\cL_{i_n}}
  (x_1,\dots ,x_n) . 
\end{split}
\end{equation}
D\"utsch, Hurth, Krahe and Scharf, \cite{SchHuDueKra94a} - 
\cite{SchHuDue95b}, \nocite{SchHuDueKra94b}\nocite{SchHuDue95a} found that 
for eqn. \eqref{GOGI} to hold in Yang-Mills theory for two arguments the 
normalization constant $C_{A,1}$ in eqn. \eqref{DA} must be 
$C_{A,1}= -\half$. \\
Unfortunately there exists no general proof that condition \eqref{N6} 
can always be accomplished or that it is compatible with \eqref{N3} 
and \eqref{N4}\footnote{At a first sight it
may seem that \eqref{N4} has nothing to do with \eqref{N6} since there is 
no generator in the time ordered products whose normalization \eqref{N6}
determines. The point is that compatibility with \eqref{N3} requires 
a set of relations among which are also some that involve time ordered 
products that contain a generator. Then \eqref{N4} could fix their 
normalization in a way that compatibility between \eqref{N3} and \eqref{N6}
is inhibited. In this sense we think that \eqref{N4} and \eqref{N6} 
shall be compatible.}. But we can show that the generalized operator
gauge invariance together with \eqref{N5} is already sufficient for 
\eqref{N6}. For the construction of solutions of \eqref{N6} under the 
assumption that generalized operator gauge invariance holds see appendix 
(\ref{WIGOGI}).\\
The proof that either the eqn. \eqref{N6} or eqn. \eqref{GOGI}
have common solutions with the other normalization conditions must be 
done in individual models. \\
As far as we know QED is the only example where this is done --- 
for the proof see section (\ref{QED}). The existence of solutions for 
eqn. \eqref{GOGI} is in QED a direct consequence of the existence of 
solutions for the electric current Ward identities \eqref{N5'}. \\
In Yang-Mills theories the solutions for eqn. \eqref{GOGI} can be 
explicitely given in first order, see section (\ref{YM}). A detailed 
study of \eqref{GOGI} with $i_1 = \dots = i_n = 0$ for Yang-Mills theory
without matter fields can be found in \cite{SchHuDueKra94a} - 
\cite{SchHuDue95b} --- this equation is called {\em operator gauge 
invariance}. This result has been generalized to Yang-Mills theory
with matter fields by D\"utsch \cite{Due96a}. They come to the result 
that operator gauge invariance holds in that theory provided a weak 
assumption concerning the infrared behaviour is satisfied. Loosely 
speaking the infrared behaviour must not be too bad. It is usually
assumed that this assumption is satisfied, otherwise not even off 
shell Green's functions would exist.\\
It should be possible to prove the generalized version of 
operator gauge invariance under the same assumption and along the same 
lines as in their calculation, but this has not been done up to now --- 
and it is probably a long winded work, the original calculation
took a series of four articles.\\
Another promising strategy to prove generalized operator gauge invariance 
is to translate the results of algebraic renormalization \cite{PigSor95}
to causal perturbation theory. The {\em descent equations} can be viewed 
as the generalized operator gauge invariance version of that framework. 
It has been proven in \cite{PigSor95} that they can be accomplished 
for Yang-Mills theories. Unlike our causal approach algebraic 
renormalization is a loop expansion, i.e. an expansion in the parameter 
$\hbar$ and not in the coupling constant. Furthermore it is a functional 
approach, in contrast to causal perturbation theory which is an operator 
approach. So in order to make the results cited above available to the 
causal theory some translational work has to be performed. This has not 
been done up to now.\\
For the equations \eqref{N6} with $\sum i_n \geq 5$ the compatibility
of normalization conditions is easy to prove: These $T$-products comply 
automatically with \eqref{N3} and \eqref{N4} since their ghost number,
which is the minimal number of field operators in the Wick products 
in the causal Wick expansion, exceeds the spacetime dimension, so we
are in the situation of the inequality \eqref{ineq} and therefore the
extension is unique and complies with \eqref{N3}, \eqref{N4} and 
\eqref{N5}. \\
We have stated altogether six normalization conditions for the 
time ordered products (where for the last it remains open whether it can
always be accomplished). One could ask whether these conditions 
suffice to make the extension of the $T^0$-products to the diagonal unique. 
Unfortunately this is not true. There remains a certain ambiguity, even 
though calculations in first order show that the normalization conditions 
restrict the freedom of the extensions severely --- in fact there are 
many examples where the above conditions suffice to make the extension 
unique. \\
The decisive point is that the normalization conditions suffice to
prove a lot of relations in the interacting theory like field equations, 
nilpotency of the interacting BRS charge and others, notwithstanding the 
remaining ambiguity. \\
Another interesting feature of these normalization conditions is that
a subsequent enlargement of the algebra $\eP$ --- by the introduction
of new basic fields or by inclusion of generators for higher derivatives 
of the 
basic fields than before --- does not change the normalization of the 
time ordered products with arguments in the original, smaller algebra. 
Moreover these normalizations do not depend on the model with regard 
to which they are considered. For example there are certain time 
ordered products that occur both in QED and in Yang-Mills theory, but 
due to our construction their normalization is the same in both cases,
provided the normalization constants $C_{\vp_i,k}$ are chosen equal.
This is of course a consequence of the fact that the normalizations 
are completely independent of the Lagrangian. The latter is in this
context a polynomial in $\eP$ not outstanding from the others. 
So the idea behind the whole construction is to determine all time
ordered products a priori, store them in a big library and fetch them
if they are needed for a certain calculation. The remaining ambiguity
of the time ordered products is certainly a handicap. Ambiguous time 
ordered products should be laid down in this library with an endorsement 
that they are ambiguous and what the allowed normalizations are.


%% file: cpt.tex
\section{Local causal perturbation theory}
\label{cpt}

This chapter is devoted to the formulation of local causal perturbation 
theory. It will establish the connection of time ordered products with 
interacting quantum field theory. In the framework of causal perturbation 
theory the S-matrix and the interacting field operators are defined
in terms of time ordered products, see below.\\
As usual the interaction will be defined by the S-matrix. But as we 
investigate local theories, there will be no interpretation of the 
S-matrix available as an operator mapping in-states onto out-states.
These asymptotic states are a global concept that looses
its meaning in a local framework. Nevertheless the S-matrix is the 
central object of the interacting theory. It determines the theory 
since the local interacting field operators are defined in terms of
it. \\
In the causal approach infrared divergences are completely independent 
of the ultraviolet divergences --- in particular there cannot be a 
cancellation of infrared with ultraviolet divergences. We circumvent the 
problem of infrared 
divergences by considering only local theories. By a local theory we mean 
the following situation: We choose an open, bounded domain $\cO\subset M$
in Minkowski space --- usually such that it is causally complete ---
in which the interacting fields are localized and consider the field
algebra generated by these fields. \\
The crucial observation that makes it possible to abandon the adiabatic
limit and therefore to avoid infrared divergences is due to Brunetti and 
Fredenhagen \cite{BruFre96}. They found that a modification of the 
interaction outside the domain $\cO$ induces only a unitary transformation 
of the field algebra. Since this does not touch the physical content
of the theory, it is in particular possible to switch off
the interaction outside $\cO$. With the coupling being a test function, 
infrared divergences cannot occur.\\
The chapter gives a short presentation of causal perturbation
theory in the formulation of Epstein and Glaser \cite{EpGlas73}. 
For the reader interested in details of the causal approach we refer 
to the textbook of Scharf \cite{Sch95}. We use here the notation of
Epstein and Glaser which is different from that in the book of 
Scharf.\\
At first we construct the S-matrix by means of time 
ordered products. In the second section we define interacting field 
operators in terms of retarded products. Advanced and causal 
products are defined in the third section. The model we consider is 
determined by an interaction Lagrangian. It is a polynomial in $\eP$, 
but not every polynomial in $\eP$ can serve as a Lagrangian. We 
postulate in the fourth section five conditions such a polynomial must 
satisfy in order to define a possible Lagrangian. 
\subsection{The S-matrix}\label{Smatrix}
The S-matrix is defined as a formal power series in terms of time ordered
products of the Lagrangian as
\beql{defSmatrix}
  S(g\cL) \defined \sum_{n=0}^{\infty} \frac{i^n}{n!} \int d^4x_1 \cdots 
  d^4x_n \,g(x_1)\cdots g(x_n)\, \TT{\cL,\dots ,\cL\vps}(x_1,\dots ,x_n)
\end{equation}
Here $g$ is the coupling ``constant'', i.e. in our approach a real test 
function in $\cD(M)$. The notation $(g\cL)$ in the argument of $S$ is 
of cause only symbolic, since the product of a test function in $\cD(M)$ 
with a symbol in $\eP$ is not defined. It means that 
the polynomials are the arguments of the time ordering which are smeared
out with the test functions. For sums the symbolic notation means
e.g.
\begin{equation}
  S(g_1\cW_1+g_2\cW_2) \defined \one + i \int d^4x_1 
  \left[ g_1(x_1) \TT{\cW_1}(x_1) + g_2(x_1)\TT{\cW_2}(x_1)\right]
  \, + \dots \quad .
\end{equation}
The S-matrix is an element in $\tilde{\C}\cdot \End{\eD}$, i.e.\ the set of 
formal power series whose elements are endomorphisms on $\eD$. This
is true because 
$\TT{\cL,\dots ,\cL}(x_1,\dots ,x_n) \in \Dist_n(\eD)$ and 
$g(x_1)\cdots g(x_n) \in \cD(M^n)$. \\
The S-matrix is also the generating functional of the time ordered 
products, i.e. the time ordered products can be recovered from the 
S-matrix by means of 
\begin{equation}
  \TT{W_1,\dots ,W_n\vps}(x_1,\dots ,x_n) = 
  \frac{\de^n}{i^n\de g_1(x_1)\cdots \de g_n(x_n)} 
  \left. S\left(\sum_{k=1}^{n} g_k W_k\right)\right|_{g_1=\cdots g_n=0}.
\end{equation}
The inverse S-matrix $S^{-1}(g\cL)$ is also a formal power series. 
From eqn. \eqref{unit} we conclude 
\begin{equation}
  S^{-1}(g\cL) = \sum_{n=0}^{\infty} \frac{(-i)^n}{n!} \int d^4x_1 
  \cdots d^4x_n \, g(x_1)\cdots g(x_n)\, \TTb{\cL,\cdots \cL\vps}
  (x_1,\dots ,x_n).
\end{equation} 
The S-matrix is also pseudo unitary, $S(g\cL)^* = S^{-1}(g\cL)$, by 
means of normalization condition \eqref{N2}. 
\subsection{Interacting fields and retarded products}\label{intfieretpro}
The interacting fields are constructed according to Bogoliubov as 
operator valued distributions by
\beql{interfields}
  \inter{W_i}{g\cL}{y} \defined S(g\cL)^{-1} 
  \frac{\de}{i \de h(y)}\left. S(g\cL + hW_i) \vps \right|_{h=0}.
\end{equation}
Here $h$ is a test function in $\cD(M)$. The corresponding localized
field operators are
\begin{equation}
  \inter{W_i}{g\cL}{f} \defined \int d^4y f(y) \inter{W_i}{g\cL}{y}
\end{equation}
where $f$ is a test function with support in the domain $\cO$ as. The 
algebra of field operators that are localized in $\cO$ is denoted as 
$\ctF(\cO)$.\\
Inserting the definition of the S-matrix the distributional field 
operators can be written as
\begin{equation}
\begin{split}
  \inter{W_i}{g\cL}{y} = \sum_{n=0}^{\infty} \frac{i^n}{n!} 
  \int d^4x_1 \cdots d^4x_n \,g(x_1)\cdots g(x_n) &\\
  \times \RR{\cL,\cdots ,\cL}{W_i\vps}(x_1,\dots ,x_n;y) &.
\end{split}
\end{equation}
This expression contains the so called retarded or $R$-products whose 
definition in terms of $T$- and $\bT$-products reads
\beql{R}
\begin{split}
  &\RR{W_1,\dots ,W_n}{W_i\vps}(x_1,\dots ,x_n;y)  \\
  &\qquad \defined \sum_{Y\subset X} (-1)^{\betr{Y}}
  \TTb{W_Y}(x_Y)\TT{W_{Y^c},W_i}(x_{Y^c},y) .
\end{split}
\end{equation}
Here $X= \set{x_1,\dots , x_n}$. For the notation we refer to eqn. 
\eqref{shortnot}.\\
According to eqn. \eqref{unit} the retarded products can be alternatively
expressed as
\beql{Ralter}
\begin{split}
  &\RR{W_1,\dots ,W_n}{W_i}(x_1,\dots ,x_n;y)   \\
  &\qquad =  \sum_{Y\in X} (-1)^{\betr{Y}}
  \TTb{W_Y,W_i}(x_Y,y)\TT{W_{Y^c}}(x_{Y^c}) .
\end{split}
\end{equation}
Causality \eqref{causality} implies that the retarded products have 
retarded support (justifying their name), i.e.
\begin{equation}
  \begin{split}
    &\supp \RR{W_{X}}{W_i}(x_X,y)  \\
    &\qquad \subset
    \set{(x_1,\cdots ,x_n,y) \in M^{n+1}:\quad  
    x_i \in \left( y + \backlc\right)\quad \forall x_i \in X} .
  \end{split}
\end{equation}
The interacting fields in $\ctF(\cO)$ therefore depend only on the 
interaction in the past of $\cO$. From the 
definition of the interacting field distributions D\"utsch and 
Fredenhagen derive in \cite{DueFre98} the commutator 
relation:
\beql{comm}
  \begin{split}
    \scomm{\inter{W^1}{g\cL}{x}}{\inter{W^2}{g\cL}{y}} =
    -\sum_{n=0}^{\infty} \frac{i^n}{n!} \int d^4x_1 \cdots d^4x_n 
    \,g(x_1)\cdots g(x_n) \times & \\
    \left\{
      \RR{\cL, \dots ,\cL ,W^1}{W^2\vps}(x_1,\dots ,x_n,x;y) 
    \right. \quad \,&\\
    \left.
      \mp\RR{\cL, \dots ,\cL ,W^2}{W^1\vps}(x_1,\dots ,x_n,y;x) 
    \right\}.&
  \end{split}
\end{equation}
\subsection{The advanced and the causal product}\label{advcausprod}
The advanced product is defined as
\beql{A}
\begin{split}
  &\AA{W_1,\dots ,W_n}{W_i\vps}(x_1,\dots ,x_n;y)   \\
  &\qquad \defined \sum_{Y\subset  X} (-1)^{\betr{Y}}
  \TT{W_{Y^c}\vps}(x_{Y^c}) \TTb{W_Y,W_i\vps}(x_Y,y) 
\end{split}
\end{equation}
or, with the alternative expression analogous to eqn. \eqref{Ralter},
\beql{Aalter}
\begin{split}
  &\AA{W_1,\dots ,W_n}{W_i\vps}(x_1,\dots ,x_n;y)  \\
  &\qquad = \sum_{Y\in X} (-1)^{\betr{Y}}
  \TT{W_{Y^c},W_i}(x_{Y^c},y\vps)\TTb{W_Y\vps}(x_Y) .
\end{split}
\end{equation}
They have advanced support, 
\begin{equation}
  \begin{split}
    &\supp \AA{W_{X}\vps}{W_i}(x_X,y) \\ 
    & \qquad \subset \set{(x_1,\dots ,x_n,y)\in M^{n+1}: \quad x_i 
    \in \left( y + \forlc\right)\quad  \forall x_i \in X} .
  \end{split}
\end{equation}
The interacting fields can also be defined in terms of advanced 
products instead of retarded products without changing the local field
algebra if we define
\begin{equation}
  \inter{W_i}{g\cL}{f} = \int d^4y \,f(y) \,\frac{\de}{i \de h(y)}
  \left. S(g\cL + hW_i) \vphantom{\sum} \right|_{h=0}\times S(g\cL)^{-1}. 
\end{equation}
This would only result in a unitary transformation on $\ctF(\cO)$ with
$S(g\cL)$ as the unitary operator.  \\
Finally we define the causal product as
\begin{equation}
\begin{split}
  &\DD{W_1,\dots ,W_n}{W_i\vps}(x_1,\dots ,x_n;y)   \\
  &\quad \defined  \RR{W_1,\dots ,W_n}{W_i\vps}(x_1,\dots ,x_n;y)
  - \AA{W_1,\dots ,W_n}{W_i\vps}(x_1,\dots ,x_n;y)
\end{split}
\end{equation}
which has obviously causal support:
\begin{equation}
  \begin{split}
    &\supp \DD{W_{X}\vps}{W_i}(x_X,y) \\ 
    & \qquad \subset \set{(x_1,\dots ,x_n,y)\in M^{n+1}: \quad x_i 
    \in \left( y + \lc\right)\quad  \forall x_i \in X} .
  \end{split}
\end{equation}

\subsection{Conditions on the interaction Lagrangian}\label{interlag}
Up to now the Lagrangian density $\cL$ that defines the model via
the S-matrix could have been an arbitrary polynomial in $\eP$. There is 
a number of restrictions that such a polynomial must satisfy before it 
can define a reasonable physical model. In this section we will collect  
these restrictions.\\
At first, it must be Lorentz invariant:
\begin{equation}\label{C1}
  \Ad{U}{p} \TT{\cL} (x) = \TT{\cL} (\Lm^{-1}x) \qquad \forall \,\,
  p=(0,\Lm) \in \cP_+^{\uparrow} .  \tag{\bf C1}
\end{equation}
The second condition it must satisfy is pseudo-unitarity:
\begin{equation}\label{C2}
  \left(\TT{\cL}\right)^*(x) = \TT{\cL} (x) . \tag{\bf C2} 
\end{equation}
Furthermore it must have vanishing ghost number,
\begin{equation}\label{C3}
  s_c \TT{\cL} (x) = 0 .  \tag{\bf C3} 
\end{equation}
A Lagrangian with non vanishing ghost number would define a strange 
theory. The individual orders in perturbation theory of an interacting 
field would have a ghost number increasing (or decreasing) with the 
order. Such a theory would be super-renormalizable, provided it is 
power counting renormalizable, see below. \\
Since the S-matrix should also be BRS-invariant, one could also
expect an equation like 
\begin{equation}\label{exeich}
  s_0 \TT{\cL} (x) = 0
\end{equation}
to hold. Unfortunately it is in general --- and specifically in QED 
and Yang-Mills-theory --- impossible to find a Lagrangian for which eqn. 
(\ref{exeich}) holds. So we must weaken the condition a little.
Therefore we demand that there exist polynomials $\cL_n^{\mu_1\dots \mu_n}$
in $\eP$, the so called $Q(n)$-vertices, such that the following 
equations hold:
\beql{C4}
  s_0 \TT{\cL_n^{\mu_1,\dots ,\mu_n}}(x) = i \dd_\rho^x
  \TT{\cL_{n+1}^{\mu_1,\dots ,\mu_n,\rho}}(x) .
  \tag{\bf C4}
\end{equation}
The $Q(n)$-vertices must be totally antisymmetric in their Lorentz 
indices. The other index indicates the ghost number 
\begin{equation}
  g\left(\cL_n\right) = n \cL_n . 
\end{equation}
In \eqref{C4} there will be only a finite number of nontrivial 
equations, i.e.\ there exists an $m\in \N$ such that $\cL_m = 0$. 
The $Q(n)$-vertices have always the same canonical dimension as the 
original vertex $\cL$, and they also contain the same number of 
generators. Therefore $\cL_5 = 0$ for power counting 
renormalizable theories (see below) since $\cL_5$ must have 
ghost number five and it is impossible to construct a polynomial with 
ghost number five and a canonical dimension not exceeding four. If the 
original vertex contains only three generators as it is usually the case 
then already $\cL_4= 0$. In Yang-Mills theory --- with or without 
matter --- even $\cL_3=0$ and in QED $\cL_2=0$. These results can be
derived by explicit calculation.\\
The last condition on the Lagrangian we want to impose is power counting 
renormalizability. Perturbation theories can be divided into three
groups according to the canonical dimension of their Lagrangian: Those
with a canonical dimension less than the spacetime dimension are
super renormalizable, that means the number of free normalization 
parameters decreases with the order and finally vanishes, so the theory
is completely determined by a finite number of such parameters. 
Power counting renormalizable theories are those where the canonical 
dimension equals the space time dimension. For those theories there 
exists for all orders in perturbation theory a common upper bound for
the number of free parameters in the extension. Non renormalizable 
theories have Lagrangians whose canonical dimension exceeds 
the spacetime dimension, and this leads to a number of free normalization 
parameters that may increase with the order. Although the predictive power 
of such theories --- perturbative gravitation is an example of those --- 
is rather poor, it is nevertheless possible to deal with them in the 
framework of causal perturbation theory.\\
For our considerations non renormalizable Lagrangians play no role and 
therefore we exclude them explicitely. As we always work in four spacetime 
dimensions, the condition for renormalizability reads
\begin{equation}\label{C5}
  \deg \cL \leq 4  , \tag{\bf C5} 
\end{equation}
where $\deg$ means the canonical dimension.


%% file: nilpo.tex

\section{The interacting theory}
\label{nilpo}

In this chapter we come back to the program for the construction
of interacting gauge theories outlined in chapter (\ref{alg}). We 
formulated at the end of section (\ref{defstab}) four requirements for 
an interacting gauge theory. With the construction of local interacting 
field theories in the last chapter and the normalization conditions in 
chapter \eqref{normal} we are now able to determine under which conditions 
these requirements can be accomplished. The first condition --- the 
condition that suitable ghost and BRS charges can be found in the free 
model --- must be verified for the individual model. This has been done 
for the free models underlying QED and Yang-Mills theory in section
(\ref{freeBRS}). In this chapter we will see that the other three 
conditions hold if all normalization conditions \eqref{N1} - \eqref{N6} 
are satisfied and if the conditions \eqref{C1} - \eqref{C5} are valid 
for the Lagrangian $\cL$ which defines the model. We assume throughout 
this chapter that these preconditions hold. \\
In the first section we collect a number of properties all interacting
fields share from their very definition. Among them are e.g.\ covariance
and locality. In addition we derive a relation between the interacting
field operators for the higher generators and those for the basic 
generators.\\
In the second section we formulate field equations for the 
interacting field operators.  These equations are determined by 
normalization condition \eqref{N4}.  \\
In the third section we come to interacting operators that are of 
particular importance in gauge theories. In this section we define
the interacting ghost current, the interacting ghost charge and the 
ghost number of interacting fields. We prove that the interacting ghost 
current is conserved and that the higher order contributions of the 
ghost charge vanish. As a consequence every interacting field has the 
same ghost number as the corresponding free field. \\
In the fourth section we define the most essential operators in an 
interacting gauge theory: the interacting BRS current, the interacting 
BRS charge and the interacting BRS transformation. We find that the
interacting BRS current is conserved only where the test function $g$ 
that defines the coupling is constant. The BRS charge is constructed 
only for spacetimes that are compactified in spacelike directions.
Otherwise its definition would not be well posed. We prove also that 
with our definitions the BRS algebra holds. This means in particular 
that the interacting BRS charge is nilpotent. \\
In the last section we examine the relation between the 
quantum field theory defined above and its corresponding classical
theory and formulate a correspondence law for these
theories.
\subsection{General properties of interacting fields}
We begin our considerations with\\
{\em C-numbers:} From the definition of the retarded products, eqn. 
\eqref{R}, we can find that they vanish if at least one of
their arguments is a multiple of the identity --- provided the total 
number of arguments is at least two, see \cite{DueFre98}. This implies 
immediately for interacting fields that are generated by $\C$-numbers 
that they possess no higher order terms:
\begin{equation}
  \inter{\al\cdot\one}{g\cL}{x} = \al \cdot \one , \qquad \al \in \C.
\end{equation}
{\em Lorentz covariance:} 
The fact that the Lagrangian is a Lorentz scalar implies, together with 
condition \eqref{N1}, the Lorentz transformation properties of the 
interacting field operators:
\begin{equation}
  \Ad{U}{p} \inter{W_i}{g\cL}{x} = 
  \inter{\eR_\Lm\left(W_i\right)}{g^p\cL}{x-a}, \qquad 
  \forall p=(a,\Lm) \in \fP^\uparrow_+
\end{equation}
where $\eR$ is the representation of the Lorentz group (or its covering 
group) defined in section (\ref{defP}) and $g^p = g(\Lm^{-1}x-a)$.\\
{\em Pseudo-hermiticity:} Due to the conditions \eqref{C2} and \eqref{N2}
the Krein adjoint of the interacting fields is given by 
\beql{ph}
  \left( \inter{W_i}{g\cL}{x}\right)^* 
  = \inter{W_i^*\vps}{g\cL}{x} \qquad \forall W_i\, \in \eP .
\end{equation}
The $^*$-involution on the right hand side is the one introduced in
section (\ref{defP}).
{\em Locality:} A very important property of interacting fields is
their locality. This means that two interacting field operators (anti-) 
commute 
with each other if they are localized in spacelike separated regions. 
This can immediately be derived from eqn. \eqref{comm}:
\beql{locality}
  \scomm{\inter{W^1}{g\cL}{x}}{\inter{W^2}{g\cL}{y}} = 0 
  \qquad \mbox{ if }  x \spacel  y .
\end{equation}
{\em Primary interacting fields:} Due to normalization condition 
\eqref{N4} the interacting fields for the higher generators may be 
expressed by those for the basic generators as:
\beql{relbashigh}
  \begin{split}
    \inter{(\vp_i)^{(n,\nu_1\dots \nu_n)}}{g\cL}{x} \,\,=\,\, &
    \dd^{\nu_1}_x \cdots \dd^{\nu_n}_x  \inter{(\vp_i)^{(0)}}{g\cL}{x}\\
    & + C_{\vp_i, n} g(x)
    \inter{\partder{\cL}{\tphi_i^{(n,\nu_1\dots \nu_n)}}}{g\cL}{x},
  \end{split}
\end{equation}
where $\tphi_i$ is the field conjugated to $\vp_i$. 
\subsection{The interacting field equations}
Now we state field equations for the interacting field theory. 
They are again already determined by condition \eqref{N4} and read
\beql{FE1}
  \sum_j D^x_{ij} \inter{\vp_j}{g\cL}{x} 
  = - g(x) \inter{\partder{\cL}{\vp_i}}{g\cL}{x} .
\end{equation}
Inserting here the definition of $D^x$ --- eqn. \eqref{defDx} and the 
following ones --- we find that this implies in particular
\beql{expl}
  \begin{split}
    K^{\vp_i,x}\inter{(\vp_i)^{(0)}}{g\cL}{x} = 
    - \sum_{n=0}^{\infty} (-1)^n \dd^{\nu_1}_x\cdots \dd^{\nu_n}_x
    \left(
      g(x) \inter{\partder{\cL}{(\tphi_i)^{(n,\nu_1\dots 
      \nu_n)}}}{g\cL}{x}
    \right) ,
  \end{split}
\end{equation}
where $K^{\vp_i,x}$ was defined in eqn. \eqref{defK}. These are exactly 
the field equations that are derived as the 
Euler-Lagrange equations for a classical field theory with a Lagrangian 
$\cL_0 + \cL$, where $\cL$ is the interaction Lagrangian and $\cL_0$ is 
the free Lagrangian that implies the free field equations 
\begin{equation}
  K^{\vp_i,x} \vp_i(x) = 0, \qquad \vp_i(x)\text{ a classical field.}
\end{equation}
But there is one important difference between the field equations in the 
classical theory and those in the quantum theory. While the classical 
field equations govern the dynamics of the system, this in not true for 
the quantum field equations. The reason is that the classical theory has 
fewer independent variables. The field equations determine the time 
evolution of the basic fields on the left hand side. Therefore the time 
evolution of the entire classical theory is determined by the field 
equations, since all variables are basic fields or products thereof. 
This is not true in the quantum theory, because the interacting fields 
for composed elements in the algebra $\eP$ are not products of those for 
the generators\footnote{A product of distributional field operators is 
not defined a priori. It can be examined in the framework of operator 
product expansions \cite{Wil69,Wil71,Zim73}, but we will not discuss 
this here.}. Therefore the time evolution of the interacting fields for 
composed elements of $\eP$ is left open by the equations above. \\
The quantum field equations are completely independent of the
normalization constants $C_{\vp_i,k}$ in eqn. \eqref{generalelem}. 
They are also independent of the normalization of time ordered products, 
provided condition \eqref{N4} applies.
\subsection{The interacting ghost current and the ghost charge}
The interacting ghost current is defined as the interacting field 
operator that is generated by the free ghost current $k^\mu$, see 
section \eqref{freeBRS}:
\begin{equation}
  \tilde{k}^\mu (x) \defined \inter{k^\mu}{g\cL}{x} .
\end{equation}
This current is conserved as is easily derived by means of \eqref{N5} 
and \eqref{C3}:
\beql{ghcurrcons}
  \dd_\mu^x \tk^{\mu}(x) = 0  .
\end{equation}
From eqn. \eqref{ph} and the fact that the free ghost current is 
anti-pseudo-hermitian we find that the interacting ghost current is 
anti-pseudo-hermitian, too:
\beql{PHk}
  \left(\tk^{\mu}(x)\right)^* = - \tk^{\mu}(x) .
\end{equation}
The interacting ghost charge is defined as
\begin{equation}
  \tQ_c \defined  \lim_{\lm \searrow 0} \int d^4y \, h_\lm(y) 
  \tk^0(y) , 
\end{equation} 
where $h_\lm(x^0,\vec{x}) = \lm h^t(\lm x^0)b(\lm \vec{x})$, see eqn. 
\eqref{defh}. Here the coordinate frame is chosen such that the origin 
$0$ is in the domain $\cO$ where the fields are localized. We restrict 
the admissible spatial test functions $b$: At first the temporal test 
function $h^t$ is selected such that $0 \in \supp h^t$ and the
following equation holds:
\begin{equation}
  \left(\supp (\dd g) \cap \left[\cO + \forlc \right] \right) \gtrsim
  \left( \supp h^t \times \R^3 \right) \gtrsim
  \left(\supp (\dd g) \cap \left[\cO + \backlc \right] \right) .
\end{equation}
Then only test functions $b$ are admitted with the following 
properties: $b(\vec{x}) = 1$ for all $\vec{x} \in \R^3$ for which an 
$x^0 \in \supp h^t$ exists such that 
\begin{equation}
  (x^0,\vec{x}) \in \left( \supp g + \forlc \right) .
\end{equation}
The question arises whether the limit in the definition of $\tQ_c$ 
exists. We will show that this is indeed true.\\
The zeroth order of the interacting ghost current is simply the 
free ghost current. For the free current we know already that the limit 
exists, so we confine our attention to the higher orders. \\
We will prove that the higher orders of the ghost charge do not
depend on $\lm$. For this purpose we calculate for the $n^{\rm th}$ order 
of the ghost charge, $n\geq 1$ and $\lm \leq 1$:
\beql{difference}
  Q_{c,\lm}^n - Q_{c,1}^n  = \int d^4x 
  \left( h_\lm(x) - h_{1}(x)\vps\right) \tk^{0,n}(x) .
\end{equation}
Here $\tk^{\mu,n}$ is the $n^{\rm th}$ order of the ghost current. 
We have for all $n \geq 1$ that 
$\supp \tk^{\mu,n} \subset \left( \supp g + \forlc \right)$ 
due to the support properties of the retarded products. \\
With our conventions for the test functions we can substitute in eqn.
\eqref{difference} on the right hand side $h^t(x^0)b(\lm x)$ for 
$h_1(x) = h^t(x^0)b(x)$ because $h^t(x^0)\left(b(\lm x)-b(x)\right)$ 
vanishes on the support of $\tk^{\mu,n}$, $n\geq 1$. Then eqn. 
\eqref{difference} becomes 
\begin{equation}
  Q_{c,\lm}^n - Q_{c,1}^n  = \int d^4x 
  \left( \lm h^t(\lm x^0) - h^t(x^0) \vps\right) b(\lm \vec{x}) 
  \tk^{0,n}(x) .
\end{equation}
There exists a test function $H_{\lm} \in \cD(\R)$ such that
\begin{equation}
  \dd_0^x H_{\lm}(x^0) = 
  \left( \lm h^t(\lm x^0) - h^t(x^0) \right) .
\end{equation}
Inserting this into \eqref{difference} we get
\begin{equation}
  \begin{split}
    Q_{c,\lm}^n - Q_{c,1}^n  &= \int d^4x 
    \left(\dd_0^x H_{\lm}(x^0)\right) b(\lm \vec{x}) \tk^{0,n}(x) \\
    &= - \int d^4x H_{\lm}(x^0) \left(\dd_i^x b(\lm \vec{x})\vps\right)
    \tk^{i,n}(x) ,
  \end{split}
\end{equation}
where we have partially integrated and used the fact that $\tk^\mu$ is 
conserved. By construction we have
\begin{equation}
  \supp 
  \left( H_{\lm}(x^0) \left(\dd_i^x b(\lm \vec{x})\right)\vps\right)
  \cap \left(\supp g + \forlc \right) = \emptyset .
\end{equation}
Comparing this with the support of $\tk^{\mu,n}$, we see that the 
integral vanishes. Therefore the higher orders of $\tQ_c$ do not depend 
on $\lm$. Even more, because of current conservation, eqn. 
\eqref{ghcurrcons}, one can choose $h^t$ such 
that the support of $h_1$ is entirely in the past of $\supp g$. Then the 
higher order terms vanish due to the support properties of the retarded 
products, so the interacting ghost charge coincides with the free ghost 
charge or, strictly speaking since $\tQ_c$ is a formal power series,
\beql{Qcfree}
  \tQ_c = (Q_c , 0, 0, \cdots ) .
\end{equation}
Since the ghost current is anti-pseudo-hermitian, the ghost charge is it, 
too:
\beql{PHQc}
  \tQ_c^* = - \tQ_c .
\end{equation}
The interacting ghost number of a localized field operator is measured by 
the following derivation:
\begin{equation}
  \ts_c\left( \inter{W_i}{g\cL}{x} \right) \defined
  \comm{\tQ_c}{\inter{W_i}{g\cL}{x}} .
\end{equation}
As the interacting ghost charge coincides with the free one, we have
\begin{equation}
  \ts_c\left( \inter{W_i}{g\cL}{x} \right) = 
  s_c\left( \inter{W_i}{g\cL}{x} \right) . 
\end{equation}
This implies immediately, due to \eqref{N5} and \eqref{C3}, that the 
interacting field operators have the same ghost number as the
corresponding free fields:
\beql{ghcuder}
  \ts_c\left( \inter{W_i}{g\cL}{x} \right) = 
  g(W_i)\inter{W_i}{g\cL}{x}  , \qquad g(W_i) \in \Z . 
\end{equation}
\subsection{The interacting BRS current, BRS charge and BRS 
transformation}
The natural choice for the BRS current, 
\begin{equation}
  \tj_B^{\mu}(x) \,= \,\inter{j^{\mu}_B}{g\cL}{x} ,
\end{equation}
is not conserved in general, so this cannot be the correct
interacting BRS current. The situation is even worse: Explicit 
calculations in first order QED and Yang-Mills theory shows that there 
exists no normalization of the time ordered products such that this
current is conserved even in first order, irrespective of our 
normalization conditions. The best one can achieve is that the
current is conserved where the coupling is constant, and even this
seemingly liberal condition fixes the normalization in first order
uniquely. \\
A direct calculation reveals that this normalization is not compatible
with the normalization conditions \eqref{N3} and \eqref{N4}. But there
is an expression for the interacting BRS current that is compatible
with the normalization conditions in first order and that is conserved 
in the sense above, not only for Yang-Mills theories but for every 
theory. Adopting this expression as the definition of the interacting 
BRS current we have
\beql{defj}
  \tj_B^{\mu}(x) \,\defined \,\inter{j^{\mu}_B}{g\cL}{x} 
  - g(x) \inter{\cM^\mu_1}{g\cL}{x} ,
\end{equation}
where $\cM^\mu_1$ is the $R(1)$-vertex, see condition \eqref{C4}.
Normalization condition \eqref{N6} implies that this current is indeed 
conserved where the coupling $g$ is constant:
\begin{equation}
  \dd_\mu^x \tj_B^{\mu}(x) \,= \, 
  (\dd_\nu g)(x) \inter{\cL^\nu_1}{g\cL}{x} .
\end{equation}
The fact that the interacting BRS current is not everywhere conserved is 
a severe drawback, since it complicates the definition of the BRS 
charge, see below. So the question arises whether a more clever choice 
for the BRS current could have yielded one that is everywhere conserved. 
But this turns out to be impossible in general. Concretely, in QED as 
well as in Yang-Mills theory the explicit calculation shows already in 
first order that no such choice exists. So in general this result cannot 
be improved. \\
By the same reasoning as for eqn. \eqref{PHk} one derives that 
$\tj^{\mu}_B$ is pseudo-hermitian
\beql{PHj}
  \left(\tj^{\mu}(x)\right)^* = \tj^{\mu}(x) .
\end{equation}
Now we come to the definition of the BRS charge. As already mentioned 
this definition is more difficult than that of the ghost charge was,
since the BRS current is not everywhere conserved. The problem can be 
seen as follows: The natural choice for the BRS charge would be 
\beql{BRSch1}
  \tQ_B = \lim_{\lm \searrow 0} \int d^4y \, h_\lm(y) \tj^0_B(y) , 
\end{equation} 
with $h_\lm$ like above. Unfortunately this expression would depend on 
the choice of $h_\lm$, unlike for $\tQ_c$, and the 
higher orders would depend on $\lm$, both because the BRS current is
not conserved. If the higher orders depend on $\lm$ the limit is no 
longer under control. \\
In order to make $\tQ_B$ independent of $h_\lm$, the support of $h_\lm$ 
must be for every $\lm$ in a region where $g$ is constant. This
would mean that $g$ is everywhere constant, i.e.\ the adiabatic limit
must be performed, and this limit does not exist in general. \\
Another possibility would be not to perform the limit and choose e.g.\ 
$h_1$ as a test function in the definition of the BRS charge. In this 
case the BRS charge would clearly become well defined, but it would also 
be a local operator, and such an operator could not annihilate states 
with finite energy due to the theorem of Reeh and Schlieder. It is 
unlikely that the cohomology defined with it has good properties, and 
therefore we exclude this possibility.\\
The way out of this seemingly pitfall was found by D\"utsch and 
Fredenhagen \cite{DueFre98}. In order to allow functions that are 
constant in 
spacelike directions as test functions, they embed the double cone 
$\cO$ isometrically into the cylinder $\R \times C_L$ with $\R$ the time 
axis and $C_L$ a cube of length $L$ sufficiently big to contain $\cO$. 
This spatial compactification does not change the properties of the local 
algebra $\ctF(\cO)$. This is why the quantization of free fields in a box, 
mentioned in chapter (\ref{free}), is important for us. For the details
of the construction we refer to \cite{DueFre98}.\\
In the compactified space $h$ and $g$ can be chosen to be test functions 
such that $g$ is constant on $\supp h$ with the same value as on $\cO$.\\
With these test functions we are able to give a definition of the 
interacting BRS current in a spatially compactified spacetime. At first, 
we choose the test function $h$ to be
\begin{equation}
  h(x) = h^t(x_0) , \qquad h^t \text{ like in \eqref{defh}}, \qquad
  \lra h \in \cD(\R \times C_L), 
\end{equation}
and the coupling $g$ such that
\begin{equation}
  \left.g\right|_{\supp h} = \left.g\right|_{\cO} = \mbox{ constant}, 
  \qquad g \in \cD(\R \times C_L) .
\end{equation}
With $g$ and $h$ now both being a test function --- on $\R \times C_L$
--- the BRS charge can be defined as 
\beql{BRSch2}
  \tQ_B \defined \int_{\R \times C_L} d^4y \, h(y) \tj^{\,0}_B(y)  . 
\end{equation}
It is easy to see by an analogous reasoning as for the ghost charge that 
this BRS charge is independent of $h$. \\
The zeroth order of this BRS charge agrees with the free BRS charge in 
the $\R \times C_L$ spacetime, $\tQ_{B,0} = Q_B$, if $h_\lm$ is replaced 
there by $h$.  Of course the limit is then not performed because
it would be void. Unlike the interacting ghost charge the interacting 
BRS charge has also non vanishing higher order contributions. The 
reasoning which showed that the higher contributions of $\tQ_c$ vanish
cannot be applied here, since $\supp h$ may not be (not even partly) in 
the past of $\supp g$ from its very definition. \\
Like the BRS current the BRS charge is pseudo-hermitian:
\beql{PHQB}
  \tQ_B^* = \tQ_B .
\end{equation}
Since $s_c Q_B = Q_B$ in the free theory, we find with eqn.
\eqref{N5int}
\begin{equation}
  \comm{\tQ_c}{\tQ_B} = \ts_c \left(\tQ_B\right) = \tQ_B .
\end{equation}
So the first part of the BRS algebra holds. The most important 
property of the BRS charge is its nilpotency, the second part of the 
BRS algebra. This will be proven next. \\
To this end we write at first $\tQ_B$ in a different form that is 
more adequate for the proof. We use for the interacting 
field $\inter{j^\mu}{g\cL}{x}$ in the definition of $\tj^\mu$ the
equation \eqref{normBRS} from the appendix and the identity 
$\tQ_c = Q_c$, eqn. \eqref{Qcfree}. With it the interacting BRS 
charge can be written as
\beql{altQB}
  \begin{split}
    \tQ_B = Q_B
    + \sum_{n=0}^{\infty} \frac{i^n}{n!} \int &d^4y d^4z d^4x_1\cdots 
    d^4x_n h(y) (\dd_{\nu} g)(z) \times \\
    &\times g(x_1) \cdots g(x_n) 
    \RR{\cL, \dots ,\cL,\cL^{\nu}_1}{k^{0}\vps}(x_1,\dots x_n,z;y).
  \end{split} 
\end{equation}
The first term in the sum on the right hand side is the free BRS charge.
Since its properties are already known, we confine our attention 
to the higher order terms. From the form of these terms one can see 
immediately that the interacting BRS charge becomes the free one in the
adiabatic limit, if this limit exists. We are here particularly 
interested in theories where the adiabatic limit does not necessarily
exist. \\
We introduce a test function $H \in \cD(\R\times C_L)$ with the
property 
\begin{equation}
  \dd_{\mu}^{y} H(y) = - \de^{0}_{\mu} h(y),  \qquad 
  H \in \cD(\R\times C_L) ,
\end{equation}
such that $H(x)=1$ for all $x$ in the past of $\supp g$ and $H(x)=0$ for
all $x$ in the future of $\supp g$. Inserting this into the expression 
above, we find by partial integration and with the help of eqn. \eqref{N5} 
the following alternative formulation for the $n^{\rm th}$ order of 
$\tQ_B$, $n\geq 1$:
\beql{altQB3}
\begin{split}
  \tQ_B^{(n)} =& \frac{i^{n-1}}{(n-1)!} \int d^4z d^4x_1\cdots 
  d^4x_{n-1}  H(z) (\dd_{\nu} g)(z) \times \\
  & \qquad \times g(x_1) \cdots g(x_{n-1}) 
  \RR{\cL, \dots ,\cL}{\cL^{\nu}_1\vps}(x_1,\dots ,x_{n-1};z) .
\end{split}
\end{equation}
The \nth order of $(\tQ_B)^2$ decomposes according to
\begin{equation}
  \left((\tQ_B)^2\right)^{(n)} = 
  \sum_{k=0}^{n} (\tQ_B)^{(k)} (\tQ_B)^{(n-k)} 
  = s_0 \left( (\tQ_B)^{(n)}\right) 
  + \sum_{k=1}^{n-1} (\tQ_B)^{(k)} (\tQ_B)^{(n-k)} .
\end{equation}
At first we will calculate $s_0 \left( (\tQ_B)^{(n)}\right)$. With the 
help of the generalized operator gauge invariance, eqn. \eqref{GOGI}, 
with $i_1=1$ and $i_k = 0$ otherwise, we get
\begin{equation}
\begin{split}
  &s_0 (\tQ_B)^{(n)} \, = \\
  &\qquad = \frac{i^{n-2}}{(n-2)!} \int d^4x_1 \cdots d^4x_{n-2} 
  \,d^4y \,d^4z \,g(x_1) \cdots g(x_{n-2}) \\
  &\qquad \quad \times (\dd_{\rho}g)(y)\, H(y)\,(\dd_{\mu}g)(z) \,H(z)\,
  \RR{\cL,\dots,\cL,\cL^{\rho}_1}{\cL^{\mu}_1\vps}(x_1,\dots,x_{n-2},y;z) \\
  & \qquad\quad + \frac{i^{n}}{(n-1)!} \int d^4x_1 \cdots d^4x_{n-1} \,d^4z 
  \,g(x_1) \cdots g(x_{n-1}) \\
  & \qquad \qquad \times 
  \left(\dd_{\rho}^z\left[ (\dd_{\mu}g)(z) H(z) \right]\vps\right)
  \RR{\cL,\dots ,\cL}{\cL^{\mu\rho}_2\vps}(x_1,\dots,x_{n-1};z) .
\end{split}
\end{equation} 
Here an additional factor $H(y)$ has been inserted in the first 
integral. This factor does not change the result due to the retarded 
support of the distribution. \\
Let us at first consider the second integral on the right hand side. 
Calculating the derivative of the square bracket in the last line, we get 
$(\dd_{\mu}\dd_{\rho}g)(z)H(z) + (\dd_{\mu}g)(z) (\dd_{\rho}) H(z)$. 
The second term vanishes since the supports of $\dd g$ and $\dd H$ are 
disjoint. The first term is symmetric in $\mu$ and $\rho$ while the 
retarded product is antisymmetric in these indices, due to the 
antisymmetry of the $Q(2)$-vertex. Therefore the entire second integral 
vanishes. \\
Now we come to the first integral. Here the test functions are also 
symmetric under permutation of $(z,\mu)$ and $(y,\rho)$. If we look at the
definition of the retarded products, eqn. \eqref{R}, we see that there 
are terms where both $\cL^{\mu}_1$ and $\cL^{\rho}_1$ appear as arguments 
in the same time ordered product or antichronological product. These
contributions vanish, because the distributions are antisymmetric in 
$(z,\mu)$ and $(y,\rho)$ due to graded symmetry ${(\bf P2)}$. The only 
contributions that remain lead to our final expression 
for $s_0 (\tQ_B)^{(n)}$:
\beql{s0QB}
\begin{split}
  &s_0 (\tQ_B)^{(n)} \, = \\
  &\quad = - \frac{i^{n-2}}{(n-2)!} \int d^4x_1 \cdots d^4x_{n-2}\, d^4y\, 
  d^4z \,g(x_1) \cdots g(x_{n-2})\, (\dd_{\rho}g)(y)\, H(y) \\
  &\qquad \times (\dd_{\mu}g)(z) \,H(z)\sum_{Y\subset X} (-1)^{\betr{Y}}
  \TTb{\cL,\dots ,\cL,\cL^{\rho}_1}(x_Y,y) 
  \TT{\cL, \dots ,\cL,\cL^{\mu}_1}(x_{Y^c},z)
\end{split}
\end{equation} 
with $X= \set{x_1,\dots,x_{n-2}}$. \\
To calculate $\sum_{k=1}^{n-1} (\tQ_B)^{(k)} (\tQ_B)^{(n-k)}$ we make use
of the two ways to express $R$-products in terms of $T$- and 
$\bT$-products, that means we use eqn. \eqref{altQB3} for the individual
orders of the BRS charge, inserting eqn. \eqref{Ralter} for the retarded 
products on the left hand side and eqn. \eqref{R} for those on the right 
hand side. Then we get after a little combinatorial analysis
\begin{equation}
\begin{split}
  &\sum_{k=1}^{n-1} (\tQ_B)^{(k)} (\tQ_B)^{(n-k)} \,= \\
  &\quad = \frac{i^{n-2}}{(n-2)!} \int d^4x_1 \cdots d^4x_{n-2} \,d^4y \,d^4z 
  \,g(x_1) \cdots g(x_{n-2}) \,(\dd_{\rho}g)(y) \,H(y)\, (\dd_{\mu}g)(z)\\
  &\qquad \times H(z)
  \left[
    \sum_{Y,Z,U,V}
    (-1)^{\betr{Z}+\betr{V}} 
    \left( 
      \TTb{\cL,\dots,\cL,\cL^{\mu}_1\vps}(x_Z,z) \TT{\cL,\dots,\cL\vps}(x_Y)
    \right)
  \right. \\
  &\qquad
  \left.
    \hspace{119pt} \times 
    \left( 
      \TTb{\cL,\dots,\cL \vps}(x_V) \TT{\cL,\dots,\cL,\cL^{\rho}_1\vps}(x_U,y)
    \right)
  \vphantom{\sum_{Y,Z,U,V} }
  \right] ,
\end{split}
\end{equation}
where the sum in the square brackets runs over all disjoint partitions 
of $X$ into four subsets $U,V,Y,Z$. These subsets may be empty. This set of 
partitions can be divided into two subsets, namely the set of those 
partitions where $Y$ and $V$ are empty and its complement. This complement 
can in turn be divided in subsets with $U$ and $Z$ fixed, yielding terms 
proportional to
\begin{equation}
  \sum_{W\subset X\setminus U\setminus Z} (-1)^{\betr{W}}
  \TT{\cL,\dots,\cL}(x_W) \TTb{\cL,\dots,\cL}
  (x_{X\setminus U\setminus Z\setminus W}).
\end{equation}
This expression vanishes due to eqn. \eqref{unit} because 
$X\setminus U\setminus Z \neq \emptyset$ according to our 
assumption. So there remains only a contribution from the partitions 
with $Y=V=\emptyset$, and since $T_0 = \overline{T}_0 = \one$,
there remains only
\begin{equation}
\begin{split}
  &\sum_{k=1}^{n-1} (\tQ_B)^{(k)} (\tQ_B)^{(n-k)} \,= \\
  &\quad = \frac{i^{n-2}}{(n-2)!} \int d^4x_1 \cdots d^4x_{n-2}\, d^4y\, 
  d^4z \,g(x_1) \cdots g(x_{n-2})\, (\dd_{\rho}g)(y)\, H(y) \\
  &\qquad \times (\dd_{\mu}g)(z) \,H(z)\sum_{Y\subset X} (-1)^{\betr{Y}}
  \TTb{\cL,\dots ,\cL,\cL^{\rho}_1}(x_Y,y) 
  \TT{\cL, \dots ,\cL,\cL^{\mu}_1}(x_{Y^c},z).
\end{split}
\end{equation} 
Obviously this is just the negative of eqn. \eqref{s0QB}, yielding
\begin{equation}
  \left((\tQ_B)^2\right)^{(n)} 
  = s_0 \left( (\tQ_B)^{(n)}\right) 
  + \sum_{k=1}^{n-1} (\tQ_B)^{(k)} (\tQ_B)^{(n-k)} 
  \stackrel{!}{=} 0.
\end{equation}
Reviewing our preconditions, we have proven that with our 
definition the BRS charge is nilpotent --- and therefore the complete
BRS algebra holds ---, provided our normalization condition \eqref{N6} 
is valid. \\
At the end of this section we come to the interacting BRS 
transformation $\ts$. It could be defined as 
\begin{equation}
  \ts\left( \inter{W}{g\cL}{x} \right) = 
  \scomm{\tQ_B}{\inter{W}{g\cL}{x}} .
\end{equation}
But it turns out to be more clever to permute 
commutation and integration, and we define
\begin{equation}
  \ts\left( \inter{W}{g\cL}{x} \right) \defined
  \int_{\R \times C_L} d^4y \, h(y) 
  \scomm{\tj^0_B(y)}{\inter{W}{g\cL}{x}} 
\end{equation}
with $h$ and $g$ as in the definition of $\tQ_B$. The advantage of 
this definition is that it remains well defined for $\ts$ acting on local
fields in $\ctF(\cO)$ even if the spacetime is not compactified and $h$ 
has compact support only in timelike directions being constant in 
spacelike directions. This is well defined because locality, eqn. 
\eqref{locality}, holds --- both $\tj^0_B$ and $\inter{W}{g\cL}{f}$ are 
local fields. Therefore the commutator 
has causal support, so the integrand vanishes in the causal complement 
of $\cO$. 
\begin{equation}
  \ts\left( \inter{W}{g\cL}{f} \right) \defined
  \int d^4y \, h(y) 
  \scomm{\tj^0_B(y)}{\inter{W}{g\cL}{f}} 
\end{equation}
is a well defined expression in Minkowski space, if $h$ is chosen such
that 
\begin{equation}
\begin{split}
  &h(x) = h^t(x_0) ,  \qquad h^t \in \cD(\R) 
  \text{ as in eqn. \eqref{defh}},\\
  &g \text{ is constant on }
  \left(\cO + \backlc \right) \cap \left( \supp h + \forlc \right).
\end{split}
\end{equation}
This expression is independent of $h$. The BRS transformation is 
nilpotent. This can be seen by direct computation --- the 
calculation is then completely analogous to that for $(\tQ_B)^2 = 0$
in the compactified spacetime. A different way to prove that $\ts$ is
nilpotent is to consider $\ts$ in a compactified spacetime --- where
$\ts^2 = 0$ follows directly from $(\tQ_B)^2 = 0$. Then let the 
compactification length $L$ tend to infinity. The resultant space will
be the Minkowski space and $\ts^2 = 0$ still holds since the algebra 
does not depend on the compactification length. Therefore
\begin{equation}
  \ts ^2 A = 0 \qquad \forall A \in \ctF(\cO) .
\end{equation}
It is important to note that this reasoning holds only for local
operators. In particular the argument of Nakanishi and Ojima 
\cite{NakOji90} that a nilpotent BRS transformation defines a nilpotent 
BRS charge can not be applied here. Their argument is as follows:
\begin{equation}
    \tQ_B \defined - \ts \tQ_c  \qquad \text{and} \qquad
    0 = \ts^2(\tQ_c) = - \ts(\tQ_B) = -2 \tQ_B^2 ,
\end{equation}
but since $\tQ_c$ is not a local operator it is not in the domain of 
$\ts$ in the framework of ordinary spacetime.\\
So we arrive at the following result: For all investigations concerning 
the state space it is necessary to compactify spacetime, 
since we need the BRS charge to define the physical state space, and 
this is only defined in a compactified spacetime. But for investigations
concerning only the algebra of local observables there is no need for a
compactification because the definition of observables requires only the BRS
transformation, not the BRS charge, and the former can also be defined in 
an ordinary spacetime. \\
Summarizing the results of this and the preceeding chapter we see that
all the preconditions that we postulated at the end of section 
(\ref{defstab}) are satisfied. The only restriction is that the BRS 
current is conserved only locally, but this is sufficient for the 
construction of the local interacting gauge theory. \\
This result was derived under the assumption that the normalization
conditions \eqref{N1} - \eqref{N6} and the conditions on the Lagrangian
are satisfied. We proved in chapter (\ref{normal}) that the first five 
normalization conditions have simultaneous solutions. So the essential
point is whether condition \eqref{N6} can be satisfied for a model.
If this is the case, the construction of the physical state space 
(in the spatially compactified spacetime) and of the local observable 
algebra can be performed.
\subsection{The correspondence between quantum and classical theory}
We have seen in section (\ref{normcon}) that in our approach the 
propagators for the higher generators are different from the corresponding 
propagators for the derivated fields in other renormalization procedures.
The propagators determine the tree diagrams, and these in turn are 
known to determine the classical limit of the theory. Therefore the
question arises whether the classical limit of our theory is different
from what one would expect from other approaches. We will see that this
is indeed the case.\\
The classical fields are functions on a manifold, in this case the 
Minkowski space. Unlike the distributional field operators they may be
multiplied at the same spacetime point. We take advantage of this 
property and define a representation $C$ of the algebra $\eP$ by 
classical fields. Unlike the representation $T$ of $\eP$ in section 
(\ref{defrep}) this is not only a linear representation but also an 
algebra homomorphism. We define 
\begin{equation}
  \begin{split}
    C: \quad \eP \to \cinfty (M), \qquad &C(a\cdot A) = a\cdot C(A)
    \qquad \forall\,a\in \C,\,\,A\in \eP,\\
    &C\left(\prod_i \vp_i \right)(x) 
    = \prod_i C\left(\vp_i \right)(x) ,\qquad \vp_i \in \cG .
  \end{split}
\end{equation}
The representatives of the basic generators are the basic classical 
fields, i.e.\ we suppose that there exists for each $\vp_i \in \cG_b$ 
a classical field $\vp_i^{\rm cl}(x)$ such that 
\begin{equation}
  C(\vp_i)(x) = \vp_i^{\rm cl}(x) .
\end{equation}
The question arises how the higher generators may be represented. The 
first attempt is to define their representatives as the derivatives of 
those for the basic generators, e.g.\ 
\beql{versuch}
  C\left( (\vp_i)^{(1,\mu)}\right)(x) = 
  \dd^\mu_x C\left( (\vp_i)^{(0)}\right)(x) .
\end{equation}
But this definition is not consistent. This can be seen by comparing 
this equation with eqn. \eqref{relbashigh}, 
\begin{equation}
  \begin{split}
    \inter{(\vp_i)^{(n,\nu_1\dots \nu_n)}}{g\cL}{x} \,\,=\,\, &
    \dd^{\nu_1}_x \cdots \dd^{\nu_n}_x  \inter{(\vp_i)^{(0)}}{g\cL}{x}\\
    & + C_{\vp_i, n} g(x)
    \inter{\partder{\cL}{\tphi_i^{(n,\nu_1\dots \nu_n)}}}{g\cL}{x} .
  \end{split}
\end{equation}
If we adopted the definition above, the left hand side and the right 
hand side of this equation would be equal on the quantum level, but they
would have different classical limits, and this cannot be true. We see
that the correct prescription for the classical limit of the higher
generators is 
\begin{equation}
  \begin{split}
    C\left((\vp_i)^{(n,\nu_1\dots \nu_n)}\right) (x) \,\,=\,\, &
    \dd^{\nu_1}_x \cdots \dd^{\nu_n}_x  C\left((\vp_i)^{(0)}\right)(x)\\
    & + g\,C_{\vp_i, n} \,\,  
    C \left(\partder{\cL}{\tphi_i^{(n,\nu_1\dots \nu_n)}}\right)(x) .
  \end{split}
\end{equation}
Here we have set the coupling $g$ constant, since in a classical theory
there is no need for the interaction to be switched off. $\tphi_i$ is 
the generator of the field conjugate to $\vp_i$.\\
The fields that correspond to the higher generators are 
labelled by the normalization constants $C_{\vp_i, n}$. This is what we
expected when we pointed out the importance of the propagators for the 
classical limit, because these propagators are also labelled by the
normalization constants. \\
With the representation $C$ now defined we can formulate the 
correspondence law. It states that the distributional interacting field
operators become products of classical fields in the classical
limit according to
\begin{equation}
  \begin{split}
    &\inter{W}{g\cL}{x} \to C\left( W\vps \right)(x) \qquad 
    \forall W \in \eP.\\
    &g(x) \to g = \text{ constant.}
  \end{split}
\end{equation}


%% file: examples.tex
\section{Two particular theories}
\label{examples}

In this chapter we will examine the consequences of our general
results derived in the preceeding chapter for two well known models: 
Quantum electrodynamics and Yang-Mills theory. 
\subsection{Quantum electrodynamics}\label{QED}
The fields involved in QED are vector bosons $A_\mu$ --- the
photons ---, ghosts and anti-ghosts $u,\tu$ and charged spinors
$\psi, \psiq$ --- the electrons and positrons. \\
For QED there exists a way to determine the physical state vector space 
without the BRS formalism --- the Gupta-Bleuler procedure. Furthermore
the ghosts do not couple to the other fields. Therefore it is not 
necessary to include the ghosts in the model. Nevertheless we do so 
because we investigate QED also as a preparation for Yang-Mills theory 
where the ghosts are indispensable.\\
The corresponding free theory for QED has been treated in section 
(\ref{freeBRS}). \\
Therefore we start directly with the interaction. The interaction 
Lagrangian for QED reads
\begin{equation}
  \cL_{QED} = A_\mu j^\mu_{\rm el} \quad \in \eP.
\end{equation}
Here $A_\mu$ is the basic generator corresponding to the photon field,
and the electric current $j^\mu_{\rm el}$ is defined as
\begin{equation}
  j^\mu_{\rm el} \defined \psiq \gamma^\mu \psi \quad \in \eP
\end{equation}
with the basic generators $\psi, \psiq$ corresponding to the electron and 
the positron field. \\
It can be easily verified that this Lagrangian satisfies our 
requirements \eqref{C1} - \eqref{C3} and \eqref{C5}. The canonical
dimension of the spinors is 3/2 and that of the photons is 1, summing
up to a total canonical dimension of 4, so the model is renormalizable.
We will show that also condition \eqref{C4} is accomplished.\\
In addition we examine an important relation that we were not able 
to prove in the general framework: The normalization condition \eqref{N6}.
We prove that the other normalization conditions, in particular the Ward
identities for the electric current, eqn. \eqref{N5'}, already imply 
\eqref{N6} in QED. The proof will be given below.\\
To begin with we determine the $Q(n)$-vertices of QED from its 
Lagrangian. For condition \eqref{C4} to hold we must find $Q(n)$-vertices
that satisfy the following equations:
\begin{equation}
  s_0 \TT{\cL}(x) = i \dd_\nu^x \TT{\cL_1^\nu}(x) ,\qquad
  s_0 \TT{\cL_1^\nu}(x) = i \dd_\rho^x \TT{\cL_2^{\rho\nu}}(x), \qquad
  \dots
\end{equation}
Observing the free BRS transformations introduced in section 
(\ref{freeBRS}), we find that these $Q(n)$-vertices exist indeed:
\begin{equation}
  \cL_1^\nu = u j_{\rm el}^\nu, \qquad \cL_i= 0 \quad \forall i\geq 2.
\end{equation}
To prove that \eqref{N6} is valid we must therefore calculate the 
following expression
\begin{equation}
  \dd_\mu^y \TT{\cL_1^{\nu_1},\dots,\cL_1^{\nu_k},\cL,\dots ,\cL,j^\mu_B\vps}
  (x_1,\dots ,x_n,y) 
\end{equation}
with $\cL=\cL_{QED}$, $\cL_1$ like above and $j^\mu_B$ as defined in 
section (\ref{freeBRS}). If we insert this time ordered product into the 
causal Wick expansion, eqn, \eqref{CWE}, we find that it can be written as
\begin{equation}
  \begin{split}
    &\TT{\cL_1^{\nu_1},\dots,\cL_1^{\nu_k},\cL,\dots ,\cL,
    (A_\rho)^{(1,\rho)}} (x_1,\dots ,x_n,y) \cdot \dd^\mu_y u(y) \\
    &- \TT{\cL_1^{\nu_1},\dots,\cL_1^{\nu_k},\cL,\dots ,\cL,
    (A_\rho)^{(2,\rho\mu)}} (x_1,\dots ,x_n,y) \cdot  u(y).
  \end{split} 
\end{equation}
Since neither $\cL$ nor $\cL_1$ contain a higher generator, conditions
\eqref{N4} reveals that this expression is equal to
\begin{equation}
  \begin{split}
    &
    \left(
      \dd^\rho_y\TT{\cL_1^{\nu_1},\dots,\cL_1^{\nu_k},\cL,\dots ,\cL,
      (A_\rho)^{(0)}} (x_1,\dots ,x_n,y) 
    \right) \cdot \dd^\mu_y u(y) \\
    &
    - 
    \left(
       \dd^\rho_y\dd^\mu_y\TT{\cL_1^{\nu_1},\dots,\cL_1^{\nu_k},\cL,
      \dots ,\cL, (A_\rho)^{(0)}} (x_1,\dots ,x_n,y) 
    \right)
    \cdot  u(y).
  \end{split} 
\end{equation}
Inserting the derivative and taking into account the field equations of 
$u(y)$, we find
\begin{equation}
  \begin{split}
    & \dd_\mu^y \TT{\cL_1^{\nu_1},\dots,\cL_1^{\nu_k},\cL,\dots ,\cL,j^\mu_B}
    (x_1,\dots ,x_n,y)  \\
    &\qquad = -
    \left(
      \dd^\mu_y \square^y \TT{\cL_1^{\nu_1},\dots,\cL_1^{\nu_k},\cL,
      \dots ,\cL, (A_\mu)^{(0)}} (x_1,\dots ,x_n,y) 
    \right) 
    \cdot  u(y). 
  \end{split}
\end{equation}
With condition \eqref{N4} this expression can be rewritten as
\begin{equation}
  \begin{split}
    & - i
    \left(
      \sum_{m=k+1}^{n} \left(\dd_\mu^y \de(y-x_m)\right)
    \right.\\
    &
    \left.\vphantom{\sum_{m=k+1}^{n} } \qquad\qquad  \times
      \TT{\cL_1^{\nu_1},\dots,\cL_1^{\nu_k},\cL,\dots, j_{\rm el}^\mu,
      \dots ,\cL} (x_1,\dots ,x_n) 
    \right) 
    \cdot  u(y).
  \end{split} 
\end{equation}
Here the vertex $j_{\rm el}^\mu$ is at the $m^{\rm th}$ position. 
Pulling the derivative out of the bracket we finally arrive at
\begin{equation}
  \begin{split}
    & i \sum_{m=k+1}^{n} \dd_\mu^m
    \left(
      \de(y-x_m) 
      \TT{\cL_1^{\nu_1},\dots,\cL_1^{\nu_k},\cL,\dots, \cL_1^\mu,
      \dots ,\cL} (x_1,\dots ,x_n) 
    \right) \\
    & - i
    \left(
      \sum_{m=k+1}^{n}\de(y-x_m)\dd_\mu^m 
      \TT{\cL_1^{\nu_1},\dots,\cL_1^{\nu_k},\cL,\dots, j_{\rm el}^\mu,
      \dots ,\cL\vps} (x_1,\dots ,x_n) 
    \right) \cdot u(y) .
  \end{split}
\end{equation}
The vertices $\cL_1^\mu$ and $j_{\rm el}^\mu$ are again in the 
$m^{\rm th}$ position. Comparing the last line with the Ward identities 
for the electric current, eqn. \eqref{N5'}, we see that this term 
vanishes since $f(\cL) = f(\cL_1) = 0$. The remaining expression is 
exactly what condition \eqref{N6} predicts, provided that all the 
$R(n)$-vertices vanish, $\cM_1 = \cM_2 = \dots = 0$. Condition 
\eqref{N6} was derived using the other normalization conditions, so it 
must be compatible with all these conditions. \\
Now we come to the definition of interacting fields. Since the 
Lagrangian contains no higher generators, the following relations
hold due to eqn. \eqref{relbashigh}
\beql{noextra}
  \inter{(\vp_i)^{(n,\nu_1\dots \nu_n)}}{g\cL_{QED}}{x} = 
  \dd^{\nu_1}_x\cdots \dd^{\nu_n}_x
  \inter{(\vp_i)^{(0)}}{g\cL_{QED}}{x} .
\end{equation}
We define $F^{\mu\nu}$ as 
\beql{equation}
  F^{\mu\nu} \defined (A^\nu)^{(1,\mu)} - (A^\mu)^{(1,\nu)} \qquad 
  \in \eP.
\end{equation}
The easiest examples of interacting fields are the ghosts and the 
anti-ghosts. They do not appear in the Lagrangian $\cL_{QED}$ and 
therefore do not interact. The causal Wick expansion, eqn. \eqref{CWE}, 
implies together with the definition of the retarded products, 
eqn. \eqref{R}, 
\begin{equation}
  \inter{u}{g\cL_{QED}}{x} = u(x)\qquad \text{and}\qquad
  \inter{\tu}{g\cL_{QED}}{x} = \tu(x) .
\end{equation}
Due to relation \eqref{noextra} we can establish the usual relation 
for the interacting photon field and the field strength tensor in 
QED:
\begin{equation}
  \inter{F^{\mu\nu}}{g\cL_{QED}}{x} = \dd^\mu_x \inter{A_\nu}{g\cL_{QED}}{x}
  - \dd^\nu_x \inter{A_\mu}{g\cL_{QED}}{x} .
\end{equation}
The field equations for QED are also the usual ones:
\begin{equation}
  \begin{split}
    & \square^x \inter{A^\mu}{g\cL_{QED}}{x} 
    = - g(x) \inter{j^\mu_{\rm el}}{g\cL_{QED}}{x} \\
    \text{and}\quad 
    & (i\dddag - m )\inter{\psi}{g\cL_{QED}}{x} = - g(x) 
    \inter{\gamma^\mu A_\mu \psi}{g\cL_{QED}}{x} .
  \end{split}
\end{equation}
Furthermore we find that the interacting ghost current and BRS current 
have a particularly easy form because the ghosts and anti-ghosts
do not interact: 
\begin{equation}
  \begin{split}
    &\inter{k^\mu}{g\cL_{QED}}{x} = k^\mu(x) \\
    \text{and}\quad 
    &\inter{j^\mu_B}{g\cL_{QED}}{x} = 
    \left( \dd^\rho_x \inter{A_\rho}{g\cL_{QED}}{x} \vps\right)
    \dd^\mu_x u(x) \\
    &\qquad\qquad\qquad\qquad
    - \left( \dd^\rho_x \dd^\mu_x \inter{A_\rho}{g\cL_{QED}}{x} \vps\right)
    u(x) . 
  \end{split} 
\end{equation}
D\"utsch and Fredenhagen \cite{DueFre98} find the following 
commutator relations 
\begin{equation}
  \begin{split}
    &\comm{\dd^\mu_x \inter{A_\mu}{g\cL_{QED}}{x}}{\inter{A_\nu}{g
    \cL_{QED}}{y}}
    = i \dd^\nu D(x-y) \\
    \text{and}\quad
    &\comm{\dd^\mu_x \inter{A_\mu}{g\cL_{QED}}{x}}{\inter{\psi}{g\cL_{QED}}{y}}
    = g(x) D(x-y) \inter{\psi}{g\cL}{y}
  \end{split} 
\end{equation}
if $x,y \in \cO$. We can use the equation for the interacting BRS 
current to find the explicit form of the interacting BRS 
transformations, for example
\begin{equation}
  \begin{split}
    \begin{split}
      & \ts\left( \inter{A_\mu}{g\cL_{QED}}{x} \right) 
      = i \dd^x_\mu u(x) \\ 
      & \ts\left( \dd_x^\mu \inter{A_\mu}{g\cL_{QED}}{x} \right) 
      = 0 \\ 
      & \ts\left( \inter{\psi}{g\cL_{QED}}{x} \right) 
      = - g(x) \inter{\psi}{g\cL_{QED}}{x} u(x) \\
      & \ts\left( \inter{\psiq}{g\cL_{QED}}{x} \right) 
      = g(x) \inter{\psiq}{g\cL_{QED}}{x} u(x) 
    \end{split}\quad
    \begin{split}
      & \ts\left( u(x) \right) = 0 \\
      & \ts\left( \tu(x) \right) 
      = -i \dd^\rho_x \inter{A_\rho}{g\cL_{QED}}{x}  \\
      & \ts\left( \inter{F^{\mu\nu}}{g\cL_{QED}}{x} \right) 
      = 0 \\ 
      & \ts\left( \inter{j^\mu_{\rm el}}{g\cL_{QED}}{x} \right) 
      = 0 ,
    \end{split}
  \end{split} 
\end{equation}
for $x \in \cO$. 
The interacting electric current and the interacting field strength 
tensor are the only nontrivial observable quantities of those. The 
other two quantities with vanishing BRS transformation are not 
observable. The ghost $u(x)$ has non vanishing ghost number, and
$\dd_x^\mu \inter{A_\mu}{g\cL_{QED}}{x}$ is a coboundary and
therefore equivalent to zero. 

\subsection{Yang-Mills-theory}\label{YM}

The basic fields in Yang-Mills theory\footnote{We consider here only pure, 
massless Yang-Mills theory, for simplicity} are Lie algebra valued vector 
bosons
$A_\mu = A^a_\mu \tau_a$, ghosts $u = u^a \tau_a$ and anti-ghosts
$\tu = \tu^a \tau_a$. The $\tau_a$ form a basis of the Lie algebra. Their 
Lie-bracket gives $[\tau_a,\tau_b] = f^c_{ab}\tau_c$. The $f^c_{ab}$ are the
structure constants of the Lie-algebra. They satisfy the Jacobi-identity
\beql{jacobi}
  f^e_{ab}f^d_{ec} + f^e_{bc}f^d_{ea} + f^e_{ca}f^d_{eb} = 0 
\end{equation}
and are assumed to be totally antisymmetric. \\
The free field operators that belong to different components 
$A_\mu^a,u^a,\tu^a$ of the fields $A_\mu,u$ and $\tu$ have trivial 
commutation relations among each other, e.g.
\begin{equation}
  \acomm{u^a(x)}{\tu^b(y)} = -i \de^{ab} D(x-y) .
\end{equation}
Therefore the free model underlying Yang-Mills theory is simply a $p$-fold
copy of free QED if $p$ is the dimension of the Lie algebra. The 
underlying free model was considered in section \eqref{freeBRS}\\
The Lagrangian of Yang-Mills theory in causal perturbation theory is
\beql{LagYM}
  \cL_{YM} = \frac{1}{2} f^c_{ab} A^a_\mu A^b_\nu F_c^{\nu\mu} - 
  f^c_{ab} A^b_\mu u^a \dd^\mu \tu_c .
\end{equation}
Here $F^{\mu\nu}_c \defined (A^\nu_c)^{(1,\mu)} - (A^\mu_c)^{(1,\nu)}$. 
Note that there is no four-gluon-vertex present. It is created in second 
order perturbation theory due to $C_{A,1}=-\half$, see 
\cite{SchHuDueKra94a} - \cite{SchHuDue95b} for further details. \\
For the interacting fields we get
\begin{equation}
  \begin{split} 
    &\inter{A_\mu^a}{g\cL_{YM}}{x} = 
    \left( \inter{A_\mu^a}{g\cL_{YM}}{x}\right)^*
    \qquad \in \widetilde{\C}\cdot \Dist_1(\eD),\\
    &\inter{u^a}{g\cL_{YM}}{x} = 
    \left( \inter{u^a}{g\cL_{YM}}{x}\right)^*
    \qquad \in \widetilde{\C}\cdot \Dist_1(\eD),\\
    &\inter{\tu^a}{g\cL_{YM}}{x} = 
    - \left( \inter{\tu^a}{g\cL_{YM}}{x}\right)^*
    \qquad \in \widetilde{\C}\cdot \Dist_1(\eD) .
  \end{split}
\end{equation}
From eqn. \eqref{relbashigh} we get for the higher generators
\begin{equation}
  \begin{split} 
    &\inter{(A_\mu^a)^{(1,\nu)}}{g\cL_{YM}}{x} = 
    \dd^\nu_x \inter{A_\mu^a}{g\cL_{YM}}{x} 
    - \frac{1}{2} g(x) \inter{f^a_{bc} A_\mu^b A_\nu^c}{g\cL_{YM}}{x} ,\\
    &\inter{(u^a)^{(1,\nu)}}{g\cL_{YM}}{x} = 
    \dd^\nu_x \inter{u^a}{g\cL_{YM}}{x} 
    + g(x) \inter{f^a_{bc} A^{b,\nu} u^c}{g\cL_{YM}}{x} .\\
    &\inter{(\tu^a)^{(1,\nu)}}{g\cL_{YM}}{x} = 
    \dd^\nu_x \inter{\tu^a}{g\cL_{YM}}{x}  .
  \end{split}
\end{equation}
The first equation implies in particular
\begin{equation}
  \inter{F_{\mu\nu}^a}{g\cL_{YM}}{x} = 
  \dd_\mu^x \inter{A_\nu^a}{g\cL_{YM}}{x} 
  - \dd_\nu^x \inter{A_\mu^a}{g\cL_{YM}}{x} 
  + g(x) \inter{f^a_{bc} A_\mu^b A_\nu^c}{g\cL_{YM}}{x} 
\end{equation}
and
\begin{equation}
  \inter{(A_\mu^a)^{(1,\mu)}}{g\cL_{YM}}{x} = 
  \dd^\mu_x \inter{A_\mu^a}{g\cL_{YM}}{x} .
\end{equation}
The first relation reproduces the usual relation between the interacting
vector boson field and the field strength tensor in Yang-Mills theories. 
From the Lagrangian we can also derive the field equations using eqn. 
\eqref{expl}:
\beql{YMfe}
  \begin{split}
    & 
    \begin{split}
      \square^x \inter{A_\mu^a}{g\cL_{YM}}{x} =
      &\dd^\nu_x 
      \left[ 
        g(x) \inter{f^a_{bc} A_\mu^b A_\nu^c}{g\cL_{YM}}{x}
      \right]\\
      &- g(x) \inter{f^a_{bc} A^{\nu,b} F_{\nu\mu}^c}{g\cL_{YM}}{x}
      + g(x) \inter{f^a_{bc} u^{b} (\tu^c)_{(1,\mu)}}{g\cL_{YM}}{x},
    \end{split}\\
    &\square^x \inter{u^a}{g\cL_{YM}}{x} = - \dd^\mu_x 
    \left[ 
      g(x) \inter{f^a_{bc} A^{\mu,b} u^c}{g\cL_{YM}}{x}
    \right],\\
    &\square^x \inter{\tu^a}{g\cL_{YM}}{x} = - g(x) 
    \inter{f^a_{bc} A_\mu^{b} (\tu^c)^{(1,\mu)}}{g\cL_{YM}}{x}.
  \end{split}
\end{equation}
The Lagrangian \eqref{LagYM} obviously satisfies the conditions \eqref{C1},
\eqref{C2},\eqref{C3} and \eqref{C5}. It is also possible to find 
$Q(n)$-vertices and $R(n)$-vertices such that condition \eqref{C4} is
valid. These vertices are 
\begin{equation}
  \begin{split}
    &\cL^\mu_1 = f^c_{ab} u^a A_\nu^b F^{\nu\mu}_c 
    - \frac{1}{2} f^c_{ab} u^a u^b (\tu_c)^{(1,\mu)},\\
    &\cL^{\mu\rho}_2 = \frac{1}{2} f^c_{ab} u^a u^b F^{\mu\rho}_c,\\
    &\cL^{\dots}_i = \cM^{\dots}_i = 0 \qquad \forall i > 2 .
  \end{split}
  \qquad
  \begin{split}
    &\cM^\mu_1 = \frac{1}{2} f^c_{ab} u^a u^b (\tu_c)^{(1,\mu)},\\
    &\cM^{\mu\rho}_2 = \frac{1}{2} f^c_{ab} u^a u^b F^{\mu\rho}_c,
  \end{split}
\end{equation}
The vertices have been chosen such that condition \eqref{N6} is 
compatible with all other normalization conditions in first order.\\
We remind the reader that the existence of solutions for condition 
\eqref{N6} has not been proven for an arbitrary number of arguments in 
Yang-Mills theory. Scharf and collaborators, \cite{SchHuDueKra94a} - 
\cite{SchHuDue95b}, have proven operator gauge invariance, i.e. eqn.
\eqref{GOGI} for $i_1=\dots =i_n=0$, see section (\ref{normcon}) for 
further details.\\
We already mentioned that the free model underlying Yang-Mills theory is 
a $p$-fold copy of free QED if $p$ is the dimension of the Lie-group. The
question arises  whether there are other interactions besides Yang-Mills 
theory with the same free model. Stora \cite{Sto97} found out that the 
number of possible Lagrangians for such a model is severely restricted by 
the conditions \eqref{C1} - \eqref{C5}. Lagrangians $\TT{\cL}(x)$ may 
differ from the Yang-Mills Lagrangian 
only by a coboundary $s_0 \TT{K}(x)$ or a derivative $\dd_\mu \TT{K^\mu}(x)$
where $K$ is a scalar polynomial with ghost number $-1$ and $K^\mu$ is a 
vector polynomial with ghost number zero. By Yang-Mills Lagrangian we mean 
here an expression like \eqref{LagYM} with arbitrary constants $f^a_{bc}$ 
that are totally antisymmetric in their indices and satisfy the Jacobi 
identity \eqref{jacobi}. In particular the Lie-group structure needs not
to be put in. The Jacobi identity for the constants $f^a_{bc}$ is a 
consequance of operator gauge invariance in second order, and operator
gauge invariance in first order implies that they are totally 
antisymmetric. Since Stora's paper is not published, we refer 
the reader to the articles of Aste and Scharf \cite{AstSch98} and Grigore 
\cite{Gri98}. \\
It is usually argued that the addition of such coboundary or derivative terms 
does not change the model because coboundaries are equivalent to zero in
cohomology and derivatives should not give a contribution in the adiabatic
limit. D\"utsch \cite{Due96a} has proven that this is correct also in 
higher orders for theories where the adiabatic limit can be performed, 
e.g.\ in massive theories. But for models where this
limit does not exist the question is still open. Concerning the coboundary 
terms we remark that it is not clear whether a coboundary in the free 
theory, $\TT{A} = s_0\TT{B}$ for some $B\in\eP$, gives a coboundary in the 
interacting theory, such that
$\inter{A}{g\cL}{x} = \ts \inter{C}{g\cL}{x}$ for some $C\in\eP$. Direct 
calculations in first order indicate that this is indeed true for suitable
normalizations, but as long as this question is not clarified coboundary 
terms in the Lagrangian must not be neglected. The same is true for 
derivated terms in these theories.\\
In the rest of the section we want to compare our results with those of 
Nakanishi and Ojima \cite{NakOji90}. Their results have been derived in 
the context of quantum
field theory, but they are also classical in the following sense: They use 
field equations derived as Euler-Lagrange equations from a classical action, 
and they deliberately neglect the distributional character of field operators
and form products of field operators at the same spacetime point. Therefore it 
is possible to compare their results with the classical limit of our results.
At first we note that Nakanishi and Ojima use a different convention for the 
anti-ghosts. Their ghosts $C$ and anti-ghosts $\overline{C}$ correspond to 
ours in the following way:
\begin{equation}
  C^a \longleftrightarrow u^a, \qquad \overline{C}^a \longleftrightarrow 
  i \tu^a. 
\end{equation}
For the comparison we will always translate their results into our language. 
To make the notation shorter we introduce the covariant derivative of a field
with Lie-algebra index, $X(x)=X^a(x) \tau_a$, as
\begin{equation}
  (D_\mu X)^a(x) \defined \dd_\mu^x X^a(x) + f^a_{bc} \class{A_\mu^b}{x} 
  X^c(x) .
\end{equation}
We have the classical fields
\begin{equation}
  C(A_\mu^a)(x) = \class{A_\mu^a}{x}, \qquad C(u^a)(x) = \class{u^a}{x},
  \qquad  C(\tu^a)(x) = \class{\tu^a}{x} 
\end{equation}
and for the higher generators the representation $C$ gives
\begin{equation}
  \begin{split} 
    &C\left((A_\mu^a)^{(1,\nu)}\right)(x) = 
    \dd^\nu_x \class{A_\mu^a}{x} 
    - \frac{1}{2} g(x) f^a_{bc} \class{A_\mu^b}{x} \class{A_\nu^c}{x},\\
    &C\left((u^a)^{(1,\nu)}\right)(x) = 
    \dd^\nu_x \class{u^a}{x} 
    + g(x) f^a_{bc}\class{A^{b,\nu}}{x} \class{u^c}{x}.\\
    &C\left((\tu^a)^{(1,\nu)}\right)(x) = \dd^\nu_x \class{\tu^a}{x}  .
  \end{split}
\end{equation}
The field equations \eqref{YMfe} become in the classical limit
\begin{equation}
  \begin{split}
    & 
    \begin{split}
      (D^\mu F_{\mu\nu}^{\rm cl})^a(x) &= 
      - \dd_\nu^x\dd^\mu_x\class{A_\mu^a}{x}\\
      &\quad
      + g f^a_{bc}\left( \dd_\nu^x \class{\tu^b}{x}\right) \cdot 
      \class{u^c}{x},
    \end{split}\\
    &\dd^\mu_x (D_\mu \classa{u})^a(x) = 0, \\
    &(D_\mu \dd^\mu\classa{u})^a(x) = 0 .    
  \end{split}
\end{equation}
Here $F_{\mu\nu}^{a,{\rm cl}}$ is the classical field strength tensor,
\begin{equation}
  F_{\mu\nu}^{a,{\rm cl}} = \dd_\mu^x \class{A_\nu^a}{x} 
  - \dd_\nu^x \class{A_\mu^a}{x} + g f^a_{bc}\class{A_\mu^b}{x}
  \class{A_\mu^c}{x} .
\end{equation}
The field equations are exactly the same as those of Nakanishi and Ojima. 
For the ghost current we get
\begin{equation}
  C\left(k^\mu\right)(x) = i \sum_{a} \left(\class{u^a}{x}\dd^\mu_x
  \class{\tu^a}{x} - (D^\mu \classa{u})^a(x) \class{\tu^a}{x}\right).
\end{equation}
This is $-i$ times the result of Nakanishi and Ojima. The factor $-i$ 
comes from a different definition of the ghost current. They require 
that the ghost current and -charge be pseudo-hermitian, so that the 
eigenvalues of the ghost charge are in $i\Z$. \\
For the classical BRS current $j_B^\mu(x)$ we have according to 
definition \eqref{defj}
\begin{equation}
  j_B^\mu(x) = \class{j_B^\mu}{x} - g \class{\cM_1^\mu}{x} .
\end{equation}
This reads in terms of the basic fields
\begin{equation}
  \begin{split}
    j_B^\mu(x) &= \sum_{a} 
    \left( \vps (D^\mu \classa{u})^a(x) 
    \dd^\nu_x \class{A_\nu^a}{x} - \class{u^a}{x}
    \dd^\mu_x\dd^\nu_x \class{A_\nu^a}{x}\right)\\
    &\quad - \frac{1}{2} f^c_{ab} \class{u^a}{x}\class{u^b}{x}\dd^\mu_x
    \class{\tu_c}{x}
  \end{split}
\end{equation}
This is again --- up to a minus sign which is pure convention --- the
same result as Nakanishi and Ojima. Therefore we realize a complete
agreement between the results of Nakanishi and Ojima and ours, apart 
from different conventions. This supports both our results at the 
quantum level and also the correspondence law. The same 
relations at the quantum level would have given different results 
if we had adopted the correspondence law \eqref{versuch}, for 
example.


%% file: summary.tex
\section{Conclusions and Outlook}

We presented a universal construction of local quantum gauge theories.
It gives an algebra of local observables that has a Hilbert space
representation. For this construction to work two preconditions must
hold: The underlying free theory must be positive in the sense
discussed in chapter (\ref{free}), and the time ordered products of
free field operators must satisfy the conditions \eqref{N1} -
\eqref{N6}. The second precondition can only be violated with respect
to condition \eqref{N6}, all other conditions can always be
accomplished. If all the normalization conditions hold, a locally
conserved BRS current and with it a nilpotent BRS transformation on
the algebra of local fields can be defined. The algebra of local
observables is then defined as the cohomology of the algebra of local
fields w.r.t. the BRS transformation. If the underlying free model is
positive, the Hilbert space representation can be constructed.
Therefore spacetime must be compactified spatially in order to allow a
nilpotent BRS charge to be defined. This compactification does not
change the algebra.  It is an open question whether two
representations that are constructed
with a different compactification length are equivalent or not. \\
The most crucial point for each model that is investigated in this
framework is whether normalization condition \eqref{N6} can be
accomplished together with the other normalization conditions. We
have proven that this holds fod quantum electrodynamics, but for 
Yang-Mills theory the question is still open. We think that methods
of algebraic renormalization can help to find a solution. To reduce the 
problem to an algebraic one it could be helpful to define a BRS 
transformation $s$ on the algebra $\eP$, such that 
$T(sA) = s_0T(A)\quad \forall A \in \eP$. This requires the introduction 
of an additional auxiliary field, the scalar Nakanishi-Lautrup field
$B \in \eP$ with the properties $s\tu = iB, \,sB=0$ and 
$T(B)(x) = -\dd^\mu A_\mu(x)$. With this definition the BRS transformation
$s$ on $\eP$ can be chosen to be nilpotent. These notions could make it 
possible to translate the language of algebraic renormalization into ours. 
Normalization condition \eqref{N6} takes on the form of the descent 
equations in algebraic renormalization. Since they are proven in Yang-Mills 
theory, this could also lead to a proof of \eqref{N6} for Yang-Mills theory.\\
The renormalization scheme underlying our construction is the one of Epstein 
and Glaser. It is formulated, unlike the other renormalization schemes, in 
configuration space. Therefore it is suitable for quantum field theories on 
curved spacetimes. Brunetti and Fredenhagen \cite{BruFre99} have shown that 
the time ordered products can also be defined in globally hyperbolic 
spacetimes. To generalize our normalization conditions to these spacetimes, 
the propagators and differential operators introduced in chapter (\ref{free})
must be substituted by suitably generalized ones. With the normalization 
conditions all relations derived from them carry over to curved spacetimes,
in particular the field equations, the conservation of ghost and BRS current
and the nilpotency of the BRS charge and the BRS transformation. So it is 
possible to define an algebra of local observables even in globally 
hyperbolic spacetimes, provided these spacetimes allow propagators and their 
corresponding differential operators to be defined. 

\newpage

\vspace*{\fill} 

\textsc{Acknowledgement:} I thank Prof. Klaus Fredenhagen for 
entrusting me with this interesting topic,
for guidance and constant support during the course of this work. 
Detailed discussions with him were always a great motivation and help.
He also managed to create an inspiring atmosphere in the research group.\\
I am also greatful to Michael D\"utsch. I benefited a lot from his great 
experience withcausal perturbation theory, many valuable discussions and 
fruitful collaboration. \\
Thanks are also due to the other colleagues in our research group for many
interesting discussions on various topics. \\
Financial support given by the DFG as part of the `Graduiertenkolleg f\"ur
theoretische Elementarteilchenphysik' is greatfully acknowledged.


%% file: proofN4.tex
\section{Proof of \eqref{N3} and \eqref{N4}}

\subsection{Proof of \eqref{N3}}\label{proofN3}

The essential point in the proof that solutions for condition \eqref{N3}
exist is to show that eqn. \eqref{N3} is equivalent to the causal 
Wick expansion \eqref{CWE}. This suffices for a proof because it was
already shown in \cite{BruFre99} that \eqref{CWE} has solutions, see 
below.\\
The proof that both conditions are equivalent for a certain
$\TT{W_1,\cdots ,W_n}$ proceeds inductively. The induction hypothesis
is that eqn. \eqref{N3} and eqn. \eqref{CWE} hold and are equivalent
for the following time ordered products: all time ordered products that 
contain fewer arguments than $n$ and all that contain a combination of 
sub monomials of the $W_i$, if at least one of these sub monomials is a 
proper one.\\
At first we prove that eqn. \eqref{CWE} implies eqn. \eqref{N3}. With
eqn. \eqref{CWE} the time ordered product on the left hand side of
\eqref{N3} can be written as
\beql{CWE1}
\begin{split}
  \TT{W_1,\cdots ,W_n\vps}(x_1, \dots, x_n) =
  \sum_{\gamma_1,\dots , \gamma_n} \om_0
  \left(
    \TT{W_1^{(\gamma_1)},\cdots, W_n^{(\gamma_n)}\vps}(x_1, \dots, x_n)
  \right) &\\
  \times \frac{:\vp^{\gamma_1}(x_1) \cdots \vp^{\gamma_n}(x_n):}{
  \gamma_1!\cdots \gamma_n!} &,
\end{split}
\end{equation}
for the notation see the formulas following \eqref{CWE}. To calculate 
the (anti-) commutator with the $\vp_i(z)$ in eqn. \eqref{N3}, we note 
that the (anti-) commutator of the Wick product with the $\vp_i(z)$
gives
\begin{equation}
  \begin{split}
    &\scomm{\frac{:\vp^{\gamma_1}(x_1) \cdots \vp^{\gamma_n}(x_n):}{
    \gamma_1!\cdots \gamma_n!}}{\vp_i(z)} =\\
    &\qquad = i \sum_{k=1}^{n} \sum_{j} \De_{ij}(z-x_k) \, \frac{:
    \vp^{\gamma_1}(x_1) \cdots \vp^{\gamma_k-e_j}(x_k) \cdots 
    \vp^{\gamma_n}(x_n):}{\gamma_1!\cdots (\gamma_k-e_j)!\cdots 
    \gamma_n!}
  \end{split}
\end{equation}
if the $\gamma_k \neq 0$, otherwise the respective term vanishes.  
Here $e_j$ is the unit vector with an entry $1$ at the $j^{\rm th}$
position and the other entries zero. Therefore we get for the 
complete commutator
\begin{equation}
  \begin{split}
    i \sum_{k=1}^{n} \sum_{j} \De_{ij}(z-x_k) 
    \sum_{\stackrel{\gamma_1,\dots , \gamma_n}{\gamma_k\neq 0}}
    \om_0
    \left(
      \TT{W_1^{(\gamma_1)},\cdots, W_n^{(\gamma_n)}\vps}(x_1, \dots, x_n)
    \right)& \\
    \qquad \times
    \left[
      \frac{:\vp^{\gamma_1}(x_1) 
      \cdots \vp^{\gamma_k-e_j}(x_k) \cdots \vp^{\gamma_n}(x_n):}{\gamma_1!
      \cdots (\gamma_k-e_j)!\cdots \gamma_n!}
    \right].&
  \end{split}
\end{equation}
This becomes after a shifting of indices
\begin{equation}
  \begin{split}
    &
    \begin{split} 
      &i \sum_{k=1}^{n} \sum_{j} \De_{ij}(z-x_k) \times\\ 
      &
      \begin{split}
        \times \sum_{\gamma_1,\dots , \gamma_n}
        \om_0
        \left(
          \TT{W_1^{(\gamma_1)},\cdots, W_k^{(\gamma_1+e_j)},\cdots,
          W_n^{(\gamma_n)}\vps}(x_1, \dots, x_n)
        \right) \times&\\
        \times
        \left[
          \frac{:\vp^{\gamma_1}(x_1) \cdots \vp^{\gamma_n}(x_n):}{\gamma_1!
          \cdots \gamma_n!}
        \right] &
      \end{split}
    \end{split} \\
    &\qquad\quad =  
    i \sum_{k=1}^{n} \sum_{j} \De_{ij}(z-x_k)\, 
    \TT{W_1, \dots , W_k^{(e_j)}, \dots, W_n\vps} (x_1,\dots , x_n).
  \end{split}
\end{equation}
The last identity is valid because eqn. \eqref{CWE1} holds for the sub 
monomials according to our induction hypothesis. This proves that 
\eqref{N3} is a consequence of \eqref{CWE1}. \\
To complete the proof of equivalence we recall that we already saw
that eqn. \eqref{N3} determines the time ordered product up to a 
$\C$-number distribution. To be precise, eqn. \eqref{N3} determines
completely 
\begin{equation}
  \TT{W_1,\dots , W_n} - \om_0\left(\TT{W_1,\dots , W_n} \right) \one
\end{equation}
and leaves
\begin{equation}
  \om_0\left(\TT{W_1,\dots , W_n} \right) 
\end{equation}
open. This is exactly the same with \eqref{CWE1}. The Wick products 
are determined anyway and the numerical distributions are determined
by the $T$-products for the sub monomials if at least one 
$\gamma_i\neq 0$. Since both equations
determine the same part of the distribution and leave the same
part open, and moreover one of them is a consequence of the other,
they must be equivalent. \\
The question arises whether the expression on the right hand side of 
eqn. \eqref{CWE1} is well defined, because there appear products of
distribution. The answer is the same as in section (\ref{indkon}): 
Epstein and Glaser's ``Theorem 0'' guarantees that the product is well 
defined.

\subsection{Proof of \eqref{N4}}
\label{proofN4}
Like for \eqref{N3} we do not prove the existence of solutions for 
\eqref{N4} itself but for its integrated version
\beql{intN41}
  \begin{split}
    &\TT{W_1,\dots ,W_n,\vp_i\vps}(x_1,\dots ,x_n,y) = \\
    &\quad = i \sum_{k=1}^{n} \sum_{j} \De_{ij}^F(y-x_k) 
    \TT{W_1,\dots , \partder{W_k}{\vp_j},\dots, W_n}(x_1,\dots ,x_n)\\
    &\qquad + \sum_{\gamma_1\cdots\gamma_n}\om_0
    \left(
      \TT{W_1^{(\gamma_1)},\cdots, W_n^{(\gamma_n)}\vps}(x_1, \dots, x_n)
    \right)
    \frac{:\vp^{\gamma_1}(x_1) \cdots \vp^{\gamma_n}(x_n)
    \vp_i(y):}{\gamma_1!\cdots \gamma_n!} .
  \end{split}
\end{equation}
At the end of the section we will prove that the two conditions are 
equivalent. \\
The right hand side of eqn. \eqref{intN41} is obviously well defined,
because the first sum is a tensor product of distributions which is 
always well defined --- the argument $y$ does not appear in 
the time ordered product --- while the second sum is simply part of 
\eqref{CWE1} which was already proven to be well defined. \\
The question is whether this expression has the correct causal 
factorization outside the diagonal. To show this we proceed again 
inductively, the induction hypothesis is that eqn. \eqref{intN41} is 
valid for all time ordered products of sub monomials of the $W_i$. \\
At first we compare the expression with \eqref{CWE1}, which reveals in 
the present case
\begin{equation}
  \begin{split}
    &\TT{W_1,\dots ,W_n,\vp_i\vps}(x_1,\dots ,x_n,y) = \\
    &\quad = \sum_{\gamma_1\cdots\gamma_n}\om_0
    \left(
      \TT{W_1^{(\gamma_1)},\cdots, W_n^{(\gamma_n)},\vp_i\vps}
      (x_1, \dots, x_n,y)
    \right)
    \frac{:\vp^{\gamma_1}(x_1) \cdots \vp^{\gamma_n}(x_n):}{
    \gamma_1!\cdots \gamma_n!}
    \\
    &\qquad + \sum_{\gamma_1\cdots\gamma_n}\om_0
    \left(
      \TT{W_1^{(\gamma_1)},\cdots, W_n^{(\gamma_n)}\vps}(x_1, \dots, x_n)
    \right)
    \frac{:\vp^{\gamma_1}(x_1) \cdots \vp^{\gamma_n}(x_n)
    \vp_i(y):}{\gamma_1!\cdots \gamma_n!} .
  \end{split}
\end{equation}
So the second sum in eqn. \eqref{intN41} is already present and we 
must only show that 
\beql{firstterm}
  i \sum_{k=1}^{n} \sum_{j} \De_{ij}^F(y-x_k) 
  \TT{W_1,\dots , \partder{W_k}{\vp_j},\dots, W_n}(x_1,\dots ,x_n)
\end{equation}
is a possible extension of
\beql{firsttermN3}
  \sum_{\gamma_1\cdots\gamma_n}\om_0
    \left(
      \TTn{W_1^{(\gamma_1)},\cdots, W_n^{(\gamma_n)},\vp_i\vps}
      (x_1, \dots, x_n,y)
    \right)
    \frac{:\vp^{\gamma_1}(x_1) \cdots \vp^{\gamma_n}(x_n):}{
    \gamma_1!\cdots \gamma_n!}
\end{equation}
to the diagonal. 
Inserting eqn. \eqref{N3} into
expression \eqref{firstterm} gives
\begin{equation}
  \begin{split}
    &i \sum_{k=1}^{n} \sum_{j} \De_{ij}^F(y-x_k) \times \\
    & \qquad 
    \begin{split}
      \left[
        \sum_{\gamma_1\cdots\gamma_n}\om_0
        \left(
          \TT{W_1^{(\gamma_1)},\cdots, W_k^{(\gamma_k+e_j)},\cdots,
          W_n^{(\gamma_n)}\vps}(x_1, \dots, x_n)
        \right) 
      \right. &\\
      \left.
        \times \frac{:\vp^{\gamma_1}(x_1) \cdots \vp^{\gamma_n}(x_n):}{
        \gamma_1!\cdots \gamma_n!}
      \right] &.\\
    \end{split} \\
  \end{split}
\end{equation}
The latter is equal to expression \eqref{firsttermN3} if 
\begin{equation}
  \begin{split}
    \begin{split}
      &i \sum_{k=1}^{n} \sum_{j} \De_{ij}^F(y-x_k) \times \\
      & \qquad 
      \left[
        \sum_{\gamma_1\cdots\gamma_n}\om_0
        \left(
          \TT{W_1^{(\gamma_1)},\cdots, W_k^{(\gamma_k+e_j)},\cdots,
          W_n^{(\gamma_n)}\vps}(x_1, \dots, x_n)
        \right) 
      \right] 
    \end{split}& \\
    = \om_0
    \left(
      \TTn{W_1^{(\gamma_1)},\cdots, W_n^{(\gamma_n)}, \vp_i \vps}
      (x_1, \dots, x_n,y)
    \right) &
  \end{split}
\end{equation}
for all $\gamma_1,\dots ,\gamma_n$ and outside the diagonal. This 
equation is obviously true if
at least one $\gamma_i\neq 0$ since eqn. \eqref{intN41} is valid for
the sub monomials of the $W_i$ according to our induction hypothesis. 
So eqn. \eqref{intN41} can be accomplished if 
\beql{extension}
  i \sum_{k=1}^{n} \sum_{j} \De_{ij}^F(y-x_k) 
  \,\om_0
  \left(
    \TT{W_1,\dots , W_k^{(e_j)},\dots, W_n}(x_1,\dots ,x_n)
  \right)
\end{equation}
is a possible extension of
\beql{extended}
  \om_0
  \left(
    T^0\left(W_1,\cdots, W_n,\vp_i\vps\right)
    (x_1, \dots, x_n,y)
  \right).
\end{equation}
To see this we smear out both expressions with a
test function $\eta$ that vanishes with all its derivatives on the 
diagonal $\De_{n+1}$. Let the test function $\eta$ 
be fix and recall the definition of the partition of unity in eqn.
\eqref{partunity}. Then we can define for each subset $Z \subset 
\set{x_1,\dots,x_n,y}$ another test function $\eta_Z \in \cD(M^{n+1})$
\begin{equation}
  \eta_Z \defined 
  \begin{cases}
    f_Z \cdot \eta \qquad &\text{outside } \De_{n+1}\\
    0 \qquad &\text{otherwise}
  \end{cases}
\end{equation}
such that 
\begin{equation}
  \supp \eta_Z \in \complement_Z \qquad\text{and} \qquad
  \sum_{Z} \eta_Z = \eta .
\end{equation}
Then the following equation is valid owing to causal factorization:
\beql{parteta}
  \begin{split}
    &\int d^4y d^4x_1\cdots d^4x_n \eta(x_1,\dots,x_n,y) 
    \TT{W_1,\dots,W_n,\phi_i\vps}(x_1,\dots,x_n,y) =\\
    &\quad =\sum_{Z\subset X} \int d^4y d^4x_1\cdots d^4x_n 
    \eta_Z(x_1,\dots,x_n,y) \TT{W_Z\vps}(x_Z)\TT{W_{Z^c},\vp_i\vps}
    (x_{Z^c},y)\\
    &\qquad + \sum_{Z\subset X} \int d^4y d^4x_1\cdots d^4x_n 
    \eta_Z(x_1,\dots,x_n,y) \TT{W_Z,\vp_i\vps}(x_Z,y)\TT{W_{Z^c}\vps}
    (x_{Z^c})
  \end{split}
\end{equation}
because $Z \gtrsim Z^c$ on $\supp \eta_Z$. Here $X=\set{x_1,\dots,x_n}$.\\
Let us investigate $\TT{W_Z}(x_Z)\TT{W_{Z^c},\vp_i}(x_{Z^c},y)$ and
assume for simplicity that $Z=\set{x_{k+1},\dots,x_n}$ and 
$Z^c=\set{x_{1},\dots,x_k}$. Due to the validity of eqn. 
\eqref{intN41} in lower orders we have
\beql{intN4lo}
  \begin{split}
    &\TT{W_Z\vps}(x_Z)\TT{W_{Z^c},\vp_i\vps}(x_{Z^c},y) = \\
    &\qquad
    \begin{split}
      = i \sum_{m=1}^{k} \sum_{j} \De_{ij}^F(y-x_k) 
      \TT{W_Z\vps}(x_Z) \TT{W_1,\dots,W_m^{(e_j)},\dots,W_k\vps}
      (x_{Z^c})& \\
      + \sum_{\gamma_1\cdots\gamma_k}\om_0
      \left(
        \TT{W_1^{(\gamma_1)},\cdots, W_k^{(\gamma_k)}\vps}(x_{Z^c})
      \right) 
      \TT{W_Z\vps}(x_Z) \times &\\
      \times\frac{:\vp^{\gamma_1}(x_1) \cdots \vp^{\gamma_k}(x_k)
      \vp_i(y):}{\gamma_1!\cdots \gamma_k!}&.
    \end{split}
  \end{split}
\end{equation}
Since $Z \gtrsim Z^c$, the product in the first sum recombines to
\begin{equation}
  \begin{split}
    \TT{W_Z\vps}(x_Z) \TT{W_1,\dots,W_m^{(e_j)},\dots,W_k}(x_{Z^c}) 
    \qquad &\\
     = \TT{W_1,\dots,W_m^{(e_j)},\dots,W_n}(x_1,\dots,x_n) &.
  \end{split}
\end{equation}
For the product in the second sum we get
\begin{equation}
  \begin{split}
    &\TT{W_Z\vps}(x_Z)\frac{:\vp^{\gamma_1}(x_1) \cdots \vp^{\gamma_k}(x_k)
    \vp_i(y):}{\gamma_1!\cdots \gamma_k!} =\\
    &\qquad
    \begin{split} 
      :\dots \vp_i(y): + \sum_{l=k+1}^{n} \sum_j 
      \TT{W_{k+1},\dots,W_l^{(e_j)},\dots,W_n\vps}(x_Z)\,\De_{ij}^+(y-x_l)&\\
      \times\frac{:\vp^{\gamma_1}(x_1) \cdots \vp^{\gamma_k}(x_k):}{
      \gamma_1!\cdots \gamma_k!}&.
    \end{split}
  \end{split}
\end{equation}
Inserting this into eqn. \eqref{intN4lo} and taking \eqref{CWE1} into 
account, the second sum becomes
\begin{equation}
  \begin{split}
    i \sum_{l=1}^{k} \sum_{j} \De_{ij}^+(y-x_k) 
    \TT{W_{k+1},\dots,W_l^{(e_j)},\dots,W_n}(x_{Z}) \TT{W_{Z^c}\vps}
    (x_{Z^c})& \\
    + :\dots \vp_i(y):& . 
  \end{split}
\end{equation}
From the definition of the Feynman propagator we find
\begin{equation}
  \De_{ij}^+(y-x_l)=\De_{ij}^F(y-x_l) - \De_{ij}^A(y-x_l) 
  =\De_{ij}^F(y-x_l) 
\end{equation}
since $(y-x_l)\neq\backlc$. Recombining the terms we finally arrive at 
\begin{equation}
  \begin{split}
    &\om_0
    \left(
      \TT{W_Z\vps}(x_Z) \TT{W_{Z^c},\vp_i\vps}(x_{Z^c},y) 
    \right) = \\
    &\qquad = i \sum_{m=1}^{k} \sum_{j} \De^F_{ij}(y-x_k) 
    \om_0
    \left(  
      \TT{W_1,\dots,W_m^{(e_j)},\dots,W_n\vps}(x_1,\dots,x_n)
    \right) \\ 
    &\qquad\quad + i \sum_{l=k+1}^{n} \sum_{j} \De^F_{ij}(y-x_l) 
    \om_0
    \left(  
      \TT{W_1,\dots,W_l^{(e_j)},\dots,W_n\vps}(x_1,\dots,x_n)
    \right) \\
    &\qquad = i \sum_{m=1}^{n} \sum_{j} \De^F_{ij}(y-x_k) 
    \om_0
    \left(  
      \TT{W_1,\dots,W_m^{(e_j)},\dots,W_n\vps}(x_1,\dots,x_n)
    \right) .
  \end{split}
\end{equation}
With the same argument we can see that 
\begin{equation}
  \begin{split}
    &\om_0
    \left(
      \TT{W_{Z},\vp_i\vps}(x_{Z},y) \TT{W_{Z^c}\vps}(x_{Z^c})
    \right) = \\
    &\qquad = i \sum_{m=1}^{n} \sum_{j} \De^F_{ij}(y-x_k) 
    \om_0
    \left(  
      \TT{W_1,\dots,W_m^{(e_j)},\dots,W_n\vps}(x_1,\dots,x_n)
    \right) .
  \end{split}
\end{equation}
Taking the vacuum expectation value of eqn. \eqref{parteta} and
inserting the expressions above, we finally get
\begin{equation}
  \begin{split}
    &\int d^4y d^4x_1\cdots d^4x_n \eta(x_1,\dots,x_n,y) 
    \om_0
    \left(
      \TT{W_1,\dots,W_n,\phi_i\vps}(x_1,\dots,x_n,y) 
    \right) \\
    &\qquad = i \int d^4y d^4x_1\cdots d^4x_n \eta(x_1,\dots,x_n,y) 
    \times \\
    &\qquad \qquad\quad \times\sum_{m=1}^{n} \sum_{j} \De^F_{ij}(y-x_k) 
    \om_0
    \left(  
      \TT{W_1,\dots,W_m^{(e_j)},\dots,W_n\vps}(x_1,\dots,x_n)
    \right) .
  \end{split}
\end{equation}
So we have proven that expression \eqref{extension} is a possible
extension of \eqref{extended}, and this implies that eqn. 
\eqref{intN41} has the correct causal factorization. From the 
construction it is clear that \eqref{intN41} is compatible with 
\eqref{CWE} and thus with \eqref{N3}. It is obvious that it respects 
the Poincar\'e transformation properties and is therefore compatible 
with \eqref{N1}. The same calculation as for the compatibility of
eqns. \eqref{N1} and \eqref{N2} reveals that it is also compatible with 
\eqref{N2}.\\
Finally we have to prove that \eqref{N4} and \eqref{intN41} are 
equivalent. Eqn. \eqref{intN41} implies \eqref{N4} immediately: 
Application of the operator $D^y$, eqn. \eqref{defDx}, from the 
left on eqn. \eqref{intN41} gives the desired result.\\
On the other hand a solution of \eqref{N4} is unique. This can best be
seen for the corresponding equation for the retarded products,
\begin{equation}
\begin{split}
  &\sum_j D^y_{ij} \RR{W_1,\dots , W_n}{\vp_j\vps}(x_1,\dots ,x_n;y) = \\
  &\qquad = i\sum_{k=1}^n \RR{W_1 ,\dots ,\check{k},\dots , W_n}{
  \partder{W_k}{\varphi_i}}(x_1,\dots ,\check{k},\dots, x_n; x_k) \,
  \delta (x_k-y)  .
\end{split}
\tag{\bf N4}
\end{equation}
The difference of two solutions of this differential equation is a
solution of the homogeneous differential equation. Due to the support 
properties of the retarded products there exists a Cauchy surface 
in the $y$-space such that all Cauchy data are zero. Therefore zero 
is a solution of that equation, and $D^y$ is an operator with a unique 
solution for the Cauchy problem, see page \pageref{Dunique}. So the 
retarded products are uniquely determined and with them the time ordered 
products. This completes the proof that \eqref{N4} and \eqref{intN41} 
are equivalent. 


%% file: proofN5.tex
\section{Proofs concerning the Ward identities}\label{Wardid}

This appendix contains in its first section the proof that the ghost 
number Ward identities have common solutions with the other 
normalization conditions and that the ghost number Ward identities
imply eqn. \eqref{N5int}. In the second section we prove that 
the validity of the generalized operator gauge invariance already
implies that there exists a solution of condition \eqref{N6}.

\subsection{Proof of the ghost number Ward identities}\label{proofN5}

We begin with the proof that the equation
\beql{N5int1}
\begin{split}
  &s_c\TT{W_{1} \cdots W_{n}}(x_1,\dots ,x_n) = \\
  &\qquad = \left(\sum_{k=1}^n g(W_k)\right)
  \TT{W_{1} \cdots W_{n}}(x_1,\dots ,x_n) .
\end{split} 
\end{equation}
is a direct consequence of condition \eqref{N5},
\beql{N51}
\begin{split}
  &\dd^{y}_{\mu}\TT{W_{1} ,\dots , W_{n}, k^{\mu}}(x_1,\dots ,x_n,y) = \\
  &\qquad = \sum_{k=1}^n g(W_k)\,\delta(y-x_k) \, 
  \TT{W_{1} ,\dots , W_{n}}(x_1,\dots ,x_n). 
\end{split}
\end{equation}
Suppose, $\cO$ is an open, bounded and causally complete region in 
spacetime such that all points $x_1,\dots , x_n$ in eqn. \eqref{N5} 
lie in $\cO$ --- obviously for every set of points such a region 
can be found. Then we choose a test function $f\in \cD(M)$ such that 
$f(x) = 1\quad\forall x \in \cO'$ with $\cO'$ another open, bounded and 
causally complete region such that $\overline{\cO} \subset \cO'$. Then 
we can find a Lorentz frame where a $\cinfty$-function $H(y)$ exists 
with the following properties:
\begin{equation}
  \begin{split}
    &H\in \cinfty(M), \qquad \exists H^t \in \cinfty (\R):\quad 
    H(y) = H^t(y^0) , \\
    &H^t(y^0) = 1 \quad \forall \,y^0 < - \epsilon, \qquad H^t(y^0) = 0 
    \quad \forall \,y^0 > \epsilon, \qquad \epsilon \in \R, 
    \quad 0<\epsilon \ll 1, \\
    &\supp (H\cdot \dd_\mu f) \cap (\forlc + \cO) = \emptyset,
    \qquad    
    \supp ((1-H)\cdot \dd_\mu f) \cap (\backlc + \cO) = \emptyset .
  \end{split}
\end{equation}
The following calculations will be done in that Lorentz frame.
Smearing out the left hand side of eqn. \eqref{N5} with $f$ gives
\begin{equation}
  \begin{split}
    &\int d^4y \,f(y) \dd_\mu^y \TT{W_{1} ,\dots , W_{n}, k^{\mu}\vps}
    (x_1,\dots ,x_n,y) = \\
    &\qquad = -\int d^4y \,(\dd_\mu f)(y)\cdot H(y) \cdot 
    \TT{W_{1} ,\dots , W_{n}, k^{\mu}\vps}(x_1,\dots ,x_n,y) \\
    &\qquad \quad -\int d^4y \,(\dd_\mu f)(y)\cdot (1-H(y)) \cdot 
    \TT{W_{1} ,\dots , W_{n}, k^{\mu}\vps}(x_1,\dots ,x_n,y) .
  \end{split}
\end{equation}
According to our assumptions about the supports of the test functions 
$(\dd_\mu f)\cdot H(y)$ and $(\dd_\mu f)\cdot (1-H(y))$ we have in the 
first integral on the right hand side $y \gtrsim x_i\,\,\forall i$ and in 
the second integral on the right hand side $x_i \gtrsim y\,\,\forall i$. 
Owing to causal factorization the time ordered product 
$\TT{W_{1} ,\dots , W_{n}, k^{\mu}}(x_1,\dots ,x_n,y)$
decomposes in the first integral according to 
$\TT{k^{\mu}}(y)\TT{W_{1} ,\dots , W_{n}}(x_1,\dots ,x_n)$ and
in the second one according to 
$\TT{W_{1} ,\dots , W_{n}}(x_1,\dots ,x_n)\TT{k^{\mu}}(y)$. Therefore 
the integral can be written as 
\begin{equation}
  \begin{split}
    &\int d^4y \,f(y) \,\dd_\mu^y \TT{W_{1} ,\dots , W_{n}, k^{\mu}\vps}
    (x_1,\dots ,x_n,y) = \\
    &\qquad = \int d^4y \,(\dd_\mu f)(y)\cdot H(y) \cdot 
    \scomm{\TT{k^{\mu}}(y)}{\TT{W_{1} ,\dots , W_{n}\vps}
    (x_1,\dots ,x_n)} \\
    &\qquad \quad -\int d^4y \,(\dd_\mu f)(y) \cdot 
    \TT{W_{1} ,\dots , W_{n}\vps}(x_1,\dots ,x_n) \TT{k^{\mu}}(y).
  \end{split}
\end{equation}
Then the second integral vanishes since $k^\mu$ is a conserved current.
Partial integration in the first integral reveals according to the 
properties of $H$ and $f$
\begin{equation}
  \begin{split}
    &\int d^4y \,f(y) \dd_\mu^y \TT{W_{1} ,\dots , W_{n}, k^{\mu}}
    (x_1,\dots ,x_n,y) = \\
    &\qquad = \scomm{Q_c}{\TT{W_{1} ,\dots , W_{n}}(x_1,\dots ,x_n)}
    = s_c \TT{W_{1} ,\dots , W_{n}}(x_1,\dots ,x_n) .
  \end{split}
\end{equation}
As the smearing of the right hand side of eqn. \eqref{N5} with $f$ is 
trivial since $f=1 \,\, \forall x_k$, we finally arrive at 
\begin{equation}
\begin{split}
  &s_c\TT{W_{1} ,\dots , W_{n}}(x_1,\dots ,x_n) = \\
  &\qquad = \left(\sum_{k=1}^n g(W_k)\right) 
  \TT{W_{1} ,\dots , W_{n}}(x_1,\dots ,x_n) .
\end{split} 
\end{equation}
The proof that the ghost number Ward identities have common solution 
with the other normalization conditions proceeds along the same lines 
as the proof of D\"utsch and Fredenhagen \cite{DueFre98} for the 
electric current. An important difference between the proofs is that
for their proof it suffices to have eqn. \eqref{N4} for the basic
generators, while it is here important to have it also for the 
higher generators since the ghost current $k^\mu$ contains also higher 
generators. \\
The proof is subdivided into two parts. At first we prove that it is 
possible to normalize $\TT{W_{1} \cdots W_{n}}$ such that it satisfies 
eqn. \eqref{N5int1}.
Then we prove the same statement for condition \eqref{N5}. This seems 
to be a detour because we just saw that \eqref{N5int1} is a consequence 
of \eqref{N5}, but \eqref{N5int1} will be needed in the proof of 
\eqref{N5}.\\
Like all these proofs this one goes by induction, so we put forward the 
induction hypothesis that both \eqref{N5} and \eqref{N5int1} hold for 
fewer arguments than $n$ and for the sub monomials of the $W_i$, provided 
that at least one sub monomial is a proper one. Then the causal Wick 
expansion --- eqn. \eqref{CWE} --- tells us 
that eqn. \eqref{N5int1} can only be violated by an unsuitable 
normalization of $\om_0\left(\TT{W_{1} ,\dots , W_{n}} \right)$. 
Applying $\om_0$ to eqn. \eqref{N5int1} and taking $\om_0 \circ s_c = 0$
into account, we see that either $\left(\sum_{k=1}^n g(W_k)\right) = 0$
or $\om_0\left(\TT{W_{1} ,\dots , W_{n}} \right)= 0$. In the first case eqn. 
\eqref{N5int1} is true for an arbitrary normalization of 
$\om_0\left(\TT{W_{1} ,\dots , W_{n}} \right)$. In the second case 
validity of \eqref{N5int1} in lower orders guarantees that 
$\om_0\left(\TT{W_{1} ,\dots , W_{n}} \right)$ vanishes outside the diagonal
but not necessarily on the entire $M^n$. Nevertheless it is always possible 
to extend a distribution that vanishes outside the diagonal by a distribution 
that vanishes everywhere, and such an extension is obviously compatible 
with all other normalization conditions. So it is always possible to find
a normalization of $\TT{W_{1} ,\dots , W_{n}}$ that is a solution of all 
normalization conditions including \eqref{N5int1}. \\
Now we come to the second part, the proof that normalizations can be
found for which the ghost number Ward identities \eqref{N5}
\begin{equation}
\begin{split}
  &\dd^{y}_{\mu}\TT{W_{1}(x_1), \dots ,W_{n}(x_n), k^{\mu}(y)} = \\
  &\qquad = \sum_{k=1}^n \delta(y-x_k) \, g(W_k)\,
  \TT{W_{1}(x_1), \dots ,W_{n}(x_n)} 
\end{split}
\end{equation}
hold such that the normalization is also in accordance with 
\eqref{N1} - \eqref{N4}, provided none of the $W_i$ is equal to $k^\mu$
or contains it as a sub monomial, and none of them contains 
generators $(u^a)^{(\al)}$ or $(\tu^a)^{(\al)}$ with $\betr{\al}\geq 2$. \\
To this end we define a possible anomaly as
\beql{anomN5}
\begin{split}
  a(x_1,\dots ,x_n, y) &= 
  \dd^{y}_{\mu}\TT{W_{1}, \dots ,W_{n}, k^{\mu}}(x_1,\dots ,x_n,y) \\
  & \quad - \sum_{k=1}^n \delta(y-x_k)\,g(W_k)\, 
  \TT{W_{1}, \dots ,W_{n}} (x_1,\dots ,x_n)
\end{split}
\end{equation}
and show that a normalization can be found --- in agreement with eqns. 
\eqref{N1} - \eqref{N4} --- such that the anomaly vanishes. Recalling our 
induction hypothesis we want to show this under the assumption that
all these anomalies vanish for the time ordered products of fewer 
arguments than $n$ and in all equations that involve the sub monomials 
of the $W_i$. The
proof will be divided into three steps.\\
{\em Step 1:} At first we commute the anomaly with the basic fields
$\vp_i(x)$ in order to find that this commutator vanishes. Thereby we 
make repeated use of condition \eqref{N3} and
the fact that according to our induction hypothesis eqn. \eqref{N5} 
is already established for the lower orders and for the sub monomials. 
The result of that calculation is 
\beql{anomcomm}
  \begin{split}
    &\scomm{a(x_1,\dots ,x_n, y)}{\vp_i(z)\vps} =  \\
    &\qquad
    \begin{split}
    = i g(\vp_i) \sum_{k=1}^{n} \sum_j \De_{ij}(x_k-z) 
    \de(y-x_k) \TT{W_1,\dots ,\partder{W_k}{\vp_j},\dots ,W_n}&\\
    + i \sum_j \left( \dd_\mu^y \De_{ij}(y-z)\right)
    \TT{W_1,\dots ,W_n,\partder{k^\mu}{\vp_j}} &\\ 
    + i \sum_j \De_{ij}(y-z) \dd_\mu^y
    \TT{W_1,\dots ,W_n,\partder{k^\mu}{\vp_j}},&
    \end{split}
  \end{split}
\end{equation}
where we have omitted the spacetime arguments of the time ordered products
because the expressions would not fit into the line otherwise. We will do 
this throughout this proof. It should not cause confusion since it is 
already clear from the arguments of the time ordered products which the 
spacetime arguments are.\\
To show that the expression above vanishes we distinguish three cases:\\
{\em Case 1:} $\vp_i(z) \neq u^a(z), \tu^a(z)$. In this case both 
$\partder{k^{\mu}}{\vp_i} = 0$ and $g(\vp_i) = 0$, so the commutator vanishes 
immediately.\\
{\em Case 2:} $\vp_i(z) = u^a(z)$. At first we note that $g(u^a) = 1$. 
Furthermore we have 
$k^\mu = i (\tu_a)^{(1,\mu)}u^a - i \tu_a (u^a)^{(1,\mu)}$, 
so we get in particular $\partder{k^\mu}{\tu_a} = -i (u^a)^{(1,\mu)}$ and 
$\partder{k^\mu}{(\tu_a)^{(1,\nu)}} = i \de^\mu_\nu u^a$. Taking
this and the definition of the commutator function $\De_{ij}$ into 
account, we get for the last two lines in \eqref{anomcomm} 
\begin{equation}
  \begin{split}
    &\left( \dd_\mu^y D(y-z)\right) \TT{W_1,\dots ,W_n,(u^a)^{(1,\mu)}}\\
    &+ D(y-z) \dd_\mu^y \TT{W_1,\dots ,W_n,(u^a)^{(1,\mu)}} \\
    & - \left( \dd_\mu^y D(y-z)\right)\dd^\mu_y \TT{W_1,\dots ,W_n,u^a\vps}.
  \end{split}
\end{equation}
According to \eqref{N4} the expression above transforms into
\begin{equation}
  \begin{split}
    &\left( \dd^\mu_y D(y-z)\right)
    \left[ 
      -i C_{u,1} \sum_{k=1}^{n} \de(y-x_k)  
      \TT{W_1,\dots ,\partder{W_k}{(\tu_a)^{(1,\mu)}},\dots ,W_n}
    \right]\\
    & + D(y-z)
    \left[ 
      +i C_{u,1} \sum_{k=1}^{n} \de(y-x_k)  
      \TT{W_1,\dots ,\partder{W_k}{\tu_a},\dots ,W_n} 
    \right.\\
    &\qquad\qquad\quad\,\,\,\, - (1 + C_{u,1}) \square \TT{W_1,\dots ,W_n,u^a} \\
    &\qquad\qquad\quad\,\,\,\,
    \left.
      - i\frac{C_{u,1}}{C_{u,2}} \dd^\al_y \dd^\beta_y 
      \left[
        \sum_{k=1}^{n} \de(x_k-y) 
        \TT{W_1,\dots ,\partder{W_k}{(\tu^a)^{2,\al\beta}},\dots,W_n}
      \right] + \dots
    \right] 
  \end{split}
\end{equation}
Since we required that the $W_i$ do not contain generators $(u^a)^{(\al)}$ 
with $\betr{\al} \geq 2$, the last line and the following terms containing
derivatives w.r.t. higher generators on the $W_i$, indicated by the dots, 
vanish. Then comparing the
remaining expression with the first line in \eqref{anomcomm} reveals that 
these expressions cancel each other if and only if $C_{u,1} = -1$. So the 
choice $C_{u,1} = -1$ is a necessary (and, as it will turn out, sufficient)
condition for eqn. \eqref{N5} to hold. \\
{\em Case 3:} $\vp_i(z) = \tu^a(z)$. The calculation for this case is 
completely analogous the the one before and reveals $C_{u,1} = -1$ as a 
necessary condition for the commutator to vanish, too.\\
So with $C_u = -1$ the commutator of the anomaly with every free field 
vanishes. Consequently $a(x_1,\dots , x_n, y)$, smeared with an arbitrary test
function, is a $\C$-number distribution. \\
{\em Step 2:} We already know that time ordered products with at least one
generator as an argument are completely determined by the time ordered
products in lower orders and those for the sub monomials. We will now
examine whether this normalization is compatible with \eqref{N5}. \\
Since the anomaly can at most be a $\C$-number distribution, it is 
sufficient to calculate its $\C$-number part $\om_0 (a(x_1,\dots , x_n, y))$. 
So we want to prove that
\begin{equation}
  \begin{split}
    &\dd_\mu^y \om_0
    \left(
      \TT{W_1,\dots, W_n, \vp_i,k^\mu \vps}
    \right) = \\
    &\qquad
    \left(
      \sum_{k=1}^n \de(y-x_k) g(W_k) + \de(y-z) g(\vp_i)
    \right)
    \om_0
    \left(
      \TT{W_1,\dots, W_n, \vp_i\vps} 
    \right) .
  \end{split}
\end{equation}
With a repeated use of eqn. \eqref{N4int} this can be transformed into 
\beql{interm}
  \begin{split}
    &i \, g(\vp_i)\sum_{k=1}^n \sum_{j} \De_{ij}^F (z-x_k)\, 
    \de(y-x_k) \,\om_0
    \left(
      \TT{W_1,\dots, \partder{W_k}{\vp_j},\dots, W_n} 
    \right) \\
    &+ i\sum_{j} \left( \dd_\mu^y \De_{ij}^F (z-y)\right)\om_0
    \left(
      \TT{W_1,\dots, W_n, \partder{k^\mu}{\vp_j}}
    \right) \\
    &+ i\sum_{j} \De_{ij}^F (z-y) \,\dd_\mu^y \,\om_0
    \left(
      \TT{W_1,\dots, W_n, \partder{k^\mu}{\vp_j}}
    \right) \\
    & \qquad = g(\vp_i) \de(y-z)
    \left(
      i \sum_{k=1}^n \sum_{j}\De_{ij}^F (z-x_k)\,\om_0
      \left(
        \TT{W_1,\dots, \partder{W_k}{\vp_j},\dots, W_n}
      \right)  
    \right) .
  \end{split}
\end{equation}
Again we can distinguish different cases here. \\
In the first case, $\vp_i \neq (u^a)^{(\al)}, (\tu^a)^{(\al)}$, we have 
again both $\partder{k^{\mu}}{\vp_i} = 0$ and $g(\vp_i) = 0$, so the 
equation holds automatically. The cases $\vp_i = (u^a)^{(\al)}$ or
$\vp_i = (\tu^a)^{(\al)}$ with $\betr{\al} \geq 2$ cannot occur because 
they were explicitely excluded. So there remain four cases where we have
to prove that the equation above is indeed valid: 
$\vp_i = (u^a),(u^a)^{(1,\mu)}, (\tu^a)$ and $(\tu^a)^{(1,\mu)}$. For 
simplicity we will treat only $\vp_i = (u^a)$, the calculation for the
other cases is analogous.\\
Remembering $g(u^a) = 1$, $\partder{k^\mu}{\tu_a} = -i (u^a)^{(1,\mu)}$ 
and $\partder{k^\mu}{(\tu_a)^{(1,\nu)}} = i \de^\mu_\nu (u^a)$ from 
the first step, we see that the sum of the second and third line on the 
left hand side of equation \eqref{interm} give, where eqn. \eqref{N4}
has been used,
\begin{equation}
  \begin{split}
    &\left(\dd^\mu_y D^F(z-y)\right)
    \left[
      i \sum_{k=1}^n \de(y-x_k) \om_0
      \left(
        \TT{W_1,\dots, \partder{W_k}{(\tu^a)^{(1,\mu)}},\dots, W_n}
      \right)  
    \right] \\
    & +D^F(z-y)
    \left[
      i \sum_{k=1}^n \de(y-x_k) \om_0
      \left(
        \TT{W_1,\dots, \partder{W_k}{(\tu^a)},\dots, W_n}
      \right)  
    \right] \\
    & - \de(z-y) 
    \left[
      i \sum_{k=1}^{n} D^F(y-x_k) \TT{W_1,\dots, 
      \partder{W_k}{(\tu^a)},\dots, W_n}  
    \right.\\
    & \qquad\qquad
    \left.
      i \sum_{k=1}^{n} \left(\dd^\mu_y D^F(y-x_k)\right)
      \TT{W_1,\dots, \partder{W_k}{(\tu^a)^{(1,\mu)}},\dots, W_n}
    \right] . 
  \end{split}
\end{equation}
Comparing this with the other lines in eqn. \eqref{interm}, we see that 
the last two lines cancel the right hand side of that equation while 
the first two lines cancel the first line on the the left hand side. So 
the equation is indeed satisfied. As we already remarked, it can be 
proven by an analogous calculation that this is also true if 
$\vp_i = (u^a)^{(1,\mu)}, (\tu^a)$ or $(\tu^a)^{(1,\mu)}$.
With this we have proven that the time ordered products with at least
one generator among their arguments satisfy condition \eqref{N5}
automatically.\\
{\em Step 3:} We know up to now that eqn. \eqref{N5} can only be violated 
by $T$-products that have no generator among their arguments, and this 
violation can be at most a $\C$-number. In addition we know that the 
anomaly must be local because of causal factorization and validity of 
\eqref{N5} in lower orders, so it can be written as
\beql{polyanom}
  a(x_1,\dots ,x_n,y) = \om_0\left(a(x_1,\dots ,x_n,y)\right) 
  = P(\dd) \de(y-x_1)\cdots \de(y-x_n) 
\end{equation}
for some polynomial of spacetime derivatives $P(\dd)$. 
To show that such an anomaly can always be removed we notice that 
\beql{intanom}
  0 = \int d^4y \,f(y)\,a(x_1,\dots ,x_n,y) = 
  \int d^4y \,a(x_1,\dots ,x_n,y) 
\end{equation}
where $f$ is a test function like in the proof of eqn. \eqref{N5int}.
The first identity is an immediate consequence of that equation.
This is the point in the proof of \eqref{N5} where it is necessary
to know in advance that \eqref{N5int1} holds. The second identity
is true since $f=1$ in a domain around each $x_k$. \\
Let us consider the Fourier transformation of the anomaly,
\beql{FT}
  \begin{split}
    \hat{a}(x_1,\dots,x_n,y) &= (2\pi)^n \int d^4x_1 \cdots d^4x_n
    a(x_1,\dots ,x_n,y) e^{i(p_1x_1+\dots +p_nx_n)} \\
    &= (2\pi)^n P(-ip_1,\dots , -ip_n)e^{i(p_1+\dots +p_n)y}.
  \end{split}
\end{equation}
For the second identity we have adopted eqn. \eqref{polyanom} for the
anomaly. Inserting \eqref{FT} back into eqn. \eqref{intanom}, 
we find that the polynomial $P(-ip_1,\dots , -ip_n)$ vanishes on the
hyperplane $p_1+\dots+p_n = 0$:
\begin{equation}
  P(-ip_1,\dots , -ip_n) \de\left( p_1+\dots+p_n\right) = 0 .
\end{equation}
Now we define $\widetilde{P}(q,p_1,\dots , p_{n-1}) \defined 
P(-ip_1,\dots , -ip_n)$ with $q \defined p_1+\dots +p_n$ and consider
its Taylor expansion around the origin:
\begin{equation}
  \widetilde{P}(q,p_1,\dots , p_{n-1}) = 
  \sum_{k=1}^{{\rm degree} \widetilde{P}} \sum_{\betr{\al}+\betr{\beta}=k}
  \frac{q^\al p^\beta}{\al!\beta!}
  \left(
    \frac{\dd^{\betr{\al}}\dd^{\betr{\beta}}}{\dd q^\al \dd p^\beta}
    \widetilde{P}
  \right)
  (0)
\end{equation}
where $p \defined (p_1,\dots ,p_{n-1})$. So the derivatives 
$\frac{\dd^{\betr{\al}}}{\dd q^\al}$ describe a variation orthogonal to 
the hyperplane $p_1+\dots+p_n = 0$, the derivatives 
$\frac{\dd^{\betr{\beta}}}{\dd p^\beta}$ a variation within it. Since 
$\widetilde{P}$ vanishes throughout the entire plane, terms with 
$\betr{\al} = 0$ must vanish. Therefore the Taylor expansion can be 
rewritten as
\begin{equation}
  \widetilde{P}(q,p_1,\dots , p_{n-1}) = q\cdot
  \sum_{k=0}^{{\rm degree} \widetilde{P}-1} 
  \sum_{\betr{\al}+\betr{\beta}=k} \frac{q^\al p^\beta}{\al!\beta!}
  \left(
    \frac{\dd^{\betr{\al}}\dd^{\betr{\beta}}}{\dd q^\al \dd p^\beta}
    \widetilde{P}_1^\mu
  \right)
  (0)
\end{equation}
with a new polynomial $\widetilde{P}_1^\mu$. Reversing the Fourier 
transformation we find
\begin{equation}
  P(\dd) = \left( \sum_{i=1}^{n} \dd_\mu^i\right) P_1^\mu(\dd) 
\end{equation}
where the polynomial $P_1^\mu$ is the Fourier transform of 
$\widetilde{P}_1^\mu$. With this expression we can write the anomaly as
\begin{equation}
  a(x_1,\dots ,x_n,y) = - \dd^y_\mu
  \left(
    n\cdot P^{\mu}_{1} \de(x_1-y)\cdots \de(x_n-y)
  \right).
\end{equation}
So the anomaly can be removed by addition of 
$n\cdot P^{\mu}_{1} \de(x_1-y)\cdots \de(x_n-y)$ to the the previous
normalization of $\TT{W_1,\dots , W_n, k^\mu}(x_1,\dots ,x_n,y)$. This
is obviously a valid normalization and so the desired normalization
has been found.\\
The question remains why we had excluded polynomials with $k^\mu$ as
a sub polynomial. The reason is that we can assure that a normalization
with the desired properties exists, but we cannot assure that time ordered
products like 
\begin{equation}
  \TT{k^\mu, k^\nu, W_1,\dots,W_n}(y,z,x_1,\dots,x_n)
\end{equation}
are symmetric under simultaneous exchange of $\mu, \nu$ and $y,z$ as they 
must. There is indeed a counterexample for the Ward identities of the 
axial current $j^\mu_A = :\psiq \gamma^\mu\gamma^5\psi:$, where it is not
possible to find a normalization of $\TT{j^\mu_A,j^\mu_A,j^\mu_A}(x,y,z)$ 
with the required symmetries. Excluding the respective polynomials from
the allowed arguments makes sure that this situation does not occur. \\
From the proof above it is clear that the normalization we have found is 
compatible both with \eqref{N3} and \eqref{N4}. But we can also immediately 
see that \eqref{N5} respects Poincar\'e transformation properties and 
therefore \eqref{N1}. Taking the adjoint of \eqref{N5} finally reveals that
it complies also with \eqref{N2} and therfore eventually with all other
normalization conditions. 

\subsection{Relation between \eqref{N6} and generalized operator gauge 
invariance}\label{WIGOGI}

It is always possible to define an operator valued distribution 
$\TT{\cL_{i_1}, \dots, \cL_{i_n},j^\mu}$ by
\beql{normBRS}
  \begin{split}
    &\TT{\cL_{i_1}, \dots ,\cL_{i_n},j^\mu}(x_1,\dots ,x_n,y)  
    \defined \\
    &\qquad \defined - s_0 \TT{\cL_{i_1}, \dots ,\cL_{i_n},k^\mu}
    (x_1,\dots ,x_n,y) \\
    &\qquad \qquad + i \sum_{m=1}^{n} \dd_{\nu}^{m} 
    \TT{\cL_{i_1}, \dots ,\cL_{i_m+1}^\nu, \dots ,\cL_{i_n},k^\mu}
    (x_1,\dots ,x_n,y) \\
    &\qquad \qquad - i \sum_{m=1}^{n} \de(x_m-y) 
    \TT{\cL_{i_1}, \dots ,\cM_{i_m+1}^\mu, \dots ,\cL_{i_n}}
    (x_1,\dots ,x_n) \\
    &\qquad \qquad + i \sum_{m=1}^{n} \de(x_m-y) \cdot i_m \cdot
    \TT{\cL_{i_1}, \dots ,\cL_{i_m+1}^\mu, \dots ,\cL_{i_n}}
    (x_1,\dots ,x_n) .
  \end{split}
\end{equation}
$\TT{\cL_{i_1}, \dots, \cL_{i_n},j^\mu}$ is at this point only 
a name for that distribution, we must still prove that it is indeed an
extension of $^0\TT{\cL_{i_1}, \dots, \cL_{i_n},j^\mu}$. Before we
do that, we point out that it implicates \eqref{N6} almost immediately. 
Of course the time ordered products on the right hand side must
satisfy eqn. \eqref{N5}. Taking the derivative w.r.t.\ the $y$ coordinate,
we find with \eqref{N5}
\beql{intermediate}
  \begin{split}
    &\dd_\mu^y \TT{\cL_{i_1}, \dots ,\cL_{i_n},j^\mu\vps}
    (x_1,\dots ,x_n,y) = \\
    &\qquad 
    = - \left( \sum_{m=1}^{n} \de(x_m-y) \cdot i_m \right)
    s_0 \TT{\cL_{i_1}, \dots ,\cL_{i_n}\vps}(x_1,\dots ,x_n) \\
    &\qquad\quad\,\,
    \begin{split}
      + 
      i \sum_{l=1}^{n} \dd_{\nu}^{l}
      \left[ 
        \TT{\cL_{i_1}, \dots ,\cL_{i_l+1}^\nu, \dots ,\cL_{i_n}\vps}
        (x_1,\dots ,x_n) \times \vphantom{\sum_{l=1}^{n}}
      \right. \qquad&\\
      \left. 
        \times 
        \left( \sum_{m=1}^{n} \de(x_m-y) \cdot i_m + \de(x_l-y)\right)
      \right]&
    \end{split}\\
    &\qquad\quad\,
    + i \sum_{m=1}^{n} \left(\dd^m_\nu \de(x_m-y) \vps\right)
    \TT{\cL_{i_1}, \dots ,\cM_{i_m+1}^\nu, \dots ,\cL_{i_n}\vps}
    (x_1,\dots ,x_n) \\
    &\qquad\quad\,
    - i \sum_{m=1}^{n} \left(\dd^m_\nu \de(x_m-y)\cdot i_m \vps\right)
    \TT{\cL_{i_1}, \dots ,\cL_{i_m+1}^\nu, \dots ,\cL_{i_n}\vps}
    (x_1,\dots ,x_n) .
  \end{split}
\end{equation}
Smearing out this equation with a test function $f$ like the one defined
following eqn. \eqref{N51} gives eqn. \eqref{N5int1}, the calculation is the 
same as at the beginning of the last section. Inserting this result into eqn. 
\eqref{intermediate} we get immediately eqn. \eqref{N6}. \\
Eqn. \eqref{normBRS} is obviously a well posed definition since all 
operations involved in it are well defined --- in-particular the time 
ordered product in the last line contains no vertex at $y$, so the 
product with the delta distribution is a tensor product. \\
The crucial question is whether the operator valued distribution has
the correct causal factorization outside the diagonal $\Diag_{n+1}$. 
Only then it is really a time ordered product of its arguments as the 
notation suggests. Basically we must do the same construction as in the
respective point for \eqref{N4}, see section (\ref{proofN4}). We give
here only a simplified version of this proof where the essential point
may be more easily understood. The detailed version can easily be 
derived from this sketch.\\
Suppose the points $x_1,\dots , x_n$ are in a relative position such that 
\begin{equation}
  \emptyset \neq I=\set{x_1,\dots,x_k} \gtrsim \set{x_{k+1},\dots,x_n,y}.
\end{equation}
This is the situation we encounter in eqn. \eqref{parteta} in the first 
sum --- if $I=\set{x_1,\dots,x_k} \lesssim \set{x_{k+1},\dots,x_n,y}$, 
corresponding to the second sum there, the argument works as well.
Then 
\beql{causfact}
  \TT{\cL_{i_1}, \dots ,\cL_{i_n},j^\mu} = 
  \TT{\cL_{i_1}, \dots ,\cL_{i_k}}\TT{\cL_{i_{k+1}},\dots,\cL_{i_n},j^\mu}
\end{equation}
where we omitted the spacetime indices for simplicity. Eqn. 
\eqref{normBRS} is valid for 
$\TT{\cL_{i_{k+1}},\dots,\cL_{i_n},j^\mu}$ since we assumed that eqn. 
\eqref{normBRS} holds already for time ordered products with fewer
arguments. Together with eqn. \eqref{causfact} this gives the following 
expression
\begin{equation}
  \begin{split}
    &\TT{\cL_{i_1}, \dots ,\cL_{i_n},j^\mu} =\\
    &\qquad = -s_0\left[\TT{\cL_{i_1}, \dots ,\cL_{i_k}}
    \TT{\cL_{i_{k+1}},\dots,\cL_{i_n},k^\mu} \right]\\
    &\qquad\quad + \left[s_0 \TT{\cL_{i_1}, \dots ,\cL_{i_k}}\right]
    \TT{\cL_{i_{k+1}},\dots,\cL_{i_n},k^\mu}\\
    &\qquad\quad + \sum_{l=k+1}^{n} \dd_\nu^l
    \left[
      \TT{\cL_{i_1}, \dots ,\cL_{i_k}}
      \TT{\cL_{i_{k+1}},\dots,\cL_{i_l+1}^\nu,\dots,\cL_{i_n},k^\mu}
    \right]\\
    &\qquad\quad - i \sum_{l=k+1}^{n} \de(x_l-y) 
    \left[
      \TT{\cL_{i_1}, \dots ,\cL_{i_k}}
      \TT{\cL_{i_{k+1}},\dots,\cM_{i_l+1}^\mu,\dots,\cL_{i_n}}
    \right] 
  \end{split}
\end{equation}
where we have omitted spacetime arguments for simplicity. For 
$s_0 \TT{\cL_{i_1}, \dots ,\cL_{i_k}}$ we may use the generalized 
operator gauge invariance \eqref{GOGI} in lower orders as long as
$k\neq n$. Unfortunately also the case $k=n$ occurs if all the 
$x_i$ coincide and only $y$ is separated from them. This is the only
case where we must know in advance that \eqref{GOGI} holds. If this 
would not be true then our definition \eqref{normBRS} would be a well 
defined operator valued distribution, but no an extension of 
$T^0(\cL_{i_1}, \dots ,\cL_{i_n},j^\mu)$ to the diagonal --- this 
means that it could differ from the $T^0$-product even outside the 
diagonal. Hence we need to assume that \eqref{GOGI} is valid  
also for $k=n$. Furthermore we may add in the last
sum the terms with $l=1,\dots,k$ since the delta distributions vanish
because $y$ and the $x_1,\dots,x_k$ may never coincide. Recombining
the products of $T$-products into a single $T$-product according to 
eqn. \eqref{causfact} one gets immediately \eqref{normBRS}. As 
already remarked the calculation comes to the same result if 
$\emptyset \neq I=\set{x_1,\dots,x_k} \lesssim \set{x_{k+1},\dots,x_n,y}$.
So \eqref{normBRS} is a well defined operator valued distribution
that agrees --- as long as \eqref{GOGI} is valid --- with 
$T^0(\cL_{i_1}, \dots ,\cL_{i_n},j^\mu)$ if smeared with
a test function that vanishes with all its derivatives on the diagonal,
so it is an extension of that $T^0$-product to the diagonal and
therefore a possible normalization of 
$\TT{\cL_{i_1}, \dots ,\cL_{i_n},j^\mu}$.\\
So we have just proven that the conditions \eqref{N6} and \eqref{GOGI}
are equivalent.
